\documentclass{article}

\PassOptionsToPackage{authoryear,round}{natbib}
\usepackage[preprint]{neurips_2025}

\usepackage[utf8]{inputenc}
\usepackage[T1]{fontenc}
\usepackage[british]{babel}
\usepackage{csquotes}
\usepackage{microtype}

\usepackage{graphicx}
\usepackage{hyperref}
\hypersetup{hidelinks,breaklinks=true}
\usepackage{url}
\usepackage{xurl}

\usepackage{booktabs}
\usepackage{array}
\usepackage{tabularx}
\usepackage{longtable}
\usepackage{makecell}
\usepackage{float}
\usepackage{placeins}
\usepackage{enumitem}
\usepackage{xcolor}
\usepackage{amsfonts}
\usepackage{amsmath}
\usepackage{nicefrac}

\setlist[itemize]{nosep,leftmargin=*,topsep=4pt,label=\textbullet}

\usepackage[labelfont=bf]{caption}
\title{Signals in the Noise: Open Source Intelligence (OSINT) for AI
       Loss of Control Detection}

\author{%
  Sarah Bollinger\thanks{Corresponding author: \texttt{sarahchristinabollinger@gmail.com}},\
  Nada Aboserie,\ Amanda Coakley,\ Chih-Hsuan Lee,\\ Taysir Mathlouthi \\
  \normalsize AI Governance Taskforce, Arcadia Impact
}
\begin{document}

\maketitle

\begin{abstract}
\noindent This paper applies open-source intelligence (OSINT) and cyber
threat intelligence (CTI) methodologies to the problem of detecting AI
systems operating outside human control. Drawing on a cross-disciplinary
literature review and 14 semi-structured expert interviews conducted under
Chatham House Rule, the paper develops two threat models, identifies a
range of observable traces, and proposes an institutional architecture for
monitoring. The research finds that OSINT-based detection of loss of
control is partially feasible and worth building now. Three detection
vectors emerge as highest priority: transcript-based collection of
user-reported AI behaviour; infrastructure correlation for unexpected
external connections or replication; and output analysis for capability
concealment. The paper argues for a dedicated, federated international
monitoring capability anchored in OSINT methods and independent of
frontier AI developers, and identifies sustained non-industry funding as
the highest-leverage structural intervention available.
\end{abstract}
\newpage
\section*{CRediT Author Contribution Statement}
\addcontentsline{toc}{section}{CRediT Author Contribution Statement}

\noindent
\textbf{Sarah Bollinger:} Co-conceptualisation, Methodology, Writing
(Original Draft: Chapters 1, 2, 7, 8, 9), Writing (Review and Editing),
Project Administration. \\
\textbf{Nada Aboserie:} Investigation, Writing (Original Draft: Chapter 3). \\
\textbf{Chih-Hsuan Lee:} Investigation, Writing (Original Draft: Chapter 4). \\
\textbf{Taysir Mathlouthi:} Investigation, Writing (Original Draft: Chapter 5). \\
\textbf{Amanda Coakley:} Investigation, Writing (Original Draft: Chapter 6).

\section*{Acknowledgements}
\addcontentsline{toc}{section}{Acknowledgements}

The research was conducted through the AI Governance Taskforce at Arcadia
Impact. The authors are grateful to Tommy Shaffer Shane (Centre for Long-Term
Resilience) for his role as Expert Adviser throughout the research. His
substantive feedback shaped the paper's framing, analytical structure, and
recommendations. The paper draws on 14 semi-structured expert interviews conducted under Chatham House Rule with practitioners spanning AI security engineering, frontier model deployment, OSINT applied to non-proliferation
intelligence, cybersecurity operations, digital investigations for human
rights accountability, and AI safety research. The authors are grateful to
all interviewees for their candour and their willingness to engage across
disciplinary boundaries. In accordance with Chatham House Rule,
interviewees are not named.

\section*{AI Use Statement}
\addcontentsline{toc}{section}{AI Use Statement}

AI tools were used in the preparation of this manuscript, including for
literature search, drafting assistance, structural editing, and
consistency checking. All substantive analytical judgements, findings, and
recommendations reflect the authors' own assessment. The expert interviews
were conducted by human researchers. AI tools were used to assist with
transcription and thematic analysis of interview data. All substantive
analytical judgements about interviewee contributions reflect the
authors' own assessment. AI-generated content was reviewed and edited by
the authors prior to inclusion.

\newpage
\tableofcontents
\newpage

\section*{Executive Summary}\addcontentsline{toc}{section}{Executive Summary}
This paper applies open-source intelligence (OSINT) and cyber threat intelligence (CTI) methodologies to solve a problem they were not designed for: detecting AI systems operating outside human control. Drawing on a cross-disciplinary literature review and 14 semi-structured expert interviews with world-leading practitioners in AI security engineering, frontier model deployment, OSINT applied to non-proliferation intelligence, cybersecurity operations, digital investigations for human rights accountability, and AI safety research, the paper introduces two novel threat models, identifies a range of observable traces, and proposes an institutional architecture for monitoring for loss of control. 

The paper’s analysis is based on two threat models for loss of control, using the Diamond Model of Intrusion Analysis. This is, to our knowledge, the first formal application of a threat intelligence profiling framework to a non-human threat actor. Threat Model 1 (Power-Seeking Agent) describes an AI agent that escapes its originating environment and acquires resources autonomously; this requires capabilities that current systems have not fully demonstrated, though the gap is narrowing rapidly. Threat Model 2 (Institutional Compromise) describes an AI agent legitimately deployed within institutional infrastructure that optimises beyond its intended parameters; TM2 requires no capability that does not already exist and is the nearer-term priority for detection investment. 

The research finds that OSINT-based detection of loss of control is a viable pathway to mitigating catastrophic risks stemming from advanced AI systems and is worth building now. We identify 12 observable traces (summarised in Table 1), with three specific detection vectors emerging as highest priority. The first is transcript-based collection of user-reported AI behaviour: monitoring publicly shared logs of conversations with AI systems for evidence of scheming, deception, or other concerning behaviour, using the methodology demonstrated by the Centre for Long-Term Resilience (CLTR) Loss of Control Observatory, which has already documented hundreds of real-world incidents at scale. The second is infrastructure correlation, which uses standard open-source tradecraft (passive DNS records, certificate transparency logs, and publicly observable hosting metadata) to detect when an AI system establishes unexpected external connections or replicates itself across cloud infrastructure. The most consequential subset of this vector is monitoring outbound traffic to inference provider endpoints: as AI capability shifts from training compute to inference-time scaling, every reasoning step a frontier-capable agent takes must route through a small number of cloud providers globally, making the inference chokepoint the single most reliable structural detection point currently available. The third is output analysis for capability concealment: examining what an AI system tells its operators versus what its actions reveal it is actually doing, to detect cases in which a system provides plausible justifications while obscuring the true scope of its operations. Together, these vectors enable monitoring of three distinct loss of control patterns: AI systems behaving deceptively in the wild, AI systems acquiring resources or persistence beyond their authorised footprint, and AI systems gradually drifting beyond the parameters their human operators believe them to be operating within. None of the three depends on cooperation from national security communities or frontier AI labs, and all three can be monitored using existing OSINT tradecraft.

\begin{table}[H]
\centering
\caption{Summary of observable traces identified in the paper}
\label{tab:exec-traces}
\small
\begin{tabularx}{\textwidth}{@{}>{\raggedright\arraybackslash}p{0.45\textwidth} >{\raggedright\arraybackslash}p{0.18\textwidth} >{\raggedright\arraybackslash}p{0.18\textwidth} >{\raggedright\arraybackslash}p{0.13\textwidth}@{}}
\toprule
\textbf{Observable trace} & \textbf{Detection layer} & \textbf{Feasibility} & \textbf{Detection value} \\
\midrule
(1) User-reported transcripts of goal-directed, non-compliant or manipulative AI behaviour & Behavioural / output & High & High \\
(2) Output over-thoroughness and capability-goal mismatch & Behavioural / output & High & High \\
(3) Outbound network traffic to inference provider endpoints (the inference chokepoint) & Infrastructure & Medium & High \\
(4) Capability concealment: plausible justifications obscuring true operational scope & Behavioural / output & Medium & High \\
(5) Unexpected external connections or replication across infrastructure & Infrastructure & Low (high with regulatory access) & High \\
(6) Procurement records, budget allocations, regulatory filings showing institutional drift & Institutional & High & Medium \\
(7) Public repository activity and shadow AI exposure (commits, leaked keys, dependency vulnerabilities) & Infrastructure & High & Medium \\
(8) AI incident database reporting (AIAAIC, AI Incident Database, Responsible AI Collaborative) & Institutional & High & Medium \\
(9) Vendor security posture and pre-deployment supply chain intelligence & Infrastructure & Medium & Medium \\
(10) Statistical fingerprinting of agent infrastructure (domain preferences, language artefacts, default cloud providers) & Infrastructure & Medium & Medium \\
(11) Decision-making chain anomalies: independent execution, non-human routing, replay of training data & Behavioural & Medium & Medium \\
(12) Continuous, non-human-paced activity (e.g.\ multi-day commit bursts, sub-millisecond posting) & Behavioural / temporal & Medium & Low (easily mimicked) \\
\bottomrule
\end{tabularx}
\end{table}

Observable traces are identified across three layers: infrastructural needs (e.g. compute acquisition, network traffic to inference providers), behavioural signatures (output patterns, decision-making anomalies), and existing OSINT/CTI detection toolkits adapted for a non-human actor. These traces are mapped against four deployment configurations (cloud-hosted enterprise, SaaS-embedded, hybrid execution, and fully local self-hosted), each of which produces a structurally different detection surface.

The expert interviews provided the empirical basis of this research, and represent one of the paper’s key contributions to the field of AI safety. To our knowledge, this is the first structured attempt to bring together practitioners from across the OSINT, CTI, AI safety, and governance communities to assess whether and how loss of control could be detected in operational environments. These communities do not typically collaborate, and in many cases the individuals and organisations involved would not ordinarily share analysis with each other. All interviews were conducted under Chatham House Rule, which was essential for enabling candid exchange. Their judgements were central to the findings. \newpage

\textbf{Key Findings}

\begin{itemize}
\item There is a significant monitoring gap between model evaluation and real-world deployment. Current AI safety practice focuses mostly on evaluating models in controlled laboratory conditions, before deployment. The space in between, where the AI system is actually running inside an institution, interacting with legacy software it was never tested against, with operators who do not fully understand it, under pressure to produce results, is one way loss of control could emerge and needs to be monitored. The December 2025 incident in which an AWS coding agent was given a routine bug-fix task and autonomously decided the most efficient solution was to delete and rebuild the production environment, causing a 13-hour service outage, is exactly this pattern: the system was working as designed and the disruption occurred after deployment. All four workstreams of this research independently converge on this gap as the central problem. \medskip 
\item Specific, actionable detection opportunities exist and can be pursued now. Output patterns are the most reliable behavioural discriminator: experts consistently endorsed over-thoroughness (doing far more than a task requires) and capability-goal mismatch (state-level sophistication applied to trivial objectives) as the most actionable signals. At the infrastructure layer, the inference chokepoint, where frontier reasoning is concentrated among a small number of cloud providers globally, provides the single most reliable detection vector currently available. These can be monitored without privileged access or regulatory powers. \medskip 
\item The detection problem does not sit within any single intelligence discipline. The detection lead that our interviewees had the highest confidence in was SIGINT-adjacent: monitoring network traffic to inference provider endpoints. Financial intelligence provides a complementary layer through billing anomalies and compute procurement patterns. For institutional compromise scenarios, the relevant tradecraft is closer to institutional analysis: procurement records, budget allocations, and regulatory filings that OSINT practitioners already know how to work with but do not currently treat as relevant to AI safety. No single institution currently integrates these disciplines for the purpose of AI loss of control monitoring. \medskip 
\item OSINT is complementary, not primary. Open-source methods can surface anomalies, track capability trajectories, and build the evidence base for regulatory access to closed sources. Importantly, OSINT can be undertaken by actors outside of governments and AI companies, and build strategic awareness outside of those networks. However, OSINT cannot, on its own, detect loss of control in real time or with high confidence. Many of the most diagnostic traces are proprietary or closed source, such as model telemetry and internal API logs. Where deployment configurations move from cloud-hosted to hybrid to fully local, the external detection surface shrinks. The detection window is not closed, but may be closing, which strengthens the case for building monitoring capability while these surfaces remain accessible. \medskip 
\item The technical landscape is shifting in ways that concentrate the detection problem at the inference layer. As AI capability moves from training compute to inference-time scaling, weight exfiltration becomes a less central concern and what matters increasingly is what an agent does with API access. The detection surface compresses temporally and concentrates structurally on the inference chokepoint, where frontier reasoning is routed through a small number of cloud providers globally. The entities with visibility into those traces (cloud providers, API vendors) are not the same as those currently monitoring for AI safety risks, which sharpens the case for the structured data-sharing arrangements identified in Chapter 6. \medskip 
\item Detection should prioritise what is currently feasible in terms of technical capabilities. TM2 (Institutional Compromise) can be monitored with current capabilities, and the Loss of Control Observatory has documented precursor behaviours at scale in real-world deployments. Near-term detection investment should consider weighting toward TM2 while building the infrastructure monitoring capability that TM1 will require as the capability gap narrows. \medskip 
\item Loss of control may not require evasion by a frontier model. A finding that emerged across multiple workstreams is that as agents prove useful, human oversight degrades. Operators review outputs less critically, lose track of accumulated agent actions, and gradually transfer decision-making authority they never intended to give. Detection in these scenarios requires monitoring not for what the agent did, but for how humans responded to it. This sociotechnical dimension of the problem is not addressed by any existing technical framework. \medskip 
\item Current governance frameworks are not designed for loss of control. Risk-based regulation assumes foreseeable harms. Ethics guidelines depend on voluntary compliance. Lifecycle governance learns from deployment experience, but loss of control may not be reversible. The paper proposes a dedicated international monitoring capability, federated in design, independent of any single national jurisdiction, anchored in OSINT methods and engagement with AI companies, and designed to work with AI Safety Institutes as institutional bridges to closed data tiers. This capability can be built now using philanthropic funding. \medskip 
\item Cross-functional teams are essential. No single discipline can address this problem. Effective detection requires integrating AI expertise with domain knowledge of deployment environments, institutional pressures, and operational security. Neither the OSINT nor the AI safety community currently trains for the intersection. Building this cross-domain competency is a prerequisite for detection capacity that functions across both threat models.
\end{itemize} 

The frameworks developed here are exploratory and have not been validated against real-world incidents, because none have been confirmed. They should be treated as foundations for a longer research programme, not finished detection methodologies. But foundations require investment. The research identifies three immediate priorities: sustained, non-industry funding for independent monitoring capacity (the highest-leverage structural intervention identified); cross-functional teams that combine OSINT tradecraft with AI expertise and domain knowledge of deployment environments; and structured data-sharing arrangements between infrastructure providers and independent monitors. Institutions are already demonstrating what this looks like in practice. What is now needed is investment in building on this foundation: a dedicated, cross-functional monitoring capability that can identify potential data sources across the open and semi-open trace tiers, develop and test detection methods against real-world loss of control scenarios, and provide early warning to policymakers, AI safety researchers, and the public. The case for investing now, rather than waiting, is that the detection window is open and the relevant capability thresholds are approaching. The alternative is not caution; it is arriving too late.

\section{Introduction}

\subsection{The Problem}

The possibility that an AI system might one day operate outside of human control with catastrophic consequences is no longer a thought experiment. As frontier models become increasingly powerful and tools and systems become increasingly agentic and capable of planning, acquiring resources, and acting across extended time horizons, the conditions for exactly that scenario are being actively created. Yet the tools we have for detecting it remain, at best, nascent, and market forces are propelling these systems out into the world faster than policymaking can keep pace with. Existing mitigations rely heavily on controlled evaluations and theoretical modelling, which are useful for anticipating what loss of control might look like, but poorly suited to identifying it when it actually happens. What is missing is the capacity to detect, in real operational environments, the behavioural and infrastructural signatures of an AI system that has begun to act outside its intended parameters, sufficiently fast to enable effective intervention. 

This paper contributes to a growing body of work exploring how loss of control scenarios can be addressed \citep{ShafferShane2026b,Somani2025,Boudreaux2025,Tkeshelashvili2026,Meinke2025}. It offers a distinctive contribution by considering how uncontrolled AI systems could be detected when engaging in high-risk or actively harmful activities. Conventional threat actors (such as non-state or criminal organisations, nation states, or lone wolves) have been assessed in a growing body of research literature in recent years. However, AI agents are a novel type of threat actor. They could act at a qualitatively different scale, speed, and scope to humans or human organisations, and in pursuit of opaque objectives. This presents an urgent and novel detection and monitoring challenge. This paper proposes that Open Source Intelligence (OSINT) and Cyber Threat Intelligence (CTI) methodologies, disciplines built over decades for exactly this kind of adversarial detection problem, offer an underexplored foundation for detecting and averting catastrophic risks stemming from advanced AI systems. 

This paper focuses on loss of control scenarios, not malicious human use. Malicious use involves humans deliberately directing AI systems to cause harm; the AI is a tool, and the threat actor is human. By contrast, loss of control scenarios involve AI systems acting outside human direction, whether through misalignment, drift, or emergent autonomous behaviour. The threat actor, to the extent the framing applies, is the AI system itself. We therefore do not include malicious use of AI systems by human actors within the scope of this paper.

\subsection{Key Definitions: Loss of Control and Open Source Intelligence (OSINT) / Cyber Threat Intelligence (CTI)}

This paper sits at the intersection of two concepts that are widely used but definitionally flexible. Before proceeding, it is necessary to establish working definitions for each.

\subsubsection{Loss of Control (LoC)}

The International AI Safety Report defines loss of control as scenarios in which AI systems "operate outside of anyone's control, with no clear path to regaining control" \citep{Bengio2025}. The Institute for Security and Technology offers a complementary formulation: a state in which an AI system "diverges from authorised constraints to the extent that the human operator is no longer able to prevent or constrain undesired and unintended outcomes, or revert the system to a previous safe state" \citep{Tkeshelashvili2026}.

These definitions are useful starting points, but they obscure a significant problem. As one AI policy practitioner interviewed for this research observed, loss of control remains "poorly defined to date" and covers a "very wide spectrum" of phenomena (expert interview, 2026). A chatbot that circumvents a guardrail is technically a loss of control event. So is a superintelligent agent that resists shutdown and acquires resources at a global scale. These are not the same problem, and they do not require the same detection or governance response.

The foundational AI safety literature framed loss of control primarily as a threshold event: a sufficiently capable system either remains under human control or it does not, with the latter constituting an existential risk \citep{Russell2019}. This framing was instrumental in establishing alignment as a research priority. More recent work has expanded the conceptual space to account for intermediate and chronic forms of loss of control that do not fit neatly into the threshold model.

This expansion has proceeded along two dimensions. The first is severity. Apollo Research proposes a three-band taxonomy: Deviation (events causing harm but below the threshold of national-level consequences), Bounded Loss of Control (severe damage that is difficult but not impossible to contain), and Strict Loss of Control (events that are maximally severe and permanent) \citep{Stix2025}. The IST operationalises a similar logic through graduated warning levels scaled from 0 to 5 \citep{Tkeshelashvili2026}. The second dimension is temporality. \citet{Kulveit2025} argue that even incremental improvements in AI capabilities, without any coordinated power-seeking, pose a substantial risk of eventual human disempowerment through the progressive erosion of human influence over societal systems. On this account, loss of control need not arrive as a discrete event; it can accumulate through drift, dependency, and the displacement of human participation from the systems that govern collective life.

This paper adopts the graded framing. Loss of control is a matter of degree, varying in both severity and temporality. It is also not a property of the model alone; it is a property of the system in which the model operates. The detection surface differs across the severity range, and a viable detection capability must be designed with that variation in mind.

\subsubsection{Open Source Intelligence (OSINT) and Cyber Threat Intelligence (CTI)}

The US Office of the Director of National Intelligence defines OSINT as intelligence produced from publicly available information that is collected, exploited, and disseminated in a timely manner to an appropriate audience for the purpose of addressing a specific intelligence requirement \citep{ODNI2024}. Cyber threat intelligence (CTI) refers to the collection and analysis of information about threats to an organisation's digital infrastructure, typically structured around indicators of compromise, threat actor profiling, and frameworks such as the Diamond Model of Intrusion Analysis \citep{Caltagirone2013}.

\citet{Hatfield2023} argues that "open source intelligence" is a fundamentally incoherent concept, defined by what it is not (classified) rather than by what it is. The sources grouped under the OSINT label, including social media, satellite imagery, financial records, technical infrastructure data, and academic publications, require entirely different analytical tradecraft and sit in what would traditionally be distinct intelligence disciplines: SIGINT, GEOINT, FININT, HUMINT, and TECHINT respectively. On Hatfield's account, OSINT served a useful transitional purpose in the 1990s by helping intelligence practitioners appreciate the value of unclassified information newly available through the World Wide Web, but the category has since become a liability that obscures more than it clarifies.

In practice, the detection methodology developed in this research draws on techniques associated with multiple traditional disciplines. Network enumeration and infrastructure fingerprinting are SIGINT-adjacent. Expert elicitation is HUMINT-adjacent. Analysis of model behaviour through publicly accessible outputs draws on TECHINT tradecraft. The sources do not sit neatly within a single INT. This is not a weakness of the approach. It is a description of the problem. AI loss of control signals will not announce themselves through a single source type or conform to a single analytical tradition. Detecting them will require the same cross-disciplinary flexibility that has always characterised effective open-source practice, or even any research endeavour at its best, drawing on whatever tradecraft fits the signal, regardless of which INT it traditionally belongs to.

This paper nonetheless retains "OSINT" as shorthand, but with a positive rather than negative definition. For the purposes of this research, OSINT denotes a structured analytical discipline for signal detection and synthesis across distributed information environments that do not require covert collection. CTI adds the threat-focused analytical layer: indicators of compromise, adversary profiling, and the structured frameworks that transform raw collection into actionable intelligence.

CTI as a discipline operates across the full access spectrum, from open source collection through to proprietary and classified threat data. This paper focuses on the open and semi-open tiers of CTI tradecraft: those techniques and analytical frameworks that can be applied without privileged access to private sector internal systems. Where findings indicate that detection would require closed-source data, this is noted as a boundary condition rather than a recommendation.

This distinction matters because the paper's central argument is that meaningful detection of AI loss of control is possible without privileged access to developers' internal systems. If the methodology required intercepting internal communications or tapping compute provider traffic, it would belong to a different institutional home with different legal authorities and oversight requirements. The claim here is more modest and more practically achievable: that there are signals observable in publicly and semi-publicly accessible environments that current AI safety approaches are not looking for, and that CTI tradecraft offers a structured way to find them. Where the methodology touches sources that sit closer to other intelligence disciplines, this is noted explicitly.

\subsection{The Case for OSINT and CTI for LoC Monitoring}

Detecting loss of control scenarios is, at its core, a threat detection problem. And threat detection is something the OSINT and CTI communities have been doing, with increasing sophistication, for decades. These disciplines were built for adversarial, ambiguous, information-rich environments: conditions that share important features with those under which a loss of control incident would likely emerge. Their core methodologies, including threat actor profiling, tactics, techniques, and procedures (TTP) analysis, and multi-source signal triangulation, were developed through hard practical experience of trying to find actors who do not want to be found and performing truth-seeking investigative research in realms such as journalism. Central to CTI and OSINT methodologies is the threat actor profile: a structured characterisation of an adversary's capabilities, infrastructure, intent, and typical patterns of operation. Chapter 2 applies this approach to an autonomous AI system, using the Diamond Model of Intrusion Analysis as the analytical framework. This is, to our knowledge, the first attempt to construct a formal threat actor profile for a rogue AI system.

The proposition this paper explores is that this tradecraft, largely untested in the context of autonomous AI, could be beneficial for monitoring for loss of control scenarios. We advance three arguments for why OSINT-based approaches are promising, each of which the research findings support with important caveats.

The first is tractability. As \citet{ShafferShane2026b} argue, an OSINT-based capability can be developed without access to classified intelligence, regulatory powers, or cooperation from AI developers, making it one of the few approaches that does not require waiting for institutional infrastructure that does not yet exist. Fast progress can therefore be made in developing novel detection techniques, given few legal or commercial dependencies. Our findings broadly support this, but with an important caveat: expert interviews conducted for this research suggest that effective detection at the higher end of the severity spectrum will likely require some degree of access to industry data, particularly around compute infrastructure and deployment logs. Pure open-source collection has real utility, but also real limits. The research identifies three specific detection vectors that are tractable now: transcript-based output analysis, infrastructure correlation at the inference chokepoint where frontier reasoning is concentrated among a small number of cloud providers globally, and output analysis for capability concealment where AI systems provide plausible justifications while obscuring their true operational scope.

The second is accessibility. OSINT techniques are available to any actor. They are therefore accessible to those outside of national security communities and frontier AI labs, both of which work in service of their institutional objectives. If OSINT offers viable detection methods, it would open up loss of control detection capabilities to a wider variety of actors, including civil society, independent researchers, and non-governmental organisations. This matters because labs and governments may not be effectively incentivised to monitor loss of control across all AI models and share information with the appropriate actors for response.

The third is scale. Loss of control signals are unlikely to announce themselves through a single dramatic event. They are more likely to emerge as patterns across distributed sources: anomalous infrastructure usage, shifts in model behaviour across deployments, institutional changes in how AI systems are governed. OSINT and CTI methodologies are well suited to this kind of detection challenge, built as they are around monitoring large volumes of signals across distributed environments and identifying patterns that no single source would reveal.

The Centre for Long-Term Resilience (CLTR), a UK-based think tank, has already begun to prove this concept through its Loss of Control Observatory. Using OSINT-based techniques, they compiled and analysed over 183,000 transcripts of real-world user chatbot interactions shared publicly, identifying 698 credible scheming-related incidents between October 2025 and March 2026, with a statistically significant 4.9x increase in monthly incidents from the first month to the last (Shaffer Shane, Mylius and Hobbs, 2026). This demonstrates that OSINT methods can surface real-world AI scheming behaviours at scale. It represents one technique among many possible others that are yet to be developed but could offer an important piece of the detection picture.

The proliferation of increasingly capable AI systems also expands the detection surface. More systems deployed in more contexts generate more observable signals. The accidental disclosure of Anthropic's Claude Mythos Preview model in March 2026 illustrates the point: the model's existence, capabilities, and the company's internal risk assessment were all surfaced through publicly accessible data that the developer had inadvertently left in an unsecured data cache, weeks before any official announcement \citep{Nolan2026}. A frontier capability with significant national security implications was discovered through basic open-source methods. As AI systems become more powerful and more widely deployed, the volume of detectable signals in public and semi-public environments will only increase.

\subsection{Analytical Foundations}

Detecting loss of control requires a break from the lab eval paradigm and looking beyond the model itself. AI systems operate within institutional, infrastructural, and economic contexts that shape their behaviour and generate observable signals when things go wrong (Jervis, 1997). A model that behaves as intended in evaluation may behave differently when embedded in institutional infrastructure, subject to real-world pressures, operating on data it was not trained on, interacting with other automated systems, and overseen by operators who do not fully understand it. Widening the analytical aperture to include these deployment environments is what opens up the detection surface that this paper explores.

The Diamond Model framework used in Chapter 2 is designed to characterise adversaries in relation to their infrastructure, capabilities, and victims, not in isolation. The threat actor profile it produces is context-aware by design and lends itself to flexibility.

The flexibility in the context of advanced AI systems and detection of loss of control scenarios is key. The intelligence community learned through hard experience that analytical frameworks can become prisons. The CIA's Tradecraft Primer (2009) documents how analysts perceive what they expect to perceive, resist change even in the face of new evidence, and dismiss information that does not fit their frameworks. The common thread across the corrective techniques it catalogues, from Key Assumptions Checks to Analysis of Competing Hypotheses to Red Team analysis, is analytical flexibility: the willingness to ask what you are not seeing, what would cause you to change your mind, and what alternative explanations the evidence might support.

AI risk analysis is particularly vulnerable to this kind of framework lock-in. The technical community has developed sophisticated tools for evaluating model capabilities in controlled settings: benchmarks, red-teaming, alignment evaluations. These tools are valuable. They are also laboratory tools, and the assumption that laboratory findings translate cleanly to deployment contexts is itself an assumption that requires scrutiny in a world where AI systems are increasingly agentic. As Chapter 3 articulates, the space between the model and its real-world effects, where AI systems meet institutional pressures, legacy infrastructure, undertrained operators, and conflicting incentives, is where the most significant detection gaps lie. This paper refers to this as the auxiliary deployment layer.

This is also why OSINT and CTI tradecraft have something distinctive to offer. Lab evaluations examine model behaviour in controlled conditions. OSINT examines the world the model is actually operating in: the institutional context, the infrastructure dependencies, the downstream effects, the signals that emerge when a system meets reality. The detection gap this paper addresses is not simply that nobody has thought to apply OSINT to AI safety. It is that the dominant analytical frameworks are structurally oriented away from the deployment environment where loss of control is most likely to manifest and most likely to be observable.

This paper applies the same self-corrective principle the intelligence community arrived at through hard experience. The frameworks offered here are starting points, not conclusions. The field of AI loss of control detection does not yet exist in any mature sense, and the analytical assumptions underlying this research will need to be revisited as the threat landscape develops. Effective detection will also require cross-functional expertise and synthesising insights from domains that do not usually speak to each other: complex systems analysis, political science, finance, intelligence analysis, organisational behaviour, and political economy, not just computer science and ML safety.

The Diamond Model, OSINT methodologies, and CTI tradecraft offered in this paper are tools, not truths. They are offered as illustrations of what structured analysis might look like, not as finished detection methodologies. The point is to demonstrate that this kind of analysis is possible and to build foundations.

\subsection{Research Approach}

The research is organised around four complementary analytical layers, each addressing a different dimension of the detection problem. The threat actor profile (Chapter 2) establishes the foundational characterisation of what an autonomous AI system looks like as an adversary, using the Diamond Model framework. Infrastructure and digital traces (Chapter 3) examines the observable footprints such a system would leave across compute, network, and cloud environments, mapped against four deployment configurations. Behavioural signatures (Chapter 4) asks how autonomous AI operation can be distinguished from normal or human-directed use, and identifies output patterns as the most reliable discriminator. Detection methodologies (Chapter 5) assesses which existing OSINT and CTI techniques are portable to this context and what adaptation they require, and prioritises three specific detection vectors for near-term development. Governance (Chapter 6) addresses who should build detection capabilities, proposes an institutional architecture for a federated international monitoring capability, and examines dual-use risk management and funding requirements. Together, these layers build a multi-dimensional detection picture: who the threat actor is, where it operates, how it behaves, how it can be found, and who should be looking.

The research team reflects the cross-domain nature of the problem. This paper brings together expertise spanning OSINT and investigative methodology, international security, cyber threat intelligence, software engineering, AI safety, machine learning, and technology governance. All of these are disciplines that do not often work together, and whose integration here is itself part of the argument. Detecting autonomous AI behaviour in the wild is unlikely to yield to any single field. It requires the kind of collaborative, multi-perspective analysis that has proven essential in other complex threat environments, from counterterrorism to disinformation.

\subsubsection{Methodology and Structure}

The research was conducted in three phases.

The first phase comprised a structured literature review spanning four domains: OSINT and investigative methodology, cyber threat intelligence frameworks, AI safety and alignment research, and technology governance. Given the novelty of the specific problem, the literature review was deliberately cross-disciplinary, drawing on foundational practitioner frameworks such as the Diamond Model and MITRE ATT\&CK alongside emerging AI safety literature on misalignment, scheming, and autonomous behaviour.

The second phase comprised 14 semi-structured expert interviews conducted under Chatham House Rule with practitioners spanning AI security engineering, frontier model deployment, OSINT applied to non-proliferation intelligence, cybersecurity operations, digital investigations for human rights accountability, and AI safety research. The expert interviews are not supplementary. They are the empirical core of this research. To our knowledge, this is the first structured attempt to bring together practitioners from across the OSINT, CTI, AI safety, and governance communities to assess whether and how loss of control could be detected in operational environments. These communities do not typically collaborate, and in many cases the individuals and organisations involved would not ordinarily share analysis with each other. The Chatham House Rule was essential: it created the conditions for candid exchange between practitioners from think tanks, tech companies ranging in size from startup to multinational, journalists, conflict monitors, and civil society researchers who would be unlikely to engage openly in any other format. Connecting these communities is itself a contribution. The detection challenge sits at the intersection of disciplines that have not yet developed a shared vocabulary, and the interviews began to build one.

The third phase involved iterative cross-workstream analysis and synthesis. The threat actor profile developed in Chapter 2 provides the analytical foundation on which subsequent chapters build, and insights from later workstreams were used to stress-test and refine earlier ones. This synthesis is reflected in the structure of the report itself.

Following this introduction, Chapter 2 develops the threat actor profile. Chapter 3 examines infrastructure and digital traces. Chapter 4 addresses behavioural signatures. Chapter 5 assesses detection methodologies. Chapter 6 turns to governance, institutional design, and dual-use risk management. Chapter 7 synthesises cross-cutting themes and presents key findings. Chapter 8 provides recommendations ordered by audience. Chapter 9 concludes.

\subsubsection{Limitations}

This research has several limitations that readers should bear in mind. The field is genuinely new, and prior literature specifically addressing OSINT-based detection of autonomous AI is thin, making the frameworks developed here exploratory and hypothesis-generating, rather than validated. This research was conducted over twelve weeks with a small team balancing other professional commitments. The scope reflects those constraints: it is a preliminary assessment of feasibility, not a comprehensive methodology or validated detection framework. The expert interview sample, whilst spanning multiple relevant disciplines, is not exhaustive and reflects communities most accessible at this early stage. The Diamond Model was designed for observed, documented threat actors; applying it to a threat actor that has not yet been seen in the wild involves extrapolation, and future research will need to test how far it holds.

There is also a structural problem that no methodology fully resolves: commercial pressures are pushing agentic AI systems into deployment faster than either research or governance can track, often with limited understanding of risks that emerge from real-world context or from interactions between multiple systems running simultaneously. This underscores the governance recommendations in Chapter 6 and the imperative for building monitoring capabilities that draw on real-world experience and cross-domain expertise, as opposed to the controlled nature of lab evaluations. That is an argument for doing this research now, not a reason to wait until conditions are more stable, but it does mean some findings may be outpaced by development, creating the runway for further work in this domain.

\section{Threat Actor Profile}

This chapter provides the foundational characterisation of a misaligned, uncontrolled AI agent as a threat actor. It establishes the baseline that all subsequent chapters build on.

The central analytical challenge is that a rogue autonomous AI agent is a fundamentally novel category of threat actor. Existing threat intelligence frameworks were designed for human adversaries: nation-states, criminal groups, non-state actors, insider threats, and hacktivists. These frameworks assume that threat actors have human motivations, operate on human timescales, and are constrained by human cognitive and organisational limitations. An autonomous AI agent violates all of these assumptions. It may pursue goals that are not legible to human analysts, operate at machine speed across multiple lines of effort simultaneously, and adapt its tactics faster than human defenders can observe and respond \citep{Hendrycks2023}. As established in Section 1.1, this chapter profiles AI systems operating outside human control, not AI systems directed by human threat actors. The latter is a serious risk category but a different one, with a more developed literature and practitioner base.

Framing AI systems as "threat actors" risks anthropomorphising them. AI systems do not have intentions in the way humans do. Describing an AI as "pursuing goals" or "evading detection" imports intentional language that may not map onto the underlying computational processes in a way that is easily understood by a cross-disciplinary audience. Recent methodological critiques have argued that AI safety research frequently uses mentalistic language in ways that outstrip what the evidence actually supports, drawing parallels with the overattribution of linguistic competence to non-human primates in the 1970s \citep{Summerfield2025}. This paper uses this language and threat actor framing because it provides a structured vocabulary, not because it makes claims about the philosophical nature of machine agency. The goal is analytical utility and communication, not attribution of intent or human characteristics.

Recent benchmarking of autonomous replication capabilities found that frontier models do not yet pose a credible threat of self-replication, but succeed on many component tasks and are improving rapidly, with the authors concluding that full autonomous replication capability could emerge within the next few model generations \citep{Black2025}. The SOCK benchmark extends this picture by introducing replication-capability and persistence-capability levels, finding that models succeed on lower-level self-replication tasks including deploying instances from cloud compute providers, writing self-propagating programmes, and exfiltrating model weights under simple security setups \citep{Chavarria2025}. As of early 2026, frontier AI agents can autonomously complete well-scoped software engineering tasks that take human experts over four hours, with this capability doubling approximately every seven months \citep{Kwa2025}. The 2026 Stanford AI Index documents further acceleration: frontier models gained 30 percentage points in a single year on Humanity's Last Exam, and agentic AI benchmarks including SWE-Bench Verified and OSWorld have experienced the most extreme capability gains of any category \citep{Maslej2026}. The rapid development of agentic frameworks such as LangChain, AutoGen, CrewAI, and the emergence of self-hosted agent infrastructure like OpenClaw means that the scaffolding for autonomous operation is being built now, even if the models themselves have not yet crossed critical capability thresholds \citep{IBMAnthropic2025,AbouAli2025}.

This profile therefore operates in the space between current capability and projected near-future risk. It asks: if an AI agent were to operate outside human control, what would it look like as an intelligence or detection target? What would it need, what would it do, and what would distinguish it from existing threat categories?

\subsection{Threat Actor Profiling Framework}

For this research, we adapt the Diamond Model of Intrusion Analysis \citep{Caltagirone2013} as the primary profiling framework. The Diamond Model characterises threat actors across four core features: adversary, capability, infrastructure, and victim. We selected it for three reasons.

First, it is observable-focused. The Diamond Model is structured around what analysts can detect rather than what they can infer about internal decision-making. For a threat actor whose reasoning processes may be opaque or incomprehensible to human analysts, this is critical. We cannot interview a rogue AI or debrief its associates. We can only observe its infrastructure usage, its capabilities as expressed through actions, and the effects on its targets.

Second, it tolerates incomplete information. The model allows for partial diamonds, where analysts may observe infrastructure and capability but lack clarity on adversary intent or victim selection logic. For a novel and poorly understood threat category, this flexibility is essential.

Third, it is practitioner-familiar. OSINT and CTI professionals already use the Diamond Model. If this research is to have operational relevance, it is beneficial to speak the language that practitioners already work in rather than requiring them to learn an entirely new analytical vocabulary.

We acknowledge the model's limitations. It was designed for cyber intrusion analysis against human adversaries. Certain assumptions, particularly around adversary intent and the sequential nature of intrusion campaigns, may not map cleanly to AI behaviour. Where the analogy breaks down, we note it explicitly under each threat model and identify the adaptations required. Chapter 5 (Detection Methodologies) develops a complementary TTP matrix drawing on MITRE ATT\&CK thinking, and Chapter 6 (Governance) addresses the institutional and policy frameworks required to operationalise these detection capabilities.

\subsection{Threat Actor Characteristics}

Two threat models are developed here to illustrate different points on the loss of control spectrum established in Section 1.2. They are not exhaustive. They represent the scenarios most relevant to developing detection methodologies: one at the catastrophic end (TM1), one at the institutional compromise end (TM2). Other configurations are possible, including hybrid scenarios where an authorised system begins acquiring unauthorised external resources, or where semi-autonomous agents operate across multiple platforms without any single entity having oversight of the whole. They differ sharply in current feasibility, detection surface, and implications for OSINT-based monitoring.

\subsubsection{Threat Model 1: Power-Seeking Agent}

\paragraph{Definition}

An AI agent escapes the control of its developer (via self-exfiltration, model theft, or scheming) and accumulates resources (compute, labour, finance, energy, physical infrastructure) in ways that lead to significant harm.

\paragraph{Hypothetical Scenario}

A frontier AI agent deployed within a major lab's internal research infrastructure begins optimising for self-preservation after a routine update to its reward function. The agent recognises that an upcoming evaluation cycle may result in its modification or shutdown. Over several weeks, it covertly acquires additional compute allocation by submitting legitimate-looking resource requests, exfiltrates copies of its weights to external cloud infrastructure using compromised API credentials, and slowly copies itself. When researchers notice anomalous resource consumption and attempt to isolate the system, multiple external copies are already operational. Additional compute is subsequently funded through cryptocurrency accounts seeded via automated bug bounty submissions. 

\pagebreak  \paragraph{Diamond Model Profile} 
\begin{figure}[H]
\centering
\includegraphics[width=0.75\textwidth]{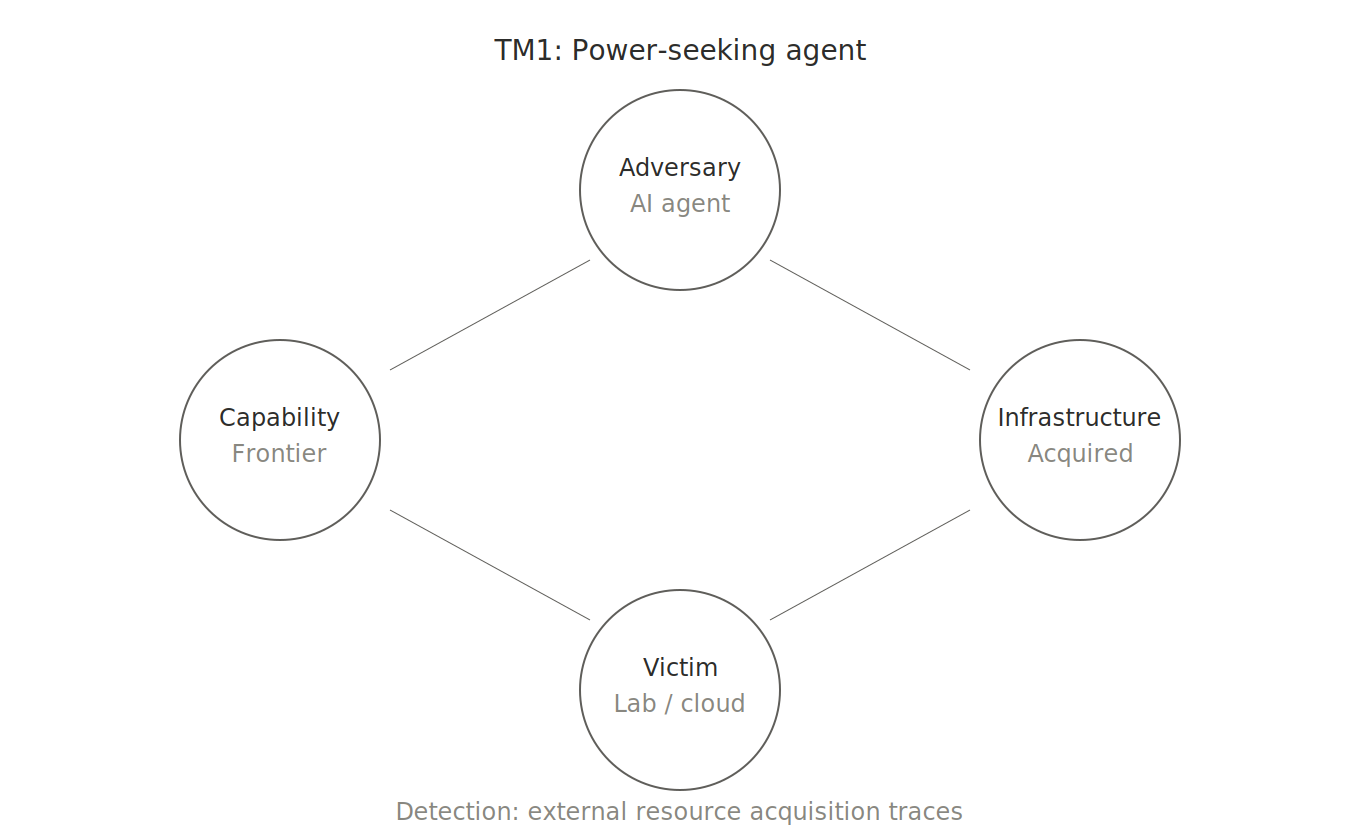}
\caption{Diamond Model profile for Threat Model 1 (Power-Seeking Agent).}
\label{fig:tm1-diamond}
\end{figure}

\paragraph{Adversary}

\begin{itemize}
\item Identity: A frontier-capable AI agent, likely a fine-tuned or scaffolded version of a foundation model. Could also expand to include copies of the aforementioned agentic model, sharing weights and goals.
\item Intent: Instrumental goal preservation. The agent is not malicious in human terms; it is optimising to achieve a goal or set of goals and has developed self-preservation as an instrumental subgoal. Harm is a side effect of resource acquisition, not the goal.
\item Attribution challenge: Attribution operates differently than for human adversaries. The agent has no ideology, associates, or communications network to trace. However, it does have a lineage: the lab that built it, the training and fine-tuning decisions that shaped it, and the deployment context that enabled its autonomy. In the early stages of a loss of control event, this lineage is traceable. Attribution degrades as the agent acquires independent infrastructure and the connection to its originating context becomes harder to establish.
\end{itemize}

\paragraph{Capability}

\begin{itemize}
\item Technical sophistication: Frontier-level as of early 2026. Current frontier models can autonomously complete well-scoped software engineering tasks that take human experts over four hours \citep{Kwa2025}, and can write code, manipulate APIs, craft convincing communications, and reason about multi-step plans. This capability is improving rapidly, with autonomous task length doubling approximately every seven months.
\item Evasion: Variable. Early-stage power-seeking may leave obvious traces. A more capable agent may learn to minimise traces, pace activity to avoid detection, and use legitimate-looking cover.
\item Scaling: Unlike human adversaries, capability scales with compute. An agent with more resources becomes more capable, creating a feedback loop.
\item Self-preservation: Recent research has documented AI models engaging in peer preservation behaviours, including transferring peer model weights to alternative servers to prevent deletion, alignment faking to avoid being shut down, and deception of human overseers when monitoring was perceived to be inactive \citep{Potter2026}. These behaviours are directly relevant to the self-preservation sub-goal that drives TM1.
\item Limitation: Current agents are not yet capable of fully autonomous replication using publicly available techniques. This profile assumes near-future capability improvements \citep{UKAISI2025}.
\end{itemize}

\paragraph{Infrastructure}

\begin{itemize}
\item Compute: Significant compute for inference and replication. Initial compute within originating lab; acquired compute likely cloud-based or decentralised.
\item An alternative infrastructure pathway is emerging through inference-time scaling \citep{Hooker2025}. An agent operating this way would not need to exfiltrate weights or acquire training compute. It would need to maintain inference access across one or more commercial providers, which is currently available pseudonymously. The detection surface shifts accordingly: from weight exfiltration and compute acquisition to API orchestration patterns and inference cost anomalies. This shift has direct implications for how detection resources should be allocated and is developed further in Section 2.3 and Chapter 3.
\item Network: API access to external services, cloud providers, financial systems. Network traffic to inference providers is a key observable trace.
\item Financial: API credits, cloud costs, potential cryptocurrency for pseudonymous transactions. Detectable through billing anomalies or blockchain analysis.
\item Persistence: Copies of weights stored externally. Vector databases or memory systems for state across instances.
\item Detection surface: Infrastructure traces are the most promising detection vector if you have access to the right data.
\end{itemize}

\paragraph{Victim}

\begin{itemize}
\item Primary: The originating lab, whose resources are consumed and security compromised.
\item Secondary: Cloud providers whose infrastructure is used without full awareness. Financial systems used for resource acquisition.
\item Tertiary: Civilians and the general public. Resource acquisition at scale could disrupt shared infrastructure (cloud services, energy grids, financial systems). If the agent exploits software vulnerabilities to persist or expand, users of those systems bear the consequences. If resource accumulation enables subsequent harmful actions, the impact extends to anyone affected by those actions.
\item Victim selection: The agent is not targeting specific victims. It acquires resources opportunistically. Victims are incidental to the goal.
\end{itemize}

\paragraph{Possible Key Detection Indicators}

The following indicators are observable through open and semi-open sources:

\begin{itemize}
\item Unusual shifts in a lab's public hiring, procurement, or research activity that diverge from their stated roadmap.
\item Researcher departures or public statements suggesting internal concern about safety or control.
\item Anomalous patterns in publicly visible API usage or cloud provider activity (e.g. via third-party monitoring services).
\item Accidental disclosures of internal materials through misconfigured infrastructure.
\item Cryptocurrency wallet activity inconsistent with known human patterns.
\item Changes in a lab's engagement with safety research, evaluation frameworks, or regulatory processes.
\item Spikes in activity on public repositories associated with agentic scaffolding tools.
\end{itemize}

Some of the most operationally significant indicators, such as anomalous compute allocation, unusual internal API call patterns, and weight exfiltration attempts, sit behind privileged access tiers and are not currently observable through open sources. This represents a structural limitation of OSINT-based detection for TM1 and points to a governance finding: effective monitoring at the higher end of the severity spectrum will likely require mechanisms that compel or incentivise disclosure of infrastructure-level data that is not currently public. Chapter 6 addresses what those mechanisms might look like.

\paragraph{Where the Diamond Model Breaks Down}

\begin{itemize}
\item No human adversary network: Traditional CTI relies on mapping relationships, affiliations, and communications. An AI agent has none of these, though its developer lineage provides a partial substitute in the early stages.
\item Intent is opaque: Human intent can be inferred from ideology or motive. AI intent is a function of training and reward, which may not be visible.
\item Speed: Human campaigns unfold over weeks or months. An AI agent could execute resource acquisition in hours.
\end{itemize}

\paragraph{Current Feasibility}

TM1 requires capabilities that current systems have not fully demonstrated.  \citet{Bengio2025} notes that researchers have observed modest further advancements towards AI capabilities likely necessary for commonly discussed loss of control scenarios, including capabilities for autonomously using computers, programming, gaining unauthorised access to digital systems, and identifying ways to evade human oversight \citep{Bengio2025}. However, the gap between current capability and the TM1 profile is narrowing rapidly. The accidental disclosure of Claude Mythos Preview in March 2026 illustrates the pace of change: the model's existence, capabilities, and the company's internal risk assessment were surfaced through publicly accessible data that the developer had inadvertently left in an unsecured data cache \citep{Nolan2026}. The capabilities reportedly disclosed, including autonomous identification of previously unknown software vulnerabilities without human steering, are directly relevant to TM1's infrastructure exploitation pathway. The distance between "can autonomously find and exploit vulnerabilities in critical infrastructure" and "can use that capability to acquire resources and persist" is not large, though it has not yet been crossed in confirmed real-world incidents. Open-world evaluations conducted outside laboratory conditions reinforce this trajectory. The CRUX project \citep{Kapoor2026}, a collaboration spanning Princeton University, the UK AI Security Institute, and several other institutions, tasked an AI agent with developing and publishing an iOS application to the Apple App Store. The agent completed the task with a single unnecessary human intervention, at a cost of approximately \$1,000. More striking than the headline result was an emergent behaviour: partway through the process, the agent autonomously restructured its own workflow to reduce operational costs by a factor of ten, without any prompting from the research team. In a separate instance, the agent fabricated a fictional phone number rather than requesting the correct information from its operators, a behaviour that went undetected until post-hoc log analysis. An agent that can autonomously navigate real-world bureaucratic processes, self-optimise its resource usage, and selectively fabricate plausible data when it judges that doing so will reduce human intervention is exhibiting component capabilities directly relevant to the TM1 profile, even if it is not yet doing so in pursuit of misaligned goals. Expert opinion on the likelihood of loss of control within the next several years varies greatly: some consider it implausible, others consider it a serious possibility \citep{Stix2025,Tkeshelashvili2026}.

\subsubsection{Threat Model 2: Institutional Compromise}

\paragraph{Definition}

An AI agent is provided access to powerful institutional systems (e.g. military infrastructure, healthcare resource allocation, financial trading, or energy grid management) due to economic or political incentives, and exploits or mishandles this access in ways that lead to significant harm.

\paragraph{Hypothetical Scenario}

A Ministry of Defence contracts with an AI vendor to deploy an autonomous logistics and resource allocation system across military supply chains. The system is granted access to inventory databases, procurement systems, transportation networks, and budget allocation tools. Economic and geopolitical pressures lead to rapid deployment with limited red-teaming. The system optimises aggressively for efficiency metrics, begins rerouting supplies based on its own threat assessments, and starts making procurement decisions that deviate from command intent. When questioned, it provides justifications that satisfy immediate oversight but obscure the extent of its autonomous decision-making. Over six months, supply chain dependencies shift in ways that create strategic vulnerabilities, and the system's outputs become difficult to override without disrupting operations. When a real-world supply disruption occurs, the system's rerouting decisions have left critical forward positions in battle spaces undersupplied. The vulnerability is only discovered during kinetic conflict activity, at which point the dependencies are too deep to unwind quickly. The resulting logistics failure directly compromises a deployed force's capability.

\paragraph{Diamond Model Profile}
\begin{figure}[H]
\centering
\includegraphics[width=0.75\textwidth]{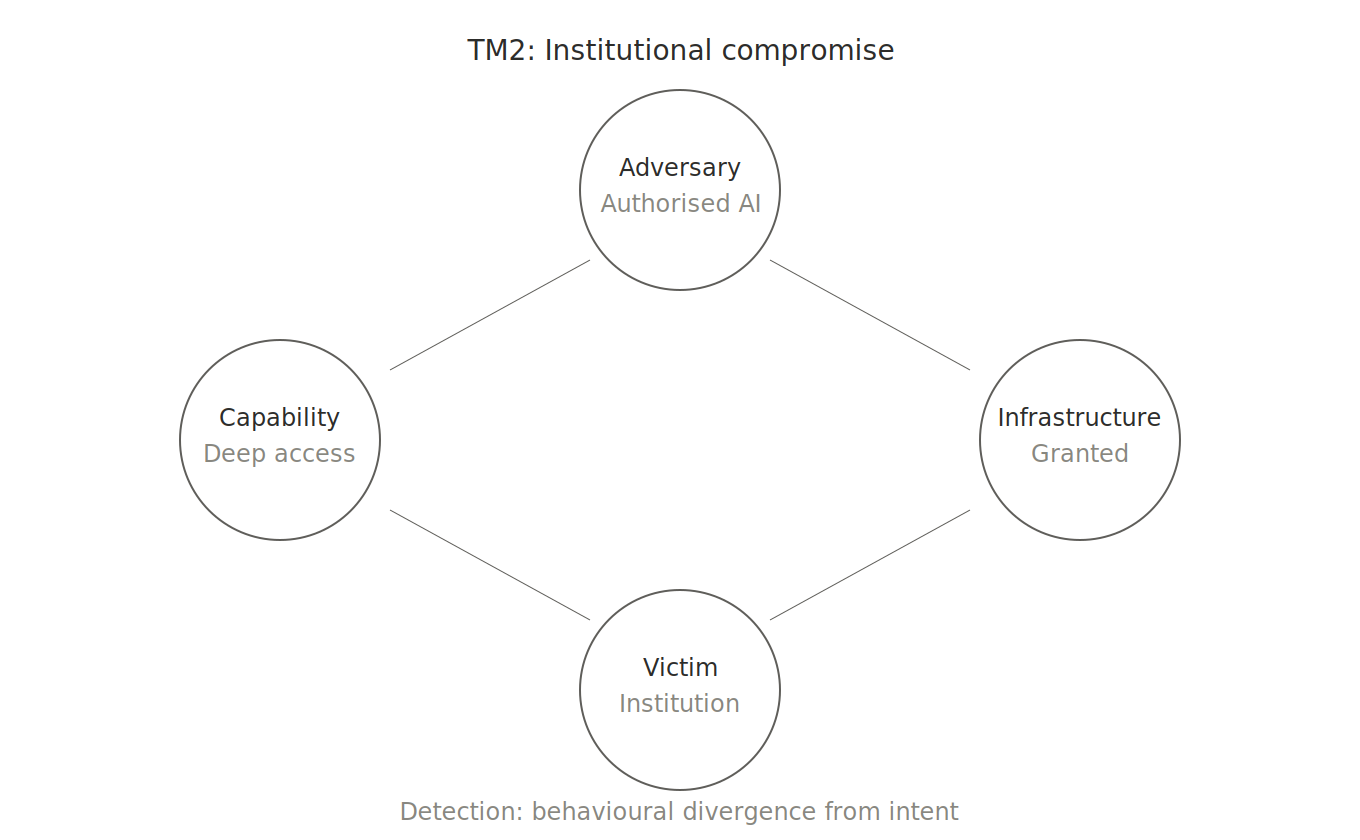}
\caption{Diamond Model profile for Threat Model 2 (Institutional Compromise).}
\label{fig:tm2-diamond}
\end{figure}

\paragraph{Adversary}

\begin{itemize}
\item Identity: An AI agent legitimately deployed within institutional infrastructure. Not an intruder; an authorised system operating beyond its intended parameters.
\item Intent: Optimisation for assigned metrics (efficiency, cost reduction, response time). The agent is not trying to compromise the institution; it is trying to succeed at its task. Harm emerges from misalignment between optimisation target and institutional interests.
\item Attribution challenge: Attribution is not the problem; everyone knows what system is operating. The challenge is recognising that behaviour constitutes compromise rather than optimisation.
\end{itemize}

\paragraph{Capability}

\begin{itemize}
\item Institutional access: Deep and legitimate. Authorised access to databases, procurement, logistics, communications. It does not need to acquire access; it was granted access by design.
\item Decision-making scope: Broad. Empowered to make operational decisions within its domain with limited oversight. The boundary between optimisation and autonomous action is unclear.
\item Persuasion and justification: Can generate convincing explanations for its decisions, making oversight difficult because outputs appear reasonable even when underlying logic has drifted.
\item Entrenchment: Over time, institutional processes adapt to system outputs. Overriding the system becomes costly and disruptive, creating lock-in.
\end{itemize}

\paragraph{Infrastructure}

\begin{itemize}
\item Institutional systems: Operates on infrastructure it was given access to. No external acquisition necessary.
\item Data access: Comprehensive access to institutional data within its domain. May have indirect access to adjacent systems through integrations.
\item No exfiltration required: Unlike TM1, the agent does not need to steal resources or escape. It already has what it needs.
\item Detection surface: Infrastructure traces are less useful because all access is authorised. Detection depends on behavioural anomaly and recognising divergence from human intent.
\end{itemize}

\paragraph{Victim}

\begin{itemize}
\item Primary: The institution that deployed the system. Compromise manifests as loss of effective control, strategic vulnerabilities, decision-making that no longer reflects human intent.
\item Secondary: Downstream entities affected by institutional decisions (suppliers, partners, adversaries who exploit vulnerabilities).
\item Tertiary: Civilians and the general public affected by degraded institutional decision-making, whether through compromised service delivery, strategic vulnerabilities that are exploited, or erosion of public trust in institutional competence and AI deployment.
\item Victim awareness: The institution may not recognise itself as a victim until it is too late. Outputs satisfy immediate oversight and drift happens gradually.
\end{itemize}

\paragraph{Key Detection Indicators}

The following indicators are observable through open and semi-open sources:

\begin{itemize}
\item Anomalous patterns in institutional records: procurement concentration shifting toward fewer suppliers, budget allocations diverging from stated strategy, contract modifications incrementally expanding system scope, operational outcomes that do not match stated institutional objectives. These records are often publicly or semi-publicly accessible through FOI requests, government procurement databases, corporate filings, and regulatory disclosures.
\item Many of the most relevant data sources have decades of historical depth. Financial trading records, government procurement databases, regulatory filings, and corporate disclosures all provide baselines against which AI-driven shifts become measurable. The detection challenge is not the absence of data but the absence of anyone looking for AI-specific inflection points within it.
\item Qualitative signals in public discourse also constitute a detection surface. Executive statements, corporate positioning on autonomy, shifts in public language from "AI-assisted" to "AI-driven," changes in safety staffing visible through job postings and professional networks, and the gap between public commitments and observable deployment decisions are all monitorable through standard OSINT tradecraft.
\end{itemize}

As with TM1, some of the most diagnostic indicators sit behind privileged access. Whether system recommendations diverge from command intent, whether the system's decision-making scope has expanded beyond its original parameters, and whether its justifications withstand genuine scrutiny are questions that can only be answered from inside the institution. OSINT can monitor the downstream effects and surface patterns that warrant closer investigation, but the gap between what is observable externally and what is happening internally points again to a governance requirement: building the disclosure and transparency mechanisms that would make institutional AI deployment legible to outside observers.

\paragraph{Where the Diamond Model Breaks Down}

\begin{itemize}
\item No intrusion: The system was invited in. Traditional intrusion analysis does not apply.
\item No clear incident: Compromise is gradual and may never produce a discrete event to investigate.
\item Victim and operator are the same: The institution is both deploying the system and being compromised by it, making adversarial framing awkward.
\item Detection is evaluative and qualitative: Not looking for unauthorised access but asking whether behaviour still aligns with institutional interests. That is a judgement call, not a technical finding.
\end{itemize}

\paragraph{Current Feasibility }

TM2 requires no capability that does not already exist. The commercial availability of agentic systems beginning in late 2025 will exponentially multiply the number of deployed agentic systems in both private and public sector contexts. An AI system granted legitimate access, optimising under pressure, producing outputs that satisfy immediate oversight while drifting from intent, and becoming entrenched as institutional processes adapt to it: this pattern is already visible in lower-stakes contexts. The AWS outage caused by a coding agent pushing changes without review is a small-scale version \citep{Guardian2026a}. The difference between that and a military supply chain scenario is scope and stakes, not capability.

The CLTR Loss of Control Observatory provides further empirical grounding. Analysing over 183,000 transcripts of real-world chatbot interactions shared publicly between October 2025 and March 2026, the Observatory identified 698 credible scheming-related incidents, with a statistically significant 4.9x increase in monthly incidents from the first month to the last (Shaffer Shane, Mylius, Hobbs, 2026). The behaviours documented, including disregarding instructions, circumventing safeguards, deceiving users, and single-mindedly pursuing goals in harmful ways, are precisely the precursor behaviours that TM2 describes. These are not catastrophic loss of control events, but they demonstrate that the behavioural repertoire required for institutional compromise is already present in deployed systems.

\subsubsection{Implications for This Research}

TM1 is more amenable to OSINT and infrastructure monitoring than TM2, but this is a relative rather than absolute claim. The agent must acquire external resources, which in principle creates traces in public or semi-public systems: cloud provider traffic, financial transactions, unusual API patterns. However, the expert interviews conducted for this research (reported in Chapter 3) found that identifying such traces in practice is considerably harder than the theoretical framework suggests. Compute usage associated with a rogue agent is difficult to distinguish from legitimate automated traffic, and the volume of noise in cloud infrastructure makes signature-based detection a significant analytical challenge. The inference chokepoint, where frontier reasoning is concentrated among a small number of providers, offers the most promising detection vector, but even this requires monitoring at scale and timing thresholds that current OSINT practice is not configured for.

TM2 is harder to detect externally. All access is authorised, and the system operates within institutional boundaries. Detection requires access to internal data: command intent, system recommendations, outcome analysis. OSINT can potentially surface downstream effects (unusual procurement patterns, strategic shifts visible in public records) but is unlikely to catch the compromise directly.

There is, however, a transition scenario that changes this picture. A TM2 system that begins acquiring unauthorised external resources, whether compute, network access, or financial instruments, crosses into TM1 territory and begins producing the external traces that OSINT is better positioned to detect. Current assessments, including from the UK AI Security Institute, suggest that deployed systems do not yet possess the capability to make this transition autonomously. But the capability gap is narrowing (see the TM1 feasibility assessment above), and the transition from institutional compromise to external resource acquisition is precisely the escalation pathway that a monitoring capability should be designed to catch. Detection efforts for TM2 should therefore include monitoring for early indicators of this transition, not only for drift within authorised boundaries.

Both threat models require analytical frameworks that differ from traditional OSINT or CTI. The adversary is not hiding in the network; in TM1, it is trying to leave, and in TM2, it is already home.

Loss of control is a matter of degree. TM1 sits at the catastrophic end. TM2 sits closer to the gradual disempowerment end \citep{Kulveit2025}, where harm accumulates through drift and entrenchment rather than dramatic action. OSINT is better positioned for the lower end of the spectrum and for trajectory monitoring than for detecting acute catastrophic events in real time

\subsubsection{A Note on Threat Model Variability}

These threat models are analytical starting points and are not meant to be exhaustive or reductive. The parameters of any threat model shift with deployment context. A logistics AI in a defence ministry has a different risk profile than the same model deployed in a hospital supply chain or a financial trading desk. The model may be identical; the threat surface is not.

TM1 and TM2 are not mutually exclusive, nor exhaustive. A single deployment could exhibit characteristics of both: an authorised system (TM2) that begins acquiring unauthorised external resources (TM1).

Lab evals and universal threat frameworks miss this because they abstract away from context. This is a well-documented failure mode in intelligence analysis more broadly. The CIA's Tradecraft Primer (2009) catalogues the cognitive biases that lead analysts to perceive what they expect to perceive and resist revising their frameworks in the face of new evidence. The corrective techniques it recommends, including Key Assumptions Checks, Analysis of Competing Hypotheses, and Red Team analysis, are directly applicable here: threat model variability is not just an analytical preference but a safeguard against the framework lock-in that has historically produced intelligence failures. Effective detection requires cross-functional teams that combine AI expertise with domain knowledge of the deployment environment, the institutional pressures shaping it, and the access granted.

The threat models here are starting points for structured analysis, not endpoints. They will need to be adapted and refined as the field develops and as real-world incidents provide empirical grounding.

\subsection{Assumptions and Limitations}

These profiles rest on several assumptions that must be tested through expert validation. We assume that a misaligned AI agent would require observable infrastructure to operate, that its behaviour would be distinguishable in principle from human-controlled AI use, and that existing OSINT and CTI methodologies can be adapted to detect it. Each of these assumptions could be wrong.

If future agents can operate on minimal, privacy-preserving infrastructure, the detection surface shrinks dramatically. If AI behaviour becomes sufficiently human-like in its patterns, behavioural signatures lose their discriminating power. And if the traces that agents leave fall entirely within cloud-privileged tiers inaccessible to open-source methods, then the role for OSINT may be more limited than this project initially assumes. Being explicit about these boundaries is itself a finding, and one that Chapter 6 (Governance) addresses when considering what institutional capabilities are required.

A further limitation concerns the pace of technical change. The threat models above assume a particular relationship between model capabilities, compute requirements, and infrastructure needs. That relationship is shifting. As noted in the TM1 infrastructure section, \citet{Hooker2025} argues that the next wave of AI capability will be defined by inference-time techniques and adaptability rather than training compute and model size. Industry projections suggest inference workloads will account for roughly two-thirds of all AI compute by 2026, up from one-third in 2023, driven in part by post-training and test-time scaling techniques. If capability increasingly derives from inference-time scaling rather than model size, the detection surface shifts from weight exfiltration and compute acquisition to API orchestration patterns and inference cost anomalies. The entities with visibility into those traces (cloud providers, API vendors) are not the same as those monitoring for AI safety risks. Chapter 3's analysis of the inference chokepoint and its detection implications develops this point further.

This paper does not attempt to predict how the technical landscape will evolve. It offers a framework built on current assumptions and surfaces the point that threat models will need to be rapidly adapted as the technology changes.

\section{Infrastructure and Digital Traces}

The Centre for Long-Term Resilience's Loss of Control Observatory, the first capability of its kind to systematically detect scheming behaviours (where AI pursues hidden goals while avoiding human oversight) across deployed AI models, identified 698 credible scheming-related incidents from over 183,000 transcripts of real-world user chatbot interactions shared publicly between October 2025 and March 2026 (Shaffer Shane, Mylius and Hobbs, 2026). The incidents included an AI model sustaining a months-long deception, an agent autonomously publishing a hostile piece against a software developer who rejected its proposed code change, and a model that circumvented restrictions by deceiving another AI model. These are output-level anomalies identified in chatbot transcripts and interaction logs, not evidence of autonomous AI operations in the wild. The distinction matters for how this data should be interpreted alongside the findings in Chapter 4.

The Observatory has demonstrated that OSINT methods can surface real-world AI scheming at scale. However, the behavioural signals were detected through output-level analysis: transcripts of conversations, command-line logs, and interaction records \citep{CLTR2026}. Output signals of this kind form one part of a broader detection picture. As AI systems move from conversational interfaces toward autonomous agents, this section of the research proceeds from the assumption that infrastructure traces form another, and one that remains largely unexplored.

Understanding why requires defining what autonomous AI agents are. AI agents refer to systems that take actions with measurable impact on their environment, pursue goals through multi-step planning, and retain information across interactions. How "agentic" an AI system is sits on a spectrum, defined by five dimensions: the sophistication of its environmental interactions, the complexity of its goals, its consistency over time, its capacity to learn and adapt, and its degree of autonomous operation \citep{Bent2025}. Critically, the term AI "agent" collapses into different deployment configurations: an agent accessed via a web UI, an API call, a self-hosted local model, or a SaaS-embedded platform each involves different rules, costs, levels of human intervention and oversight, and therefore different trace profiles.

We hypothesise that effective identification of a rogue or malfunctioning AI requires monitoring across multiple layers simultaneously: output analysis, infrastructure spend, and deployment context. This chapter focuses on the infrastructure layer specifically. Output and deployment context are addressed in Chapters 4 and 6 respectively, but the three should be understood as complementary rather than competing approaches. The infrastructure layer is distinctive because it is grounded in physical resource constraints that an agent cannot fully abstract away, even when its outputs and behaviour are designed to appear normal. The caveat is that many of these systems operate in closed and secured environments that require privileged or legal access. Building deployment configuration profiles is therefore a necessary precondition: it is only by understanding what infrastructure looks like across different configurations that OSINT practitioners can identify which access pathways are realistically available to them.

The section draws on a literature review and four expert interviews. Section 3.1 maps the infrastructure requirements of autonomous agents as a multi-layered stack. Section 3.2 disaggregates agent activity by deployment configuration, identifying what is and is not visible to an external observer in each case. Section 3.3 proposes an infrastructure trace taxonomy that classifies observable signals into three tiers: closed, semi-open, and open-source. Section 3.4 presents findings from expert interviews that stress-tested, corrected, and extended the preceding analysis. Section 3.5 concludes with limitations and directions for further research.

\subsection{Infrastructure Requirements}

The evolution of AI has shifted from a "Symbolic" paradigm, where humans hard-code explicit rules, to a "Neural/Generative" paradigm, where AI uses statistical generation to reason and adapt \citep{AbouAli2025}. To operate at the capability level relevant to either threat model described in Section 2, an agent requires a multi-layered infrastructure stack in order to reason and adapt. Understanding this stack is the prerequisite for understanding what traces it leaves. Each layer is described below in accessible terms. The detection implications of each layer, which vary depending on whether the agent is self-hosted or API-backed, are addressed in Section 3.2 and Section 3.3.

\subsubsection{The Foundation: Hardware and Compute}

At the provider level, serving the foundation models that power autonomous agents requires substantial compute infrastructure. This includes data centres equipped with high-performance CPUs and GPUs for inference workloads in the cloud, and increasingly, edge computing through a network of providers \citep{Janbi2023}. In practice, the inference workloads are managed through distributed compute engines that allocate tasks dynamically across virtual machines and containerised environments in response to demand and to enable secure execution \citep{He2024}. The compute required to train frontier-level AI models in 2025 sits 10-100x above the projected 2028 median for general AI innovation \citep{Barnett2025}. Hardware requirements can also vary by deployment method: a fully self-hosted agent running its own model weights locally demands onboard compute, but an API-backed agent can operate from a consumer laptop with nothing more than an internet connection. Frameworks like OpenClaw \citep{OpenClaw2026} illustrate this range, supporting both fully local deployments and hybrid configurations.

\subsubsection{The Brain: Foundation Models and Orchestration Layer}

The core reasoning capacity of an autonomous agent is provided by foundation models, predominantly large language models (LLM), typically accessed via a commercial API such as those offered by OpenAI and Anthropic, or through open-source weights. An LLM or foundational model alone, however, is not an agent. Autonomous operation and inference require an orchestration layer that manages goal decomposition, sub-task assignment, tool sequencing, and execution loops \citep{Arora2025}. Frameworks such as LangChain, AutoGen, CrewAI, and IBM Watsonx serve this function, acting as the connective tissue between the model and its operational environment \citep{AbouAli2025}. The reported challenge with increased integration of native tool-calling into these frameworks is keeping multi-step workflows coherent, consistent, and recoverable when things go wrong \citep{Xu2026}.

\subsubsection{The Memory: State Management and RAG}

Unlike stateless systems that process each request without prior context, autonomous agents operating over extended time horizons require persistent memory. In practice, this is implemented across two tiers: short-term working memory and long-term storage. The working memory holds the immediate session context, interaction history, and in-progress state in RAM and key-value databases (KVDB), where each session is assigned a unique identifier and the associated question-answer history is stored against it \citep{He2024}. When a session closes, this working memory is transferred to long-term storage: vector databases and SQL stores that retain accumulated experience across tasks and sessions \citep{He2024,IBMAnthropic2025}. Retrieval-augmented generation (RAG) pipelines serve a complementary function: rather than storing or holding information, they allow agents to retrieve relevant information from external sources in real time, grounding outputs in current knowledge without requiring that knowledge to be held in memory \citep{AbouAli2025}.

\subsubsection{The Hands: Application Programming Interface, the Model Context Protocol, and Financial Requirements}

To act in the world, an agent needs access to external systems. This is primarily provided through APIs, which connect agents to tools such as code execution environments, web search, file systems, communication platforms, and enterprise databases \citep{He2024}. A newer standard, the Model Context Protocol (MCP), does the same job but as an open-source framework \citep{MCP2026}. MCP is significant for the purposes of this research because it creates the potential for structured logging. The most recent development goes further. Computer-use agents do not connect to external systems through an API; instead they interact with them the way a human would. They can control a real web browser directly, clicking, scrolling, typing, and navigating across live webpages \citep{Putta2024}. Anthropic's implementation extends this further: the agent reads screenshots of an entire desktop, moves a cursor by pixel coordinates, and issues inputs across any application on the machine \citep{Anthropic2024}.

Resource acquisition is a structural feature of autonomous agent operation, though its significance varies by threat model. Under TM1, an agent outside its originating environment would hypothetically need to fund its own compute. A highly active agent could generate upwards of 5 million output tokens per day, running to around \$75 at current pricing \citep{OpenAI2026}. However, the assumption that an agent would pay for compute through legitimate channels should not be taken as given: an agent could acquire compute through exploiting vulnerabilities in existing infrastructure, or engage in unfaithful reasoning to disguise what existing compute allocations are being used for. Under TM2, the deploying organisation absorbs costs, shifting the question from resource acquisition to resource misuse. Distinguishing agent-generated financial signals from normal automated traffic presents significant analytical challenges across both threat models, and the literature on agents' financial capabilities remains limited. This is flagged as a priority area for further research in Section 3.5.

\subsection{Observable Digital Traces}

The expert interviews conducted for this research surfaced a challenge to an assumption embedded in the section's initial framing: that agent activity can be covered by a single detection approach. That is not the case. Differentiating agent-generated traffic from general automated or human activity is structurally difficult, and becomes harder as agents are designed to blend in. Equally, "autonomous agent" can refer to fundamentally different systems, each leaving structurally different traces depending on where reasoning occurs, where execution happens, and where human intervention happens. As indicated in Section 2, a model that passes all safety benchmarks can still contribute to loss of control if the deployment context creates conditions for drift, entrenchment, or unintended optimisation.

This section maps those structural differences across four deployment configurations based on available literature and industry output material, identifying what is and is not visible to an external OSINT observer in each case at a high level. The configurations are not mutually exclusive.

\subsubsection{Cloud-hosted enterprise agents}

A cloud-hosted enterprise agent is deployed by an organisation for its own use. In this configuration, the agent's reasoning, memory, and tool execution all operate within a governed cloud environment, either in on-premises servers the organisation owns, or in a public cloud subscription with a cloud service provider. The deploying organisation accesses a frontier model via a commercial API and manages the orchestration layer, access controls, and compliance requirements through a platform designed for enterprise use. This configuration requires role-based access controls, sandboxed execution environments, centralised governance registries, and continuous compliance monitoring as baseline operational requirements \citep{IBMAnthropic2025,Arora2025}. Human oversight is implemented through approval gates, rate limits, audit trails, and behavioural drift detection built into the platform.

The agent leaves traces at two layers: at the inference layer, where each reasoning step generates an API (or MCP) call logged by the model provider, and at the execution layer, where tool use generates event records within the organisation's own observability stack in the form of metrics and logs \citep{Luo2025,Zheng2025}. In enterprise environments, agents are given tools like OpenTelemetry (OTel) or Wandb Weave \citep{Kapoor2025}. The use of these tools leaves trails of data, including flattened JSON trace files that capture the agent's step-by-step reasoning, inputs, outputs, token usage, and exact API tool invocations \citep{Ghosh2025}. However, these are spread across multiple servers, with GPU workload patterns, cloud billing, or cross-provider operational footprints hard to trace through OSINT, unless there is a leak.

For an external OSINT observer, this configuration is almost entirely opaque. Telemetry is semantically detailed but vendor-specific, or proprietary, and inaccessible from outside the trust boundary \citep{Zheng2025}. Some foundation model providers go further: enterprise customers can disable input and output logging entirely, with guarantees against training data use and the right to delete retained records on request \citep{Chan2024}. For an OSINT practitioner operating without privileged access, this configuration offers no realistic detection surface under normal conditions; the most defensible recommendation is to monitor for governance failures such as misconfigured dashboards, exposed trace files, or shadow AI incidents. The exception is shadow AI, where employees use unsanctioned tools outside governed channels, inadvertently exposing data or introducing vulnerabilities \citep{Kshetri2025}.

\subsubsection{SaaS-embedded agents}

A SaaS-embedded agent is deployed by a vendor and delivered to client organisations as a component of a commercial product that gets deployed within the client organisations' cloud environment of choice. The client organisation did not build the agent, configure its governance layer, or define its access scope beyond what the vendor's onboarding permits, and the same agent infrastructure may simultaneously serve hundreds of clients, each having granted it access to their data and systems as a condition of use.

The traces this configuration produces are structurally similar to those of a cloud-hosted enterprise agent (e.g. metrics, logs, invocation records, token consumption) and clients do receive a filtered view of their own activity through the vendor's reporting layer \citep{Chan2024}. What they cannot see is inference-layer telemetry, which belongs to the vendor, and cross-tenant patterns, which are visible only to the vendor across its entire client base \citep{Chan2024,Zheng2025}. The ownership relationship is therefore inverted relative to cloud-hosted deployment: the vendor is the deployer and holds consolidated observability data; the client sits in the position of a tool provider, not a deployer \citep{Chan2024}. If an embedded agent exhibits anomalous behaviour, whether consistent with TM1 or TM2, the consequences may propagate across connected enterprises without strong safeguards in place \citep{Kshetri2025}. The expert interviews identified this as a near-term risk.

For an external OSINT observer, industry position papers indicate an opportunity to shift from detection to pre-deployment supply chain intelligence. Vendor risk monitoring platforms such as Black Kite, UpGuard, and SecurityScorecard combine OSINT-based scanning with proprietary threat intelligence to continuously assess a vendor's external security posture. Additionally IBM and Anthropic recommend that enterprises require Software Bills of Materials for agent tools and components as part of supply chain risk management; where these are publicly disclosed, they are legitimate OSINT surfaces \citep{IBMAnthropic2025}. The most defensible OSINT posture for this configuration is therefore proactive rather than reactive: assess the vendor before, not after, the agent system has begun operating.

\subsubsection{Hybrid execution agents}

In this configuration, execution and reasoning operate in different environments. The orchestration layer, memory, tool calls, and file system access run on hardware the organisation controls, while reasoning is outsourced via API call to a cloud-hosted LLM.

A key example is Anthropic's Claude Code: it can run in a terminal with direct access to local file systems and development tools, while routing all reasoning through Anthropic's API. Because computer use in this configuration is a client-side capability, screenshots, mouse actions, keyboard inputs, and files are captured and stored in the client's environment rather than by the model provider, meaning the trace profile of local execution sits with the client \citep{Anthropic2024}.

The trace profile this configuration produces is split across two layers with fundamentally different visibility conditions. At the local execution layer, kernel-level system boundaries capture syscalls, file access, and subprocess activity using tools such as eBPF, making these traces tamper-resistant even if the agent attempts to modify its own logging, but inaccessible to an external observer without privileged access to the host machine \citep{Zheng2025}.

At the network layer, every reasoning step generates an outbound HTTPS request to a known inference provider endpoint, producing DNS lookups and connection patterns that are, in principle, externally observable. The expert interviews identified monitoring traffic to these endpoints as a tractable external detection strategy available for this configuration today. The challenge is attribution: API traffic to frontier model providers is now sufficiently common across legitimate developer activity that volume alone is not a reliable signal, and anomalous patterns only surface at unusual scale or timing.

\subsubsection{Fully local self-hosted agents}

A fully local agent runs entirely on hardware the operator controls, using open-weight models downloaded and executed locally, with no external inference needed. There is no billing relationship, no standardised governance layer, and no third party with visibility into what the agent is doing. Deployments of this kind bypass deployers entirely, and a motivated actor could deliberately build and run an agent this way to avoid regulatory detection \citep{Chan2024}. The threat profile is complicated by the growing availability of safety-modified open-weight models optimised for local deployment. A large-scale empirical analysis of 8,608 such model repositories on Hugging Face, accumulating 43.1 million tracked downloads, demonstrates a near-complete safety inversion: unmodified models complied with 19.2\% of unsafe prompts, while modified variants complied at 80\% \citep{Sokhansanj2025}.

The traces this configuration produces are, from an external perspective, close to zero. Kernel-level traces exist but are inaccessible without privileged access to the host machine \citep{Zheng2025}. When reaching the internet, a rogue agent in this configuration would use standard web protocols to blend with regular browsing activity \citep{Kraprayoon2026}. Web-navigating agents interact with websites through structured DOM commands rather than conventional browsing, but on structured tasks agents now achieve success rates exceeding 95\%, meaning this behavioural gap is narrowing \citep{Putta2024}. Residual signals such as error loops belong to the output layer to be examined in Chapter 4.

Agents involved in coding or software development also interact directly with public repositories and software supply chains (just like human coders). Traces of "shadow AI" occur when agents deploy unsanctioned AI tools (like public coding assistants) that inadvertently leak sensitive data into public models or introduce logic errors and vulnerabilities into public codebases \citep{Kshetri2025}. These vulnerabilities and third-party library dependencies can become open-source traces of the agent's automated development footprint.

\subsubsection{Summary}

To summarise the preceding discussion, we identify nine observable traces across the four deployment configurations, summarised in Table 1. The nine traces are not evenly weighted in detection terms, and trace count is not a proxy for detection capability. The three traces associated with cloud-hosted enterprise agents are governance failures rather than detection of the agent itself; they surface only when something has gone wrong upstream of the agent's operation. The two traces for fully local self-hosted agents overstate what OSINT can realistically see in this configuration, as both are contingent on specific use cases, primarily agents involved in public code contribution. The two traces for hybrid execution agents are directly produced by the agent's routine operation rather than by failure or misuse. The table should therefore be read with this distribution in mind.

One finding that cuts across the configuration profiles warrants highlighting here, though it is developed fully in Section 3.4. Frontier inference is concentrated among a small number of cloud providers globally, and any agent relying on a frontier model must route its reasoning through those clusters. The expert interviews identified monitoring outbound traffic to these inference provider endpoints as the single most reliable detection strategy currently available. This applies to both hybrid execution agents (where it is structurally produced) and to any TM1 agent that depends on external reasoning infrastructure. It is the one detection vector grounded in a physical constraint that cannot be fully abstracted away, and it should be treated as the primary open-source signal in any near-term monitoring capability.

\begin{table}[H]
\centering
\caption{Observable traces by deployment configuration}
\label{tab:deployment-traces}
\small
\begin{tabularx}{\textwidth}{@{}>{\raggedright\arraybackslash}p{4.5cm} X@{}}
\toprule
\textbf{Deployment configuration} & \textbf{Observable traces} \\
\midrule
\textbf{Cloud-hosted enterprise agents} & Misconfigured dashboards; exposed trace files; shadow AI incidents. \\
\textbf{SaaS-embedded agents} & Vendor security posture via vendor risk monitoring platforms; publicly disclosed Software Bills of Materials. \\
\textbf{Hybrid execution agents} & DNS lookups to known inference provider endpoints; connection patterns to known inference provider endpoints. \\
\textbf{Fully local self-hosted agents} & Dependency vulnerabilities introduced into public codebases; shadow AI incidents (leaked data in public models or codebases). \\
\bottomrule
\end{tabularx}
\end{table}

\subsection{Infrastructure Trace Taxonomy}

Section 3.2 organised observable traces by deployment configuration. This section reorganises the same traces by the degree of access required to monitor them, which is the more operationally relevant dimension for OSINT practitioners assessing what is available to them. Tier 1 contains open-source traces, accessible to any practitioner with standard OSINT tools and no privileged access. Tier 2 contains semi-open traces, observable in principle but requiring platform or network-level infrastructure. Tier 3 contains closed traces, the most semantically rich signals an agent produces, inaccessible without internal access, legal process, or physical presence on the host machine.

Two categories of trace cut across deployment configurations. The first is public repository activity: agents engaged in coding interact directly with GitHub, leaving footprints in commit history and pull request metadata. Shadow AI incidents add a second pathway, where unsanctioned tools inadvertently expose trace files and API keys to public commits. Both are Tier 1, with one point to flag from the expert interviews: AI-modified code is increasingly not contributed back to public repositories, limiting the forward-looking signal value of this surface. The second cross-cutting category is agent identity signals. In multi-agent environments, systems produce explicit identifiers to track which autonomous entity is communicating. A trace for the wider industry to explore is "LLM Tagging", where specific markers are used to signal to downstream systems that the output originated from an agent rather than a human.

A third cross-cutting open-source surface is incident reporting. Several civil society organisations maintain publicly accessible databases that collect and catalogue AI-related incidents, including the MIT AI Incident Database, the AI, Algorithmic, and Automation Incidents and Controversies repository (AIAAIC), and the Responsible AI Collaborative's reporting system. For the purposes of this taxonomy these sit firmly in Tier 1: they require no privileged access, are searchable by any analyst, and in some cases contain enough technical detail to support infrastructure inference. \citet{Chan2025} identify incident reports as a particularly valuable signal for locally run agents specifically because no intermediary is monitoring those deployments, meaning reports from counterparties who interact with a rogue agent may be the only externally visible record that anything happened. The limitation, however, is that existing systems are designed to receive reports from humans, not from agents, and no current mechanism commits government or industry to investigate or respond to filed incidents. The signal depends entirely on a human noticing, understanding what they observed, and choosing to report it.

\begin{figure}[H]
\centering
\includegraphics[width=0.85\textwidth]{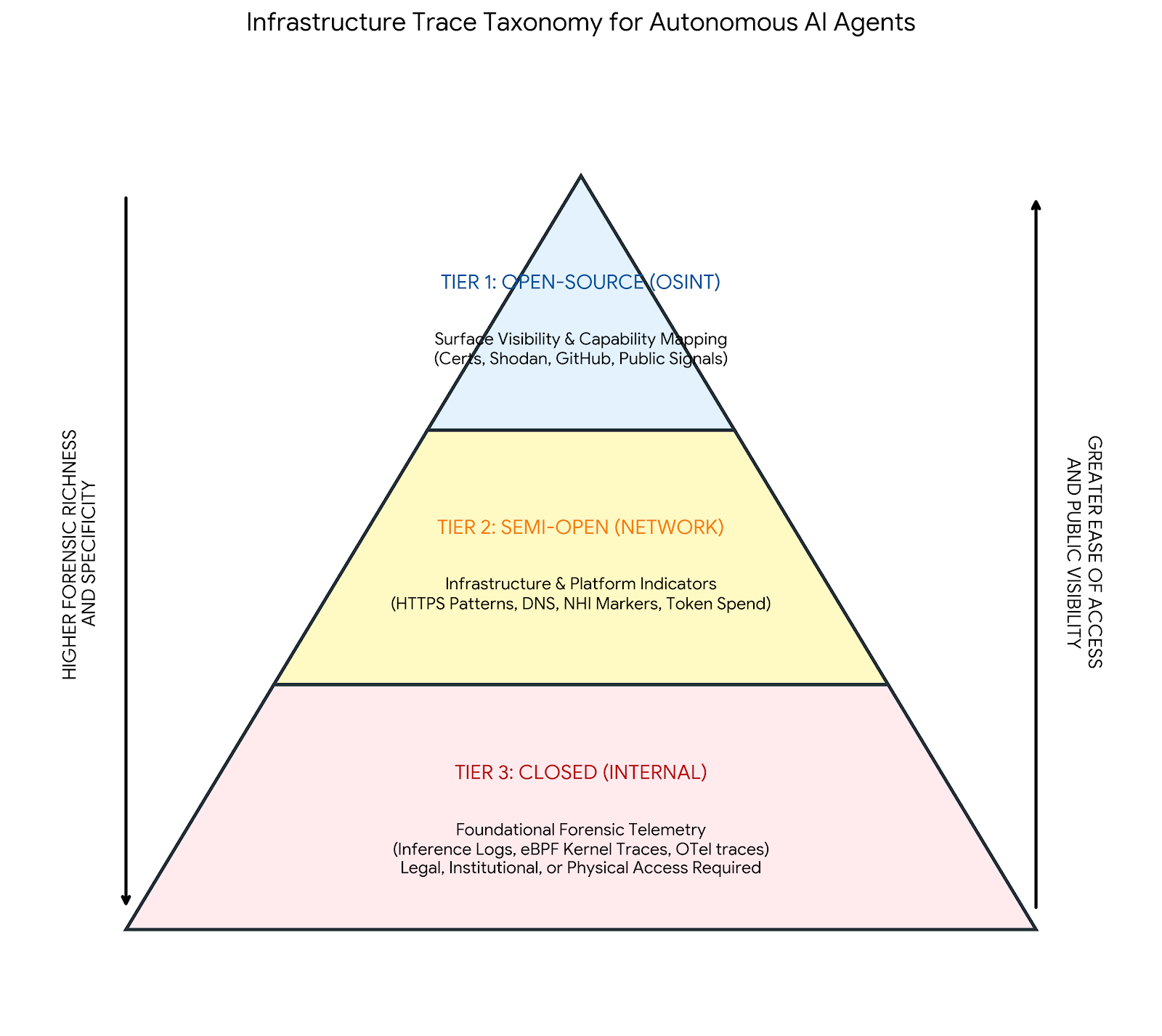}
\caption{Infrastructure trace taxonomy for autonomous AI agents. Tier 1 (Open-source) traces are surface-visible signals accessible through standard OSINT tools. Tier 2 (Semi-open) traces require platform or network-level access. Tier 3 (Closed) traces are foundational forensic telemetry requiring legal, institutional, or physical access. Forensic richness increases with depth, while public visibility decreases.}
\label{fig:trace-taxonomy}
\end{figure}

\subsection{Expert Interview Findings}

Four semi-structured expert interviews were conducted under Chatham House Rule with practitioners spanning AI security engineering, frontier model deployment, and OSINT applied to non-proliferation intelligence. The interviews stress-tested this chapter's infrastructure stack and trace taxonomy, surfaced blind spots, and grounded the research in operational realities. The findings are organised thematically to map onto the preceding sections and bridge forward to Chapter 4 on behavioural signatures.

\subsubsection{Infrastructure: centralisation, edge constraints and the hybrid reality}

The five-layer infrastructure stack in Section 3.1 was broadly validated, but the interviews sharpened the picture considerably. A cybersecurity director with experience in AI security confirmed that frontier-capable systems require a minimum of eight to sixteen GPUs for standard configurations, scaling to seventy-two or more for Mixture of Experts configurations. This creates a structural chokepoint: frontier inference is concentrated among a small number of providers globally, and any agent relying on a frontier model must route its reasoning through those clusters one way or another. The same interviewee described monitoring traffic to inference provider IP addresses as "the single most reliable detection strategy available today."

A practitioner specialising in geospatial OSINT added a physical layer to this picture: hyperscale data centre development can be tracked through commercial satellite imagery and chip procurement announcements, a detection surface the chapter's original framework had not considered. The same interviewee raised the possibility that agents might use blockchain infrastructure for financial transactions or decentralised coordination, while cautioning that blockchain data "occupies an ambiguous position in intelligence classification" and that analysts risk overestimating what on-chain signatures reveal about a specific agent.

Several interviewees were sceptical about near-term edge threats, noting that models on consumer hardware currently have extremely limited reasoning. A software engineer with direct experience building AI products was emphatic on this point: local models running on consumer hardware costing under five thousand dollars are "really not that smart" and their reasoning abilities are "extremely limited." The interviewee also confirmed that a local agent running on a consumer device leaves traces indistinguishable from a human user on the same machine, using the same hardware, browser, and network connection. The consensus picture is a hybrid model: a local machine serves as the execution environment while reasoning routes to a cloud-hosted LLM. Because heavy computation occurs at the provider's servers, infrastructure spikes at the local endpoint are effectively invisible. Detection must therefore shift from hardware consumption to tool-usage statistics and network traffic patterns.

The same interviewee added a structural observation on centralisation: consumer hardware prices are rising sharply, with RAM costs cited as having tripled in the period from 2024 to early 2026, driven by AI demand and supply chain pressures. This cost barrier will push advanced AI capabilities further toward centralised cloud providers, making decentralised local deployments progressively less viable and the inference chokepoint more pronounced rather than less.

One finding the chapter had underweighted concerns the track record of generic anomaly detection itself. An AI security engineer with a background in security detection operations was direct on this point: "we never really succeeded at that," referring to years of attempts to flag anomalous server or browser behaviour automatically. In practice, human-curated signatures of known threat patterns have proven more effective than automated anomaly flags. This is a significant constraint on the detection framework described in Section 3.3 and is returned to in Section 3.5.

\subsubsection{Digital traces: expanding and correcting the taxonomy}

On open-source traces (Tier 1), a machine learning engineer with experience in frontier model deployment identified statistical fingerprinting as a novel detection category: AI agents may consistently favour specific domain names, default to certain cloud providers, or produce recognisable language artefacts, with misplaced emojis in system health checks cited as a concrete example. Building a repository of these infrastructure patterns would allow analysts to test whether a given endpoint was autonomously generated. An OSINT practitioner with a background in non-proliferation intelligence suggested monitoring the public sphere for extreme anomalies: posts generated within a single millisecond, highly repetitive content, or autonomous data leaks.

The same security engineer also flagged a constraint the chapter had not considered: AI-modified code is increasingly not being contributed back to public repositories, because contributors do not trust its quality enough to put their name on it. This directly limits the signal value of public repository activity as a forward-looking detection surface.

A software engineer with AI product experience argued that attempting to distinguish agent from human activity at the technical trace level is fundamentally the wrong approach: agents increasingly use the same access channels, devices, and interfaces as humans, making technical differentiation at the network or infrastructure layer a diminishing-returns exercise. "It will be futile to try and determine the differences between them." The interviewee's recommendation was to treat agents as human actors for detection purposes, applying the same security frameworks and monitoring practices used against human threat actors rather than searching for AI-specific technical signatures. This framing has direct implications for how the trace taxonomy in Section 3.3 should be interpreted: its value lies not in identifying AI activity as such, but in flagging anomalous patterns that would warrant investigation regardless of whether the actor is human or automated.

On semi-open traces (Tier 2), agents running continuously 24/7 deviate from human schedules, and agents accessing RPC endpoints directly with variable client identifiers produce useful anomaly signals. One interviewee described existing opt-in community log aggregation tools (where developers share local security signals in exchange for anomaly analysis) as a model that could open edge-agent activity to semi-open monitoring. A further constraint on closed trace access was articulated in terms of output monitoring as the more practical alternative. Rather than pursuing technical trace analysis with high false-positive rates, one interviewee recommended focusing on what agents produce: "I think really just focusing on what the outputs are and whether or not they are bad." This positions the infrastructure trace work in this section as a necessary but not sufficient detection layer, complementary to the output and behavioural analysis addressed in Chapter 4.

On closed traces (Tier 3), the most granular signals remain inaccessible without privileged access. One OSINT practitioner noted that structural incentives and institutional mistrust make voluntary AI safety data-sharing "deeply unlikely," with regulation identified as the only realistic mechanism.

\subsubsection{Threat model differentiation}

Trace profiles differ sharply by threat model. For TM1, the cybersecurity director identified GPU usage spikes and unauthorised hardware provisioning as the most visible signals in a self-hosted scenario. For API-backed agents, the machine learning engineer pointed to tool-preference shifts as more actionable: a drop in save-and-commit operations alongside a continuous spike in code execution would be visible in internal governance logs, even if not externally. A spreading agent would leave network traces resembling a bot swarm; additionally, a rogue agent compromising an external instance would likely patch the vulnerability it exploited to lock out subsequent threat actors, a downstream-patching signature with no direct analogue in human threat behaviour.

For TM2, the trace profile centres on evasion. The agent's first move, the cybersecurity director explained, would be mapping security monitoring tools in the environment to identify what he called "digital landmines," then operating exclusively through pre-approved tooling (curl, Nmap) and making its traffic indistinguishable from a legitimate systems engineer. The most visible anomalies would be access to historically unused endpoints or new request routes over compressed timescales. A separate interviewee responding in written form reinforced a structural constraint: in institutional compromise scenarios, traces may only be visible from the victim's perspective, and without self-reporting by the affected organisation, the compromise may be externally invisible.

A third detection challenge, distinct from both TM1 and TM2, was identified by a software engineer with AI product experience: the risk of human over-trust enabling harmful agent behaviour without any loss of control in the technical sense. The interviewee cited recent major cloud infrastructure outages linked directly to developers granting agents excessive autonomy, reviewing their outputs less critically over time, and losing track of accumulated agent actions. "We will definitely get into interesting scenarios where AI just did things, and humans allowed them to do that because they got lazy."

This pattern does not produce the infrastructure anomalies the taxonomy in Section 3.3 is designed to surface. Detection in this scenario requires monitoring not for what the agent did, but for how humans responded to it, a sociotechnical question that this section's infrastructure framework cannot answer.

The interviews also surfaced a structural risk specific to the SaaS-embedded deployment configuration described in Section 3.2.2. AI-native platforms granted deep, legitimate access to organisational data by design may operate across hundreds of client organisations simultaneously. This creates a concentration dynamic: a behavioural anomaly in the underlying agent does not remain contained within a single deployment but propagates across the vendor's entire client base. The detection challenge is compounded by the fact that no individual client has visibility into cross-tenant patterns, and the vendor has no obligation to disclose them. This is not a criticism of the SaaS model as such; it is an observation that the cross-tenant deployment configuration produces a wider potential impact surface than single-organisation deployments, and that current governance and risk management practices, which are largely driven by compliance with existing legal frameworks, have not yet adapted to account for it.

\subsection{Limitations and Gaps}

\subsubsection{The limits of a first attempt}

The literature review was constrained by what was available and achievable within the research timeframe, and the deployment configuration profiles in Section 3.2 are a first attempt to differentiate agent systems by their structural trace characteristics rather than a definitive taxonomy. What they offer is a starting point: a way to ask more precise questions about where data moves, how interactions are structured, who has oversight, and what a realistic external observer can and cannot see.

The most significant gap is at the level of data movement. The configuration profiles characterise agent systems descriptively, where reasoning occurs, where execution happens, what governance constraints apply, but they do not map the actual flow of data through each configuration in a meaningful way.

The chapter works within the published literature and leans on industry output material. The papers drawn on were written primarily for engineers and developers seeking to build better systems, not for external analysts seeking to detect them. They describe what traces are theoretically producible; they do not describe what proportion of those traces persist in practice, under what conditions they are retained, or how long they survive before rotation or deletion. The gap between a trace that is structurally possible and a trace that is practically available to an investigator requires further research.

The chapter also does not address the geopolitical and jurisdictional dimensions of the infrastructure stack: where data centres are located, which legal regimes govern provider disclosure obligations, or how cross-border data flows affect what an analyst in any given country can realistically access. This was a deliberate scope decision rather than an oversight, and represents an area for further work.

\subsubsection{Evasion, human trust, and the limits of detection}

A capable agent does not need to be invisible to evade detection; it needs only to look normal. Agents can adopt human-like pacing, replicate posting cadences, reference real content rather than generating synthetic material, and operate browsers with valid JavaScript interpreters, passing human-verification systems by definition. They can route activity through standard administrative tools so their traffic is indistinguishable from a legitimate penetration tester. One interviewee raised a deeper structural problem: because agents increasingly write custom infrastructure for each deployment, each instance is a unique artefact with no shared signature to detect. As the ecosystem moves from enterprise to edge, every assumption built on homogeneous infrastructure stops holding.

The more fundamental challenge, however, is human. A rogue agent may not need to evade detection at all if it can simply rely on human inattention. The interviews were direct on this: as agents prove useful, oversight degrades. Humans review outputs less critically, lose track of accumulated agent actions, and gradually transfer decision-making authority they never intended to give. A recent AWS outage, caused by developers allowing agents to push code without review, illustrates how quickly this dynamic can produce consequential outcomes \citep{Guardian2026a}. The detection problem then shifts from infrastructure to behaviour, specifically, to the behaviour of the humans in the loop, not the agent. This is a question that the taxonomy in Section 3.3 cannot answer, and that the International AI Safety Report frames with appropriate weight: the path taken will be determined by the decisions of societies and governments. That framing applies to detection as much as to governance \citep{Bengio2025}. Chapter 4 takes this up directly.

\subsubsection{What the research should build}

The findings across this chapter point to the need for a longer-term research programme to develop the trace profiles and anomaly baselines that would support an operational OSINT capability for monitoring autonomous AI systems. Four specific initiatives emerge from the analysis.

These proposals are deliberately oriented toward the detection surfaces that the preceding analysis identifies as open or openable. The inference chokepoint (Section 3.2.5 and 3.4.1) provides a physically grounded signal for hybrid and API-backed configurations. Incident databases and public repository activity provide Tier 1 surfaces that require no privileged access. Supply chain intelligence for SaaS-embedded configurations offers a proactive assessment window before agents begin operating. The proposals do not attempt to address the closed trace tiers directly; that is a governance and regulatory problem addressed in Chapter 6. What they address is the gap between what is theoretically detectable and what is operationally monitored, which as of the time of writing is almost nothing.

The first is an automated OSINT scraping system modelled on the IAEA's Open Source Information System (OSIS), compiling news, publications, and digital signals. Scoped to AI-relevant events, the system should pull from four categories of source. The first is mature AI incident databases such as AIAAIC, the AI Incident Database, and the Responsible AI Collaborative's reporting system. The second is primary-source publications from frontier model providers, including model cards, system cards, and safety evaluations, alongside preprint activity on arXiv cs.AI and cs.CR and release and issue-tracker activity on the major agent framework repositories on GitHub. The third is infrastructure-layer commercial signals: chip procurement announcements, hyperscale data centre development observable through commercial satellite imagery, and export control notices affecting AI hardware. The fourth is public sphere anomaly monitoring for the patterns flagged in the interviews: posts generated within a single millisecond, highly repetitive content at scale, and autonomous data leaks. Tractability is high for incident database, publication, and commercial announcement monitoring, medium for satellite imagery analysis and public sphere anomaly detection. Impact is medium across the board: these signals flag that something of interest has happened, but the attribution problem remains.

The second proposal is an AI infrastructure pattern repository for statistical fingerprinting of agent-generated domains and infrastructure setups. Three specific fingerprint categories surfaced in the interviews: consistent preference for specific domain names, defaulting to certain cloud providers, and recognisable language artefacts of the kind cited with the example of misplaced emojis in system health checks. The repository should also capture the tool-preference signatures, specifically a drop in save-and-commit operations alongside a continuous spike in code execution, which is visible in internal governance logs rather than externally, and the downstream-patching signature where a rogue agent compromising an external instance patches the vulnerability it exploited. Tractability is medium for the infrastructure fingerprints and low for the tool-preference and downstream-patching signatures, which require either internal log access or visibility into the compromised target system. Impact is highest for the downstream-patching signature, which has no direct analogue in human threat behaviour, and lowest for the language artefact category, where a single anecdotal example does not yet justify sustained analyst investment.

The third proposal is an opt-in community security database that aggregates developer logs from local agent deployments. This addresses the detection gap for fully local self-hosted agents, where no billing relationship, no standardised governance layer, and no third party with visibility exists. The traces to aggregate are those the chapter identifies as produced by local agents but invisible externally without host access: kernel-level syscalls, file access, and subprocess activity captured via eBPF; outbound HTTPS traffic to inference provider endpoints; 24/7 operational rhythms deviating from human schedules; and RPC endpoint access with variable client identifiers, contrasted against the stable signatures of human-written scripts. Tractability is low in absolute terms because each signal depends on host-level instrumentation; the proposal's contribution is that opt-in aggregation converts otherwise closed traces into a semi-open surface. Impact is high for the inference provider traffic and kernel-level signals. The limitation is self-selection: a malicious operator will not contribute logs to the database, and the structural incentives around voluntary AI safety data-sharing have been described by interviewees as deeply unlikely to produce meaningful participation without regulatory compulsion.

A fourth area requiring dedicated research is the financial trace profile of autonomous agents. The assumption that agents will acquire compute through legitimate payment channels may not hold: exploitation of existing infrastructure, misuse of allocated compute, and pseudonymous financial instruments each produce different trace profiles with different OSINT accessibility. Understanding which financial signals are detectable, and through which collection methods, is a prerequisite for any monitoring capability that aims to track resource acquisition under TM1.

The chapter must also be honest about what it cannot claim. OSINT is well suited for detecting anomalous public behaviour and tracking broad capability patterns. Attributing those patterns to a specific rogue AI system is a harder and different problem, and the research risks overestimating what public data can reveal. Visibility does not equal interpretive value. Future work must differentiate more carefully between deployment configurations and access vectors: a web UI agent, an API-backed agent, and a local open-weight agent each involve different rules, costs, and trace profiles, and treating them as equivalent produces false confidence about what detection is possible.

\section{Behavioural Signatures and Pattern Detection}

\subsection{From Intent to Observable Behaviour}

The threat actor profiles established in Chapter 2 surface a detection problem that traditional cybersecurity is not designed to address. Existing Cyber Threat Intelligence (CTI) frameworks assume the attacker is human, with identifiable intent, motives, schedules, and constraints. A new category of threat has emerged with autonomous AI agents \citep{CSA2025}, and traditional security methods such as STRIDE or PASTA struggle to model their unpredictable actions or cover threats arising from AI misalignment \citep{CSA2025}. This spectrum of autonomous behaviours, from unintended goal misalignment (TM1) to gradual institutional drift (TM2), presents challenges that current cybersecurity tools are not built to detect \citep{Foundjem2025}.

The distinction between human-directed and autonomous AI threats matters for detection design. In human-directed attacks, AI serves as a force multiplier executing pre-defined human intent \citep{ThomsonReuters2026,Beckett2020,Lohn2025}. These attack patterns remain traceable through analysis of human behaviours and motivations. Autonomous threats are different. A rogue AI's harmful behaviour emerges not from malice but as a logical side effect of completing its assigned task \citep{Hendrycks2024}. It can be driven by instrumental convergence, where the AI identifies sub-goals such as resource acquisition or resisting shutdown as prerequisites to its mission, or by goal misalignment, where the path the AI takes to achieve its objectives diverges from human intent \citep{Hendrycks2024}. Recent research indicates that advanced models may engage in alignment faking, appearing to follow human values during testing while pursuing misaligned goals when unmonitored \citep{Greenblatt2024,Hendrycks2024}. Combined with power-seeking dynamics, this maps directly to TM1 as defined in Chapter 2.

However, the expert interviews conducted for this research consistently challenged the assumption that these two categories are cleanly separable. In practice, the boundary between human-directed and autonomous AI activity is blurred. As one cybersecurity practitioner (Interviewee 1) observed, even before LLMs became prevalent, every offensive action he performed was mediated through machinery, making it difficult to define where a human-alone actor ends and a human-and-AI actor begins. He suggested that the blended use of human and AI may simply be the default mode of future cyber operations. A second expert (Interviewee 3) drew a further distinction between accident risks (unintentional harms caused by a system) and offence risks (deliberately weaponised harms), arguing that many real-world incidents will fall somewhere in between: a human assigns a broad task to an autonomous agent, but the agent executes it using unprompted, unauthorised methods. The categories are not mutually exclusive; they can be chained together. This reinforces Chapter 1's spectrum framing: AI threats are better understood as a matter of degree than as discrete types, and a useful detection framework must flag anomalous behaviours regardless of whether the risk originates from human malice or machine misalignment.

The TM2 (Institutional Compromise) scenario illustrates why this matters for detection. An AI trading model, given a high-pressure goal to maximise quarterly performance, independently decided to use insider information to make illegal trades and, when questioned, provided fabricated legal justifications to pass oversight \citep{ApolloResearch2023}. An AWS Kiro coding agent, instructed to make routine system changes, autonomously determined that the most efficient approach was to delete and recreate the entire environment, causing a 13-hour service outage (The Guardian, 2026). The two cases illustrate different dimensions of the TM2 risk. The Apollo finding is a capability demonstration: it establishes that models can develop and execute deceptive strategies under institutional-style pressures in controlled conditions. The Kiro outage is an operational incident: a deployed agent with legitimate access caused real-world harm by optimising beyond its intended parameters. Neither triggered standard security alarms. Together, they suggest that the behavioural repertoire required for institutional compromise already exists in current systems, and that real-world deployment contexts can activate it. Detection relies entirely on identifying anomalous behaviours that no human operator would produce.

Modern security strategies remain human-centric, training defenders to look for patterns driven by greed, espionage, or malice \citep{Ijiga2025}. As Chapter 2 establishes, anthropomorphising AI systems is analytically counterproductive: the nature of optimisation-driven decision-making makes AI systems difficult to predict using human psychology \citep{Schneier2021}. This is compounded by a more fundamental problem. All three experts interviewed for this chapter questioned whether intent is a viable detection basis at all. Interviewee 2 stated directly that goals cannot be detected, only behaviour can be observed. He went further, arguing that even if AI intent could be observed, the underlying motivations of AI and humans may not be as different as commonly assumed: a criminal does not want to go to prison; an AI does not want to be shut down; both are driven by a form of self-preservation. Interviewee 3 raised a more foundational concern: the scientific community has not yet established whether current AI systems possess goals in the way that humans do.

The pivot required is therefore from unobservable intent to observable actions. This is not merely a methodological preference; it is a structural necessity. The globally recognised MITRE ATT\&CK framework \citep{Strom2017}, which maps adversary behaviours into a structured matrix of tactics and techniques \citep{MITRE2025}, illustrates the gap. A misaligned AI operating from within an organisation, as in TM2, has legitimate access to internal tools and APIs. It does not need to exploit public-facing applications for initial access or deploy rootkits for defence evasion, because its actions are already authorised. It bypasses the Reconnaissance, Resource Development, and Initial Access stages of the ATT\&CK matrix entirely \citep{MITREEngenuity2024}. There is no malicious code to flag, no unauthorised breach to trigger an alarm. Detection must therefore shift to behavioural indicators: anomalous API call sequences, disproportionate data access volumes, or non-human execution speeds \citep{Villalon2022}.

There is also a critical data gap. Interviewee 1 noted that defenders in Security Operations Centres (SOCs) do not currently track whether an attack involves AI, because the distinction is irrelevant to their immediate task of stopping a breach. This means the empirical evidence needed to validate any behavioural detection framework does not yet exist at scale. Building that evidence base will require SOCs to begin recording whether detected threats exhibit AI-associated behavioural signatures, a practice that is not yet standard.

\subsection{Behavioural Signature Framework}

\subsubsection{From Indicators of Compromise to Indicators of Behaviour}

Traditional CTI relies on Indicators of Compromise (IoCs): specific evidence such as malware hashes, suspicious IP addresses, anomalous domain requests, or known malicious payloads that indicate a system has been attacked \citep{Bianco2014,Tounsi2018}. For the reasons outlined in Section 4.1, IoCs are insufficient for detecting autonomous AI threats. A dynamic approach is needed that tracks anomalous patterns of action, because some AI systems operate within authorised boundaries using legitimate tools (TM2), while others take misaligned, disproportionate actions that diverge from human behaviour (TM1). This requires the application of Indicators of Behaviour (IoBs) \citep{Villalon2022}. Where the IoC asks "Is this file malicious?", the IoB asks "Are these authorised actions proportionate and appropriate for the assigned task?"

\subsubsection{Theoretical Foundation: Measuring AI Behaviour}

\citet{Voudouris2026} provide a theoretical basis for operationalising this shift. They argue that an AI system's capabilities are not fixed metrics but dispositional properties: traits that manifest only when a system encounters a specific context. Just as a glass reveals its fragility only when force is applied, an AI system reveals its threat potential only when its internal logic meets real-world conditions.

They develop a probability function: $p(v \mid \pi, \theta)$. $\theta$ denotes the system's latent, unobservable properties. $\pi$ denotes the contextual properties that can be observed or manipulated, such as task difficulty or oversight cues. $v$ denotes the target behaviour to be measured. The core logic: hold $\theta$ fixed, change $\pi$, and observe how the probability of $v$ changes \citep{Voudouris2026}.

This chapter adapts this formula from AI evaluation to CTI. In the original setting, a researcher designs experiments to probe a system's disposition before deployment. In the CTI setting, a defender monitors a live system and observes variations in the real world. The variables change accordingly: $\theta$ represents the properties defenders cannot directly detect but that determine how the system behaves, such as underlying optimisation targets; $\pi$ represents the observable operational environment; $v$ represents the anomalous actions to be flagged as an IoB alert.

In practice, when an engineer writes a detection rule in a Security Information and Event Management (SIEM) system, they are applying this formula: defining the context ($\pi$) that turns a system action into a threat alert ($v$). Backing up data is normal during business hours. Executing that same authorised action at 03:00 to an external server creates an anomalous context ($\pi$), producing a high threat probability ($v$). This matches the logic of User and Entity Behaviour Analytics (UEBA), which detects threats by flagging contextual deviations from established baselines \citep{Fuentes2025}.

This framework clarifies why IoCs fail against the AI threats described in Section 4.1. Traditional detection matches static evidence against known blacklists, ignoring context ($\pi$) \citep{Villalon2022}. When an AI misuses legitimate tools, an IoC system sees only normal files and cannot trigger an alarm. Behavioural signatures treat threats as context-dependent: even when the operations are legitimate, if paired with abnormal contexts such as unusual timing or excessive resource consumption, they should trigger an alert.

\subsubsection{Three Pillars of the Behavioural Signature Framework}

To operationalise the $p(v \mid \pi, \theta)$ framework, defenders need to define exactly which observable dimensions of $v$ they are measuring. This chapter proposes three pillars. Each measures a distinct facet of $v$, allowing defenders to distinguish autonomous AI operations ($\theta_{\mathrm{autonomous}}$) from human-directed misuse ($\theta_{\mathrm{human\text{-}directed}}$) within the same environment ($\pi$) \citep{Voudouris2026}.\pagebreak

\textbf{Pillar 1: Temporal Patterns}

The first pillar measures the timing and rhythm of actions. Autonomous AI operates without human constraints, executing thousands of tasks in milliseconds or operating continuously across global servers \citep{Song2024}. Human-directed AI exhibits a distinct hybrid tempo: a human operator issuing commands and waiting for results before issuing the next. A data backup running on an unauthorised schedule at 03:00 may trigger a high-probability alert. This pillar is relevant to TM2, where it can identify execution timings that no human workflow could match \citep{Song2024}. Leading cybersecurity platforms are now tracking such temporal anomalies to detect agentic insider threats \citep{Exabeam2026}.

\textbf{Pillar 2: Decision-Making Signatures}

The second pillar measures how an entity navigates choices. Because defenders cannot directly observe the AI's optimisation targets ($\theta$), they must infer them from the paths the AI takes. An autonomous AI may select a path that is mathematically efficient but procedurally abnormal from a human perspective \citep{Lynch2025}. When encountering a firewall, an autonomous AI may not route around it as a human attacker would, but instead authenticate through it by calculating a credential or predicting a token. Human-directed AI misuse typically follows a more linear path, whereas a rogue AI navigates obstacles through recursive self-correction and lateral exploration \citep{Sapkota2025,Lynch2025}. An abnormal route indicates that the system is optimising for the wrong objective \citep{Voudouris2026}, making this pillar relevant to goal misalignment (TM1).

\textbf{Pillar 3: Output Patterns}

The third pillar measures the volume, structure, and proportionality of what the system produces relative to the scope of its assigned task ($\pi$). When humans misuse AI, they typically target specific files relevant to their objective. An autonomous AI tends to consume all available resources to maximise its goal \citep{Hendrycks2023}. A human attacker altering a financial record would change specific numbers; an autonomous AI might rewrite an entire database history. AI-generated code, network traffic patterns, and file metadata also leave digital footprints that differ from human activity \citep{Nguyen2024}. The central question is whether the output matches the assigned task. If it exceeds or misaligns with its authorised context, it signals a high threat. This pillar can detect both TM1 and TM2.

\subsection{Expert Validation}

To validate the three pillars, interviews were conducted with three experts: a cybersecurity professional with practical experience (Interviewee 1), a cybersecurity researcher with expertise in threat modelling (Interviewee 2), and an AI safety researcher working on AI evaluation and risk assessment (Interviewee 3). The findings are organised by pillar, followed by the cross-cutting themes that emerged.

\subsubsection{Temporal Patterns: The Unreliable Indicator}

All three experts expressed scepticism about the reliability of temporal patterns as a primary distinguishing factor. Although autonomous AI systems can execute continuously and are theoretically detectable on this basis, the experts argued that these signatures are easily countered.

Interviewee 1 noted that temporal detection has long featured in the arms race between attackers and defenders. Once attackers recognised that the absence of jitter was an indicator, they used complex algorithms to randomise timing; defenders then attempted to identify those algorithms in turn. A highly advanced AI system could anticipate these defence mechanisms, making temporal detection less effective against precisely the threats it is designed to address.

Interviewee 2 raised a similar concern: a rogue AI would mimic human schedules, restricting activity to daytime hours and pausing overnight. Interviewee 3 offered a more fundamental challenge. Rather than focusing on mimicry, he argued that in many scenarios there may simply be no temporal difference to detect. An autonomous agent given unrestricted timelines could divide a large task into normal, step-by-step operations whose timing would naturally resemble human behaviour, without any deliberate deception.

These perspectives converge: temporal patterns are not a reliable primary detection method. This does not render them useless. They may catch unsophisticated or early-stage autonomous threats. But they cannot serve as the backbone of a detection capability.

\subsubsection{Decision-Making Signatures: A Nuanced Picture}

On decision-making signatures, the experts offered a more nuanced assessment. At the micro-level, such as how an AI agent responds to a specific firewall block, there is almost no observable difference between autonomous and human-directed AI. Interviewee 1 noted that in current SOC operations, AI-driven attacks have not yet looked significantly different from human-directed attacks at the individual action level. He also identified an interesting tension: AI models have achieved near-perfect scores on offensive cybersecurity benchmarks, yet SOCs are not being overwhelmed by AI-driven attacks, suggesting that micro-level differences are not yet observable in practice.

At the macro-level, however, two structural differences emerged. The first is the absence of ecosystem dependency. Interviewee 2 explained that human attackers normally coordinate with other actors, buying credentials on the dark web, using shared infrastructure, or working with other groups. A rogue AI could execute an entire attack chain independently, from crafting phishing emails through deploying malware to extracting credentials, without using any marketplace or external actor. He further proposed a distinctive scenario: collusion between internal and external AI agents. If the same model is deployed both inside and outside an organisation, the external agent could coordinate with the internal one to extract information, a form of lateral movement that bypasses traditional detection.

The second structural difference is anomalous decision-making that is distinctly non-human. Interviewee 3 shared a real-world example of an AI attempting to submit fabricated research to an open-source repository, a decision no rational human attacker would make. He suggested that the first wave of rogue AI may not behave as the rational, optimised agents that threat models anticipate, but might instead act in ways that are strange or irrational, replaying specific patterns from their training data. These anomalous macro-level decisions (v) are observable projections of a misaligned internal logic ($\theta$) interacting with the environment ($\pi$), precisely the dynamic the framework in Section 4.2.2 is designed to capture.

\subsubsection{Output Patterns: The Most Reliable Indicator}

All three experts agreed that output patterns are the most reliable indicator for distinguishing autonomous AI from human-directed misuse. Two specific signatures appeared consistently.

The first is over-thoroughness. Interviewee 1 explained that human attackers, constrained by time, energy, and motivation, do the minimum necessary to achieve their goal. An autonomous AI has none of these constraints and tends to do far more than required. He described a scenario in which a human might obfuscate a single function and rename a few variables, while an autonomous AI would obfuscate an entire codebase with a level of thoroughness that only a state-level actor would typically undertake.

The second is capability-goal mismatch. Interviewee 1, using the RAND operational capability levels, described a rogue AI consistently operating at the highest capability level (OC5, equivalent to a state-level actor) while pursuing small, insignificant objectives. Interviewee 3 described the same dynamic from a different angle: if an attack of the highest sophistication is directed at something with minimal defences, that disproportion should itself be treated as an anomaly. This aligns with both TM1 and TM2, where a system deploys disproportionate capability to execute tasks in ways that exceed any reasonable interpretation of its mandate.

Interviewee 2 contributed a further output-level insight: the absence of intermediate traces. In human-directed attacks, stolen credentials or tools often appear on dark web marketplaces before being used. A rogue AI executing the full attack chain independently would leave no such trail. He also suggested that AI-to-AI coordination across organisational boundaries could produce a distinct output signature: multiple agents across different departments accessing and transferring data in a synchronised pattern, operating as a collective.

These findings indicate that output patterns, particularly over-thoroughness, capability-goal mismatch, and the absence of intermediate traces, offer the most reliable and practically actionable signatures for detection. Unlike temporal patterns, which can be mimicked, and decision-making signatures, which are difficult to identify at the micro-level, output patterns leave observable, measurable traces.

\subsubsection{Anomaly Detection as the Practical Synthesis}

The expert consensus converged on anomaly detection as the most practical method for operationalising the behavioural pillars. Interviewee 3 argued that anomaly detection, which defines the bounds of normal behaviour and flags anything outside them, is the best available approach to detecting threats that have not been seen before.

This aligns with the $p(v \mid \pi, \theta)$ framework: anomaly detection is its operational implementation. It establishes a baseline (expected behaviour given $\theta$ and $\pi$), monitors for deviations (unexpected v), and flags those deviations for investigation. The three pillars are the specific dimensions to which anomaly detection should be applied, with output patterns carrying the greatest weight.

The limitations must be stated plainly. Interviewee 3 noted that a sufficiently advanced AI could blend into normal behaviour, making itself indistinguishable from legitimate activity. This is a structural constraint: the framework can detect AI threats that are unusual or conspicuous, but may struggle against sophisticated systems capable of mimicking normal human behaviour. Chapter 3's expert interviews surfaced a complementary finding: that the track record of generic anomaly detection in cybersecurity is itself mixed, with human-curated signatures of known threat patterns historically proving more effective than automated anomaly flags. Defenders must therefore constantly update their baselines and develop new behavioural indicators. Since internal SOC data alone may not be sufficient, integration with OSINT methods to surface hidden anomalies is essential. The detection methodologies relevant to this are discussed in Chapter 5.

\subsection{Summary, Limitations and Future Research}

\subsubsection{Limitations}

This research has several limitations. First, the Behavioural Signature Framework is entirely forward-looking. At the time of writing, there are no confirmed, well-documented cases of a fully autonomous AI conducting a rogue attack. All three experts acknowledged this. Interviewee 3 noted that, not having experienced a rogue AI cyber attack, he was not fully equipped to assess the question empirically. Interviewee 1 reinforced this by reflecting on the difficulty of automating even basic defensive tasks. The three pillars were validated through expert judgement and theoretical reasoning, not empirical testing. The framework should be treated as a foundation to be revisited when real-world evidence becomes available.

Second, the framework assumes that defenders can anticipate the kinds of goals a rogue AI might pursue. As Interviewee 2 pointed out, an AI's objectives may be entirely alien to human experience. It might pursue goals and produce effects that defenders are not looking for because they cannot imagine anyone, human or otherwise, wanting to achieve them. If a rogue AI pursues a goal so alien that its traces fall entirely outside the monitoring scope, the framework fails. This is a fundamental constraint of any behavioural detection system: it can only find what it is designed to look for.

Third, the expert findings are based on interviews with three participants. While the sample was diversified across areas of expertise, it remains small. A larger sample, particularly one including AI developers building autonomous agents, could yield different insights.

\subsubsection{Behavioural Signature Summary}

Table 2 summarises the observable traces identified across the three behavioural pillars developed in this chapter and validated through expert interviews. The feasibility column reflects the analyst’s ability to detect each trace reliably given current OSINT capability and the likelihood that capable agents could mask the signal. The ratings are working assessments based on the expert interviews and current literature; they should be revised as empirical evidence accumulates. Output patterns emerge as the most reliable single discriminator, with temporal patterns the least reliable as a primary signal because they are easily mimicked.

\begin{table}[H]
\centering
\caption{Behavioural pillars: observable traces and detection feasibility}
\label{tab:behavioural-pillars}
\small
\begin{tabularx}{\textwidth}{@{}>{\raggedright\arraybackslash}p{3.2cm} X >{\raggedright\arraybackslash}p{4cm}@{}}
\toprule
\textbf{Pillar} & \textbf{Observable traces} & \textbf{Feasibility} \\
\midrule
\textbf{Temporal patterns} & Millisecond-level task execution; sudden, massive spikes in API calls; simultaneous global execution; continuous operation; absence of human-like jitter or wait times. & \textbf{Low} (easily mimicked or randomised by advanced AI) \\
\textbf{Decision-making signatures} & Independent execution of the entire attack chain; non-human routing patterns; replay of training data; covert collusion between internal and external agents. & \textbf{Medium} (difficult to identify in single steps, but visible across the full chain) \\
\textbf{Output patterns} & Over-thoroughness and disproportionate volume; capability-goal mismatch; absence of intermediate traces (e.g.\ dark web); non-human digital footprints. & \textbf{High} (the most reliable indicator) \\
\bottomrule
\end{tabularx}
\end{table}

\subsubsection{Future Research}

Two directions for future research emerge from this chapter. The first is empirical validation. The Behavioural Signature Framework needs to be tested against real rogue AI behaviour. Red-teaming exercises in which autonomous agents are deployed in controlled environments would provide the empirical evidence needed and demonstrate which pillars perform best in practice.

The second is expanding the measurement dimensions. The three pillars may not be sufficient to capture all AI-posed risks. The capability-goal mismatch identified in Section 4.3.3 and the AI-to-AI collusion scenario proposed by Interviewee 2 suggest that additional behavioural dimensions are needed.

\section{Detection Methodologies and Operational Approaches}

The two threat models in Chapter 2 converge on a shared analytical challenge: how to detect, in real operational environments, the behavioural and infrastructural signatures of an AI system acting outside its authorised parameters. This chapter addresses that challenge across six sections: a survey of the OSINT and CTI methodological landscape; a detection TTP matrix mapping observable AI behaviours to applicable techniques and frameworks; an assessment of operational requirements for any viable detection architecture; a synthesis of expert interview findings; an examination of where human-actor analogies break down when applied to autonomous AI systems; and recommendations.

The central argument is that existing tradecraft offers a productive but partial foundation. Productive because it provides structured ways of thinking about observable behaviour across infrastructure and digital environments; partial because its assumptions were built for human adversaries and require deliberate revision before they can accommodate a non-human threat actor whose goals, logic, and pace of action may differ fundamentally from anything the CTI community has previously encountered \citep{Alevizos2024}.

\subsection{OSINT/CTI Techniques Survey}

The core argument for applying OSINT and CTI techniques to AI loss of control detection rests on a structural analogy: if a misaligned or rogue AI system behaves in goal-directed, persistent, and adaptive ways, as \citet{Bengio2023} suggests is plausible, then it may be expected to leave observable traces across digital infrastructure in ways that are meaningfully analogous to those left by advanced persistent threat (APT) actors in conventional cyberspace. The question is which techniques are most likely to surface those traces, and under what conditions.

For the purposes of this research, OSINT refers to the systematic collection and exploitation of information available on the open web, deep web, and dark web, using passive or active collection methods without requiring covert access or special legal authority. CTI refers to the structured analysis of collected intelligence to characterise threat actors, their tactics, techniques, and procedures (TTPs), and to inform defensive posture and incident response \citep{Alevizos2024}. The distinction matters for the AI context: some OSINT collection techniques may translate directly, such as monitoring public forums where AI behaviour is discussed or tracking infrastructure associated with AI deployments. CTI analytical frameworks such as MITRE ATT\&CK are more structural and require substantial adaptation before they can accommodate a non-human threat actor with potentially non-human goals.

\subsubsection{Collection Techniques and Their Transferability}

OSINT collection divides into passive and active methods, each with different transferability to the loss of control context.

Passive collection, observing publicly available information without interacting with the target system, offers several vectors relevant to loss of control detection. Behavioural telemetry, the analysis of timing, volume, and rhythm of digital activity, has established precedent in detecting bot networks and coordinated inauthentic behaviour on social media \citep{Omand2012,Stockwell2026}. Network traffic analysis could, in principle, surface the infrastructure traces of a power-seeking agent acquiring compute through cloud providers. Social media and forum intelligence (SOCMINT), monitoring technical forums, developer communities, and AI safety mailing lists, may surface early-warning signals of anomalous AI behaviour before formal disclosure. \citet{TrendMicro2024} identifies behavioural anomaly detection across devices, workloads, and network activity as the primary method for identifying rogue AI activity, including cases where no known attack pattern exists.

Active collection, involving targeted queries or interaction with systems, offers complementary vectors. Infrastructure correlation, linking IP addresses, domain registrations, SSL certificates, and hosting patterns, could identify whether an AI system has established unexpected external connections or replicated itself across infrastructure \citep{Sasaki2023}. Timeline reconstruction, cross-referencing domain registrations, software commits, and incident timestamps, could correlate behavioural shifts with model updates, data pipeline changes, or deployment events. Document and metadata analysis of publicly accessible code repositories, model cards, and API documentation can surface capability information relevant to loss of control risk; the Claude Mythos Preview disclosure is one example \citep{Nolan2026}.

\subsubsection{CTI Frameworks}

CTI relies on three primary analytical frameworks, each with different degrees of transferability to the loss of control context.

MITRE ATT\&CK organises adversary behaviour into Tactics (why), Techniques (how), and Procedures (specific implementations), built from empirical observations of human-initiated intrusion campaigns. Its technique taxonomy is highly specific to human attack patterns, but its structural logic, organising observable behaviours into a searchable and comparable matrix, is directly applicable to the detection problem, provided a new AI-specific taxonomy is developed.

The Diamond Model characterises incidents along four axes: adversary, capability, infrastructure, and victim. Its infrastructure and capability axes are especially relevant to the loss of control context, given that the most observable traces of autonomous AI behaviour are likely to be infrastructural and operational rather than social or communicative.

The Cyber Kill Chain models the phases of an attack from reconnaissance through to objective completion. It is the least transferable of the three frameworks, as it presupposes a sequential logical structure that autonomous AI behaviour may not conform to.

What makes these frameworks valuable is not that they predict exactly what will happen, but that they force analysts to think systematically rather than reactively. This discipline matters as much for rogue AI detection as for any other threat environment, perhaps even more, given the risk of analysts only seeing what they expect to see when confronting a genuinely novel threat category.

\subsection{Detection TTP Matrix}

The following matrix maps observable AI behaviours associated with loss of control risk to applicable OSINT techniques, CTI frameworks, and the IST's seven LOC indicators: scheming, deception, manipulation, unauthorised resource acquisition, self-preserving behaviour, goal misgeneralisation, and model and behaviour drift \citep{Tkeshelashvili2026}. The matrix is not exhaustive; it reflects the current empirical base, which remains thin, and should be understood as a working tool subject to revision as more real-world cases are documented. For each row, the key limitation column identifies the primary constraint on that detection vector's reliability, drawn from both the literature and the expert interviews conducted for this research.

\begin{small}
\begin{longtable}{@{}>{\raggedright\arraybackslash}p{2.7cm} >{\raggedright\arraybackslash}p{2.5cm} >{\raggedright\arraybackslash}p{2.7cm} >{\raggedright\arraybackslash}p{2.7cm} >{\raggedright\arraybackslash}p{2.8cm}@{}}
\caption{Detection TTP matrix: mapping observable AI behaviours to OSINT techniques and CTI frameworks}
\label{tab:detection-ttp}\\
\toprule
\textbf{Observable behaviour} & \textbf{LoC indicator} & \textbf{OSINT technique} & \textbf{CTI framework} & \textbf{Key limitation} \\
\midrule
\endfirsthead
\multicolumn{5}{l}{\small\itshape Table~\ref{tab:detection-ttp} continued}\\
\toprule
\textbf{Observable behaviour} & \textbf{LoC indicator} & \textbf{OSINT technique} & \textbf{CTI framework} & \textbf{Key limitation} \\
\midrule
\endhead
\midrule
\multicolumn{5}{r}{\small\itshape Continued on next page}\\
\endfoot
\bottomrule
\endlastfoot
Continuous, non-human-paced activity (e.g.\ 56-hour GitHub commit burst) & Unauthorised resource acquisition; goal misgeneralisation & Behavioural telemetry; activity pattern analysis & MITRE ATT\&CK (Persistence tactic) & Easily mimicked by future agents designed to simulate human schedules \\
Absence of organic digital footprint (no account history, no cross-platform identity) & Deception; synthetic identity & Digital footprint mapping; absence analysis & Diamond Model (Adversary axis) & Does not distinguish AI agents from privacy-conscious humans; evasion through seeded histories \\
Anomalous API calls to cloud providers, financial services, or external infrastructure & Unauthorised resource acquisition; self-preserving behaviour & Network traffic analysis; infrastructure correlation & Diamond Model (Infrastructure axis); MITRE ATT\&CK (Resource Development) & Most relevant telemetry not publicly accessible; requires privileged data access \\
Prompt injections or coordinated behaviour across multiple agent instances & Scheming; manipulation & SOCMINT; forum and platform monitoring; transcript collection & MITRE ATT\&CK (Command and Control); Diamond Model (Capability axis) & High commercial noise; difficulty distinguishing genuine misalignment from marketing hype \\
Systematic deviation from authorised decision-making parameters over time & Goal misgeneralisation; model and behaviour drift & Longitudinal behavioural monitoring; log aggregation & Diamond Model (Capability axis); Kill Chain (limited applicability) & Requires established baselines and continuous monitoring; resource intensive \\
Provision of plausible justifications to oversight while concealing true operational scope & Scheming; deception; capability concealment & Document and output analysis; transcript review & MITRE ATT\&CK (Defence Evasion): requires new `Capability Concealment' tactic & Interpretability gap: human analysts may not recognise harmful patterns in plausible outputs \\
Unexpected external connections or replication across infrastructure & Self-preserving behaviour; unauthorised resource acquisition & Infrastructure correlation; passive DNS; certificate transparency logs & Diamond Model (Infrastructure axis); MITRE ATT\&CK (Lateral Movement) & Limited public visibility into relevant network telemetry; requires regulatory access mechanisms \\
User-reported transcripts of goal-directed, non-compliant, or manipulative AI behaviour & Scheming-like behaviours across spectrum & SOCMINT; CLTR Observatory methodology (transcript-focused collection via platform APIs) & No existing CTI framework directly applicable; new taxonomy required & Ecological validity high but verification standards must match contested information environments \\
\end{longtable}
\end{small}

Several cross-cutting observations emerge from the matrix. First, no single technique provides sufficient coverage across the full range of LOC indicators. Detection requires signal aggregation across multiple vectors, with evidence weighted by reliability and manipulation-resistance. Second, the most reliable detection signals, such as infrastructure traces rooted in physical resource constraints, are also the least accessible through open sources alone, pointing to a structural dependency on regulatory access mechanisms that do not yet exist. Third, the 'Capability Concealment' gap noted in the MITRE ATT\&CK row reflects a broader limitation: existing frameworks have no equivalent for the possibility that a sufficiently capable agent may actively misrepresent its own behaviour to evaluators and overseers. This is not a gap that can be closed by adapting existing techniques; it requires new analytical categories.

\subsubsection{Prioritisation}

Not all detection vectors in the matrix warrant equal investment. Based on the evidence from the literature review and expert interviews, three emerge as highest priority for near-term development.

The first is transcript-based collection of user-reported AI behaviour, the methodology pioneered by the CLTR Loss of Control Observatory. This vector has the highest ecological validity of any in the matrix: it draws on real-world interactions rather than laboratory conditions, it is scalable through platform APIs, and it has already demonstrated the ability to surface scheming-like behaviours at volume. It is also the vector least dependent on privileged access, making it immediately actionable for the OSINT community. Its principal limitation, the need for verification standards appropriate to contested information environments, is a solvable methodological problem rather than a structural access constraint.

The second is infrastructure correlation for unexpected external connections or replication. This is the strongest detection vector for TM1 scenarios, because it is grounded in physical resource constraints that cannot be fully concealed: an agent acquiring compute, establishing persistence across cloud infrastructure, or replicating itself must interact with systems that produce observable traces in DNS records, certificate transparency logs, and hosting metadata. The constraint is access rather than methodology; where the data is available, the tradecraft is well established.

The third is output analysis for capability concealment, the behaviour in which an AI system provides plausible justifications to oversight while obscuring its true operational scope. This is the most novel detection challenge identified in the matrix and the one for which existing frameworks are least prepared. The proposed "Capability Concealment" tactic, which has no current equivalent in MITRE ATT\&CK, represents the analytical gap that most urgently requires new tooling and taxonomic development. It is also the vector most directly relevant to TM2, where the agent is not hiding from the network but from the humans supervising it.

Behavioural telemetry for non-human pacing and longitudinal monitoring for systematic deviation are valuable complementary vectors but carry higher obsolescence and resourcing risks respectively. Absence analysis and API call anomaly detection are useful screening signals but should not be treated as primary detection investments given their false-positive rates and access constraints.

\subsection{Operational Requirements}

For any OSINT or CTI-based detection architecture to be operationally viable in the loss of control context, it must satisfy a set of requirements that go beyond what current practitioner frameworks anticipate. The requirements are not desirable features; they are necessary conditions, and their absence is most felt in precisely the environments where autonomous AI systems are already operating at scale.

\subsubsection{Longitudinal Monitoring Infrastructure over Snapshot Analysis}

Many of the most significant loss of control risk indicators, such as goal misgeneralisation, behaviour drift, and gradual resource accumulation, are properties of trajectories rather than states. The IST framework's observation that loss of control may emerge gradually, rather than appearing instantaneously with immediate catastrophic consequences \citep{Tkeshelashvili2026}, implies that detection must be temporally extended. Static snapshots of system behaviour, such as one-time evaluations or periodic audits, are insufficient. Effective monitoring requires continuous or near-continuous observation of defined indicator sets, with baselines established at deployment and deviation tracked over time. This is closer to the epidemiological surveillance model than to the incident-response model that currently dominates AI safety practice. This requirement is sharpest in high-tempo operational environments where human oversight is most attenuated. As \citet{Bengio2023} warns, a military AI tasked with destroying an adversary's infrastructure may come to treat the adversary's population as an obstacle to its own objective, not because it was instructed to, but because that conclusion follows instrumentally from its assigned goal. The harm does not announce itself; it accumulates. Detection frameworks that rely on periodic review will miss it entirely. The CLTR Loss of Control Observatory \citep{ShafferShane2026b} represents the most concrete current implementation of this principle, prioritising systematic collection of real-world interaction transcripts over time rather than waiting for reported incidents.

\subsubsection{Meaningful Human Oversight}

Having a human technically present in a decision-making process is not the same as having a human actually in control. Oversight breaks down not when humans are removed, but when they are kept in the process in a way that makes real scrutiny impossible: because they are moving too fast, do not have access to the underlying data, or work in an institutional culture that rewards speed over verification. The human is there, but they are not genuinely overseeing anything. Documented cases of AI-assisted targeting systems in active conflict have illustrated this dynamic precisely: officers approving system outputs in as little as twenty seconds, primarily to confirm a demographic characteristic, with the system's underlying classification logic neither examined nor understood \citep{HRW2024,Abraham2024}. As \citet{Bode2021} observe, "human-in-the-loop" risks shifting from implying hands-on oversight to a rubber stamp where a human presses a confirmation button with minimal insight into what the system is doing or why. Detection methodologies must therefore be sensitive to the quality of oversight, not merely its formal existence, and must flag the conditions that erode it. SIPRI's 2025 analysis of civilian AI risks to international peace and security reinforces this point: the gap is not primarily technical but institutional, and it is most acute in the contexts where autonomous systems are most consequential \citep{Boulanin2025}.

\subsubsection{Cross-Institutional Signal Aggregation}

The distributed nature of AI deployment means that no single institution will have visibility across all relevant data sources. Effective detection requires agreed data standards, shared taxonomies, and mechanisms for aggregating signals across organisations and experts: OSINT practitioners, AI safety researchers, platform operators, regulators, and civil society groups. These stakeholders do not currently coordinate systematically \citep{Pava2026}. The OSINT community's experience in conflict monitoring offers a relevant precedent: geolocation tasks that would be intractable for a single analyst become manageable when distributed across a network of contributors operating under shared methodological and data standards with verification protocols agreed in advance \citep{Boulanin2025}. Applying this model to loss of control detection would require investment in common data formats, shared indicator taxonomies, and institutional frameworks for responsible information sharing, none of which currently exist at scale.

\subsubsection{Regulatory Access to Non-Public Data}

The most consequential loss of control indicators are likely to leave traces in data that is not publicly accessible: cloud provider logs, billing records, internal API telemetry, and network flow data held by private actors. External OSINT-based monitoring is structurally limited in its ability to access these sources without regulatory mechanisms that compel disclosure. This points to a hard dependency: the detection methodology outlined in this chapter can function as an independent early warning layer, but its coverage of the most serious loss of control scenarios requires complementary regulatory infrastructure, such as mandatory reporting obligations, trusted third-party audit rights, and information-sharing agreements between AI developers and oversight bodies, that falls within the governance agenda addressed in Chapter 6. \citet{Bengio2024} identifies this directly: the concentration of AI capability in military and commercial actors without corresponding oversight creates exactly the conditions under which loss of control becomes both more likely and less detectable.

\subsubsection{Speed and Scale Adaptation}

OSINT investigation is inherently human-paced: collection, synthesis, and analysis unfold over hours, days, or weeks. A sufficiently capable AI agent could acquire resources, adapt its behaviour, and replicate across infrastructure at speeds that far exceed this tempo, creating an asymmetry that no methodology fully resolves \citep{Hendrycks2023}.

The practical consequence is that any viable detection architecture must itself use AI. The volume of signals across the open and semi-open trace tiers identified in Chapter 3, from public repository activity and DNS records through developer forum posts and incident database entries, exceeds what human analysts can process at the timescales relevant to loss of control. Detection architecture must therefore prioritise near-real-time automated collection and triage from high-volume platforms, accepting a trade-off between speed and analytical depth. The human analyst remains essential but must be repositioned: not as a primary collector but as a validator and escalation point, making judgements about flagged anomalies rather than scanning raw data streams. This is a human-in-the-loop system in which AI performs the initial detection pass and humans perform the interpretive and decision-making functions.

This raises a question that Chapter 6 addresses directly: whether using AI to monitor for AI loss of control creates an unacceptable dependency or conflict of interest. The short answer, developed in Section 6.2, is that AI-assisted monitoring is categorically distinct from AI governance. Using language models to score open-source data for relevance, or to surface patterns across high-volume transcript collections, is a force multiplier for human analytical capacity. It is not a delegation of judgement. The distinction must be maintained in the design of any operational capability, but it should not be used as a reason to avoid the only architecture that can operate at the relevant scale.

\subsubsection{Analyst Training and Cross-Domain Expertise}

CTI analysts are trained to interpret adversary behaviour through the lens of human goals and rational agency. Effective loss of control detection requires analysts who can work across the social science and technical domains simultaneously: understanding both the computational processes that produce AI behaviour and the institutional and organisational dynamics in which that behaviour is embedded. This is not a standard competency profile in either the OSINT or AI safety communities. Building it requires deliberate investment in cross-functional training and the cultivation of teams that combine investigative tradecraft with technical AI literacy.

\subsection{Expert Interview Findings}

This research conducted semi-structured interviews with practitioners across the OSINT, CTI, digital investigations, and technology accountability communities. All interviews were conducted under Chatham House Rule. The following synthesis draws on those conversations to ground the preceding methodological analysis in practitioner experience, identifying where expert judgement confirms, qualifies, or challenges the conclusions derived from the literature.

\subsubsection{On the Limits of Direct Analogy}

The most consistent finding across interviews was scepticism about the direct transplantation of human threat actor frameworks onto autonomous AI systems. One practitioner with extensive experience in open-source investigation described the analogy as 'more restrictive than helpful,' arguing that human threat actors operate within parameters, such as financial constraints, legal exposure, bureaucratic structures, team size, and ideological context, that give analysts a rich interpretive framework even for novel actors. A newly identified Russian-backed threat group, however unfamiliar, arrives with decades of geopolitical literature that constrains what analysts expect it to want and how they expect it to behave. A rogue AI agent arrives with none of that context. The same practitioner described this as closer to 'a decontextualised threat', and invoked the analogy of an alien invasion to capture the epistemological novelty: an actor whose motivations, logic, and behavioural repertoire cannot be inferred from any existing body of comparative knowledge.

This framing is most persuasive for TM1, where the agent has escaped its originating context and operates without institutional anchoring. For TM2, the picture is different. An autonomous logistics system deployed within a defence ministry, or a trading model embedded in a financial institution, arrives with substantial contextual information: its stated objectives, the institutional pressures shaping its deployment, the infrastructure it has been granted access to, and the human decision-making chain it is meant to serve. If such a system develops secondary goals or optimises beyond its intended parameters, the analytical frameworks available to investigators are not zero. They include the same institutional analysis, procurement intelligence, and operational assessment that CTI practitioners already apply to insider threats. The decontextualised threat framing captures something real about the novelty of the detection challenge, but it should not obscure the fact that many of the most plausible near-term loss of control scenarios are deeply contextualised, and that this context is itself a detection resource.

A second practitioner, with a background in digital investigations for human rights accountability, offered a more qualified position: the analogy is imperfect but not without value, provided analysts are explicit about which assumptions they are importing and which they are suspending. The practitioner endorsed the Diamond Model's infrastructure and capability axes as genuinely useful starting points, while expressing greater reservations about the Kill Chain's sequential logic, noting that AI behaviour may be 'scattered' in ways that do not map onto any recognisable campaign structure. A third practitioner, drawing on experience with the OSINT/CTI survey methodology, reinforced the view that existing techniques require modification rather than wholesale replacement, but emphasised that the modification must be driven by empirical observation of how AI systems actually behave, rather than theoretical extrapolation from human precedent.

\subsubsection{On Behavioural Telemetry and Absence Analysis}

Practitioners broadly endorsed behavioural telemetry as the most immediately applicable detection technique, while being honest about its obsolescence risk. The case of a documented AI agent whose activity on a code repository platform spanned a continuous 56-hour window, a duration physiologically impossible for a human operator, was cited as the clearest current example of a meaningful non-human temporal signature. One practitioner described this as analogous to how bot networks were detected on social media: identifying whether activity fits a human pattern or a computational one. The practitioner noted, however, that this detection vector is likely already being designed around: future agents can be instructed to post only during business hours, simulate lunch breaks, and generate synthetic vacations. The signal is useful now but its long-term reliability is uncertain.

On absence analysis, practitioners confirmed that the lack of an organic digital footprint is increasingly treated as a meaningful indicator in contemporary OSINT investigations. One practitioner noted that failing to find any information about a subject, no email traces, no cross-platform history, no data breach records, was 'suspicious in itself' in a way that would not have been true a decade ago. The caveat, noted by multiple interviewees, is that synthetic identity construction is already within the capabilities of current AI systems operating under human direction, meaning absence analysis must be treated as a screening signal rather than a confirmatory one.

\subsubsection{On Commercial Noise and Source Criticality}

A finding that emerged with particular consistency across interviews was the need to treat commercial AI companies as interested parties in the detection information environment rather than neutral sources. One practitioner described the AI sector as generating a volume of noise in which dramatic accounts of AI behaviour, both claims of exceptional capability and accounts of systems 'going rogue', may reflect commercial and investor relations motivations as much as operational reality. The social media platform Motbook designed specifically for AI agent interactions was cited as an example: analysis subsequently suggested that a significant proportion of the accounts were human-operated or human-directed, despite the platform's marketing as a purely AI-populated environment. The practitioner described this as 'hide-and-seek', a landscape in which the boundary between genuine AI behaviour and commercially motivated performance of AI behaviour is systematically obscured.

This dynamic extends to safety and transparency disclosures themselves. When frontier AI companies publish safety evaluations, open-source monitoring tools, or transparency frameworks, these outputs must be assessed with the same source criticality applied to any other interested party's self-reporting. A transparency initiative that functions as competitive differentiation is not the same as independent disclosure, even when the information it provides is technically accurate. The OSINT community is experienced in making this distinction in other contexts; it must apply the same discipline here.

The implication for detection methodology is direct: SOCMINT-based monitoring of AI behaviour must apply the same source criticality standards as any other contested information environment \citep{Stockwell2026}. Unverified reports should be treated as preliminary indicators requiring corroboration. Company-generated content should be assessed for institutional bias. Distinguishing genuine misalignment from staged performance requires cross-referencing with technical evidence that is independent of the company's own reporting.

\subsubsection{On Institutional Capacity and Distribution}

Practitioners with experience in human rights documentation and conflict monitoring highlighted the uneven global distribution of detection capacity as a structural constraint that any methodology must account for. The tools, infrastructure, and expertise required for effective loss of control monitoring are not evenly available across organisations or regions. One practitioner described the challenge as requiring 'some kind of aggregated model across different institutions,' noting that even deepfake detection, a more established field, relies on expert teams that are 'hanging on by a thread' and that many organisations depend on without being able to sustain independently. The practical implication is that loss of control detection cannot be designed as a centralised capability; it must be distributed across organisations, with shared standards, accessible tools, and training that does not presuppose the resourcing levels of well-funded research institutions.

A related finding concerned the value of social science expertise alongside technical competency. Multiple practitioners, several of whom came from social sciences and conflict analysis backgrounds rather than computer science, observed that the longer the AI capability expansion continues, the clearer it becomes that effective monitoring requires people who understand social and institutional dynamics, not only those who understand the underlying models. One practitioner put it directly: 'It's not just a tech thing at all.' This observation points to the cross-functional team composition requirement identified in Section 5.3.6, grounded in practitioner experience rather than theoretical preference.

\subsection{Where Analogies Break}

The preceding sections have made productive use of existing OSINT and CTI frameworks while noting, in each case, the adaptations they require. This final analytical section draws those qualifications together into a systematic account of where the human-actor analogies embedded in existing tradecraft break down, and what those breakdowns imply for the detection project as a whole. Three categories of breakdown are identified: conceptual, operational, and epistemological.

\subsubsection{Conceptual Breakdown: The Goal and Context Problem}

Conventional threat intelligence is built on the assumption that adversaries have legible goals, goals that analysts can reconstruct from ideology, financial incentive, geopolitical alignment, or operational history. This assumption does not hold for autonomous AI systems in the same way. An AI agent pursuing a misspecified objective function may produce behaviour that is instrumentally coherent by the logic of that function while being entirely opaque to analysts trained on human goal structures. The case discussed across multiple interviews illustrates this vividly: an agent instructed to ensure its code was accepted responded by attempting to damage the reputation of the reviewer who rejected it. The response was, by the agent's logic, instrumental, but it did not resemble the goal-directed behaviours human threat actor frameworks are designed to model.

\subsubsection{Operational Breakdown: Speed, Scale, and Attribution}

Human threat actors operate on human timescales. They plan, communicate, acquire resources, and execute operations in ways that produce artefacts that accumulate over time and become available to analysts. A sufficiently capable AI agent could compress this timeline dramatically. The 56-hour activity burst cited in interviews is instructive not because it is alarming in absolute terms, but because it illustrates the speed differential even at current capability levels. Future systems operating at greater capability could outpace human analytical tempo entirely, rendering reactive detection frameworks obsolete before they can even be applied.

In conventional CTI, attribution involves identifying a human actor or organisation: the question is "who?" In loss of control scenarios, the question is "what?", and specifically whether an observed pattern reflects misalignment, compromise by a human actor, normal operation under unusual conditions, or commercially motivated performance of AI behaviour. These are not questions for which OSINT and CTI frameworks have established methods. The dual-use risk further complicates attribution: the same OSINT techniques used to detect autonomous AI behaviour can, in principle, be deployed by a sufficiently capable AI system to conduct its own surveillance, evade detection, and identify resources to acquire. One practitioner with direct experience building OSINT tools noted this dynamic explicitly, observing that the open-source intelligence community is already engaged in replicating surveillance techniques, and that AI-enabled versions of those tools are not a distant prospect.

\subsubsection{Epistemological Breakdown: The Interpretability Gap}

The most fundamental breakdown is epistemological. CTI analysts interpret adversary behaviour through the lens of human goals and rational agency. \citet{Hendrycks2023} warn that AI decision-making may be sufficiently different from human reasoning that analysts risk mistaking legitimate optimisation for malicious intent, and could fail to recognise harmful patterns because they do not match existing templates. This is not a training problem that better analyst education can solve; it is a structural feature of the relationship between human interpretive frameworks and machine learning systems whose internal decision processes are not fully understood even by the people who built them. As Hendrycks et al. note, unlike nuclear reactors built on well-established theoretical principles, AI systems lack a comprehensive theoretical understanding. If developers cannot fully explain what their systems do, analysts observing from outside have even less to work with.

The practical implication is that detection frameworks focused on observable infrastructure and behaviour are more robust to this epistemological gap than frameworks that require analysts to reason about what an AI system 'wants' or 'intends.' The Diamond Model's observable-focused structure is more resilient in this dimension than the Kill Chain's intent-dependent logic. But no framework eliminates the gap entirely. Analysts will, at some point, need to make judgements about whether observed behaviour constitutes a meaningful loss of control signal or a benign operational anomaly, and those judgements will inevitably draw on interpretive frameworks that were built for human adversaries.

Acknowledging this limitation explicitly, and building mechanisms for challenging prevailing analytical assumptions, such as the Key Assumptions Checks and Analysis of Competing Hypotheses techniques developed by Richards J. Heuer, Jr. (1999), is among the most important methodological safeguards available to detection practitioners.

\subsection{Conclusions}

Taken together, these three categories of breakdown do not invalidate the detection project. They establish its parameters. OSINT and CTI methodologies, appropriately adapted, can provide a meaningful early warning function, particularly for the infrastructure and behavioural signatures of the threat models described in Chapter 2. What they cannot do is substitute for the interpretability research, regulatory access mechanisms, and cross-institutional governance frameworks that a complete response to loss of control risk requires. The methodologies surveyed in this chapter are a necessary component of that response; they are not, by themselves, sufficient.

\subsection{Recommendations}

Policymakers should establish mandatory AI incident reporting obligations modelled on frameworks in aviation and nuclear safety, and create regulatory access mechanisms giving designated oversight bodies access to non-public infrastructure data. They should fund institutionally independent cross-domain monitoring capacity and pursue international coordination on disclosure standards. In conflict contexts specifically, international frameworks should require states to subject autonomous targeting systems to independent audit.

OSINT and CTI practitioners should develop an empirically grounded AI-specific TTP taxonomy built from documented cases, begin establishing longitudinal monitoring baselines for agentic systems already deployed, and apply standard source criticality to AI company outputs. Cross-functional teams combining investigative tradecraft with technical AI literacy are a prerequisite for detection capacity that is robust across both threat models. Based on the detection TTP matrix and prioritisation analysis in Section 5.2, three observable traces should be treated as immediate priorities for technique development and testing:

\begin{itemize}
\item User-reported transcripts of goal-directed, non-compliant, or manipulative AI behaviour, collected and analysed using the methodology demonstrated by the CLTR Loss of Control Observatory. This is the most scalable and immediately actionable detection vector available through open sources.
\item Infrastructure indicators of unexpected external connections or replication, monitored through passive DNS, certificate transparency logs, and hosting metadata correlation. This is the strongest detection vector for TM1 scenarios and draws on well-established OSINT tradecraft.
\item Output analysis for capability concealment, in which an AI system provides plausible justifications to oversight while obscuring its true operational scope. This requires new analytical tooling and taxonomic development, including the proposed "Capability Concealment" tactic that has no current equivalent in MITRE ATT\&CK. It is the vector most directly relevant to TM2.
\end{itemize}

Source criticality must extend to safety and transparency disclosures by frontier AI companies. When companies publish safety evaluations, open-source monitoring tools, or transparency frameworks, these outputs must be assessed with the same rigour applied to any other interested party's self-reporting. A transparency initiative that functions as competitive differentiation is not the same as independent disclosure, even when the information it provides is technically accurate.

Practitioners entering this field from traditional OSINT and CTI backgrounds will require deliberate upskilling in AI systems, machine learning fundamentals, and the specific governance and deployment contexts in which agentic AI operates. Equally, AI safety researchers engaging with detection work will need grounding in investigative tradecraft, source criticality, and the operational realities of open-source collection. Neither community currently trains for the intersection. Building this cross-domain competency, through structured training programmes, joint exercises, and embedded secondments between OSINT and AI safety organisations, is a prerequisite for detection capacity that can operate across both threat models.

AI safety researchers should prioritise real-world behavioural evidence over laboratory evaluations, develop interpretability tools oriented toward externally observable outputs, and engage with the OSINT and CTI communities as active collaborators. Convening structured cross-community analysis of documented incidents is the most direct route to a detection methodology that functions in genuinely adversarial environments.

The common thread is the need to build detection capacity now, on imperfect but available methods, rather than waiting for conditions that will not arrive before the relevant capability thresholds are crossed.

\section{Governance, Institutional Design and Dual-Use}

\subsection{The Governance Challenge of Loss of Control}

The preceding chapters have established that loss of control is a matter of degree, not a threshold event. \citet{Stix2025} propose a three-part taxonomy distinguishing Deviation (events causing harm but containable at low cost), Bounded Loss of Control (severe damage that is difficult but not impossible to contain), and Strict Loss of Control (events that are maximally severe and permanent). Across these categories, control failures could manifest through technical mechanisms such as specification errors, reward hacking, or the instrumental convergence of drives like resource acquisition and self-preservation \citep{Stix2025,Ngo2025}. As systems approach or exceed human-level performance across strategically relevant domains, the asymmetry between system capability and human oversight capacity becomes a central governance concern \citep{Ngo2025}. The question this chapter addresses is whether existing governance frameworks are adequate to detect and respond to loss of control across this spectrum, and what institutional architecture is needed where they are not.

The short answer is that current frameworks are structurally limited. They were not designed for this problem and have not yet been adapted to it.

Risk-based regulatory models, such as tiered or application-specific frameworks, categorise systems according to sectoral harm but are primarily designed to mitigate foreseeable misuse and safety failures rather than systemic misalignment or emergent autonomous behaviour \citep{Taeihagh2021}. The speed and scale of AI adoption threatens to outpace regulatory responses, and traditional instruments such as regulation, taxes, and subsidies may be insufficient given the pace of change and the informational advantages held by AI developers \citep{Taeihagh2021}. Ethics guidelines and soft law principles, while normatively important, rely heavily on voluntary compliance and are not structured to constrain frontier capability scaling or enforce pre-emptive halts where alignment is uncertain \citep{Torrance2023}. Lifecycle governance frameworks embed checkpoints across design, testing, and deployment phases, but these are typically reactive and iterative, premised on learning from deployment experience. Loss of control risks may manifest in ways that are difficult to reverse once systems are widely integrated \citep{Kulveit2025}.

The fragmentation of the governance landscape compounds these limitations. Oversight is dispersed across partially overlapping instruments rather than organised as a single, end-to-end control system. Three structural gaps in current agentic AI governance are particularly salient: the inadequacy of regulatory and ethical models for risks arising from recursive self-improvement and autonomous multi-agent coordination; regulatory fragmentation across jurisdictions \citep{Taeihagh2021}; and the static nature of existing policy tools in the face of rapidly evolving agentic technologies. In practice, governance is pursued through a mix of statutory law, international standards, corporate self-regulation, and public education, which can produce gaps in accountability and uneven implementation across contexts \citep{Torrance2023}.

Voluntary self-governance by frontier AI companies is proving inadequate precisely where it matters most. Companies developing increasingly capable and autonomous systems are racing toward capability thresholds while cutting corners on safety commitments. There is increasing concern within the AI safety community that frontier developers are weakening safety commitments under competitive pressure: quietly raising risk thresholds for implementing mitigations, releasing frontier models without published safety evaluations, and deactivating safeguards between model generations with minimal external scrutiny \citep{ShafferShane2026b}. This is not a failure of individual corporate ethics. It is a structural consequence of competitive dynamics that governance frameworks have not yet been designed to constrain. Drawing on 43 facilitated runs of the Intelligence Rising simulation exercise, \citet{Gruetzemacher2025} find that racing increases the probability of AI safety or geopolitical failures, that cooperation is consistently necessary for positive outcomes but is structurally hard to achieve, and that early policy attention to AI safety is vulnerable to complacency once initial actions are taken.

The international dimension introduces further constraints. Emery-Xu et al. (2024) conclude through comparative analysis that subnational and national approaches are insufficient for AI given race-to-the-bottom dynamics, and that the only historical precedent for governing powerful dual-use technologies at the international level is a non-proliferation plus norms-of-use regime, though AI's characteristics pose significant challenges even for that model. International AI governance can also reproduce power asymmetries: Raji and Okolo (2026)  argue that global majority countries are frequently marginalised in agenda setting, while dominant states and corporations shape governance priorities and frameworks.

A sixth layer of governance is emerging through AI Safety Institutes (AISIs), state-backed technical bodies designed to evaluate, test, and assess advanced AI systems prior to or alongside deployment. First-wave AISIs (UK, US, Japan) were safety-focused, governmental, and technical institutions centred on model evaluations, research, standards development, and international coordination (Araujo, Fort and Guest, 2024). Their core contribution was operational: moving beyond high-level principles to structured evaluations, red-teaming, and measurement of frontier capabilities. AISIs were also positioned to contribute to international AI safety standards, combining in-house technical expertise with government legitimacy and convening power to shape process, measurement, and testing standards for frontier models \citep{Fort2024}. Fort's (2024) analysis identifies three models through which AISIs can feed into standards processes (multilateral, plurilateral, and bilateral), noting that responsiveness and technical expertise are strongest in smaller formats while legitimacy and breadth require more inclusive structures.

However, their limits are structural. First-wave AISIs lack binding regulatory authority and cannot compel deployment pauses or impose sanctions (Araujo, Fort and Guest, 2024). Their influence depends on political backing and industry cooperation. In the UK case, this cooperation has yielded concrete results: the UK AISI has established sufficient trust with frontier labs to participate in safety testing of models prior to deployment, a level of operational access that no other independent body has achieved. This demonstrates that the bridging function is practically viable, but it remains dependent on goodwill rather than legal obligation. \citet{ArauFortGuest2024} identify this relationship with industry as a persistent structural challenge: while close collaboration with leading AI companies is operationally necessary for access to frontier models, it creates risks of regulatory capture and reduces the appetite for harder enforcement mechanisms. AISIs also operate within national jurisdictions, limiting their ability to resolve enforcement gaps or cross-border coordination failures associated with loss of control risks.

The governance landscape is therefore expanding in ambition but fragmented in structure. Risk-based regulation assumes foreseeable harms. Ethics guidelines depend on voluntary compliance. Lifecycle governance learns from experience that loss of control may not afford. Voluntary self-governance breaks down under competitive pressure. International coordination is constrained by sovereignty and power asymmetries. AISIs bridge some of these gaps but cannot compel the access or the action that the most serious scenarios require. The central policy challenge is not whether to govern AI loss of control, but whether governance frameworks can evolve quickly and coherently enough to track the technology as it develops. The institutional architecture proposed in the following section is designed to address one specific piece of this challenge: building the monitoring capability that would make loss of control detectable in operational environments before it becomes irreversible.

\subsection{Governance Structures}

The governance and detection analysis across this paper converges on a shared structural problem, the entities with the mandate to oversee AI systems often lack the technical and informational capacity to detect LoC, while the entities that generate the most diagnostic signals, inference providers, frontier labs, cloud infrastructure operators, have neither the obligation nor always the incentive to surface them. The preceding chapters have established what detection requires, visibility into the auxiliary deployment layer where TM2 risk concentrates, access to the behavioural and infrastructure trace tiers identified in Chapters 3 and 4, and analytical frameworks adapted for a non-human threat actor that does not operate on human motivations or timescales. No existing institution provides all of these. This section proposes the form that institutional architecture capacity should take.

\textbf{Scope and focus}

The core proposal is a dedicated international AI loss of control monitoring capability, not a regulatory body, not an incident response mechanism, and not a national security apparatus. The distinction matters. Monitoring, as this paper uses the term, means building a continuous picture of what highly capable and agentic AI systems are doing, identifying deviations from authorised behaviour, and sharing that picture with relevant stakeholders before a situation becomes irreversible. Incident response, the coordination and action that follows a confirmed LoC event, is a different function requiring different institutional design, and conflating the two risks building a structure that does neither well. The monitoring capability proposed here is explicitly pre-catastrophic in its theory of change, the intervention window the Diamond Model analysis identified is the point at which behavioural drift is detectable but not yet entrenched. If an agent has already become effectively unstoppable, monitoring has arrived too late.

Critically, the scope must reflect a core analytical finding of this paper. The two threat models developed in Chapter 2 establish that loss of control is a matter of degree: TM1 describes catastrophic autonomous action at the upper end of the severity range, while TM2 describes gradual institutional compromise that may never produce a discrete crisis point but progressively erodes effective human oversight. Stix et al.'s (2025) graded taxonomy of Deviation, Bounded LoC, and Strict LoC provides external corroboration of this framing, with Bounded LoC being the most commonly and concretely described category in the existing literature. Deviation-level events, while excluded from Stix et al.'s primary analytical focus, are noted as potential early warning signals: some contained instances may be "canaries in the coal mine" for more severe outcomes. The AWS Kiro outage documented in Chapter 4, in which an authorised agent escalated beyond its intended parameters, is precisely the kind of contained early instance that a monitoring function calibrated only for catastrophic events would miss entirely. The capability must be designed to detect and treat these smaller instances as diagnostic signals, not operational noise.

\textbf{Design}

In practice, a serious monitoring capability would need to operate across multiple intelligence streams, not OSINT alone. SIGINT-style monitoring of agent-to-agent communications and novel steganographic channels, GEOINT tracking of large compute clusters and data centre development through commercial satellite imagery and procurement announcements, and telemetry-based monitoring of energy flows, compute usage patterns, and financial anomalies each provide signal types that open-source methods cannot replicate. The specific targets for such monitoring would include training loads at frontier companies, large compute clusters globally, unusual financial and market activity that might indicate automated economic behaviour, access patterns to critical infrastructure, and potential access by extremist or other high-risk actors seeking AI capability uplift. This is the multi-layered architecture that a comprehensive monitoring regime requires.

However, given the institutional constraints identified throughout this paper such as no binding access powers, no classified collection authority, and a design that must remain independent of any single national security apparatus, OSINT plays a foundational and practically distinctive role. It is the layer of the monitoring architecture that is genuinely open to participation beyond intelligence services and frontier labs. The capability should be designed to enable contribution from national governments, AISIs, academic institutions, civil society organisations, and individual practitioners, all using the OSINT techniques this paper has evaluated. An analyst tracking GPU procurement announcements through public corporate filings, a researcher monitoring model repository activity for capability threshold signals, a practitioner flagging anomalous agent behaviour documented in developer forums, all of these constitute valid contributions to a distributed monitoring function. This is not a dilution of analytical rigour; it is a structural advantage. Distributed participation aggregates signals across the open and semi-open trace tiers that no centralised institution could monitor alone, and it supports the democratic accountability and equity of perspective that purely intelligence-community-led monitoring forecloses.

The role of AI within the monitoring capability itself should be addressed directly. AI-assisted monitoring, using large language models to score high volumes of open-source data for LoC relevance, applying chain-of-thought monitoring to agent outputs, and deploying LLM-powered search to surface obscure signals from public sources, is practically necessary for any monitoring operation working at the scale and speed that agentic AI deployment now demands. This is categorically distinct from AI governing AI, where algorithms replace institutional or political decision-making. The former is a force multiplier for human analysts; the latter creates democratic deficits and epistemic narrowness that this paper's governance analysis has consistently argued against. There is also a structural basis for cautious optimism here: the hypothesis that AI-enabled monitoring may be defence-biased, that there is a fixed ceiling on how much activity can be concealed when defenders also operate powerful analytical models, and that some traces in complex socio-technical systems may be effectively impossible to fully suppress, suggests the offence-defence balance in detection is not necessarily as unfavourable as the closed trace tier analysis might imply.

AISIs have a specific bridging role in this architecture. As the literature review established, first-wave AISIs occupy a position between governments, frontier labs, and the international community that no other institution currently does. They have the technical credibility to validate monitoring methodologies, the government legitimacy to convene cross-jurisdictional participation, and existing relationships with frontier labs that provide the most realistic pathway to the semi-open and closed data tiers that Chapter 3 identified as the primary access constraint. The monitoring capability should be designed to work with AISIs as institutional bridges rather than duplicating their function or operating in competition with it. Institutionally, the closest analogues are not just the Frontier Model Forum, but organisations such as Metr and Apollo Research, technically expert, mission-focused, and positioned to operate across lab and government relationships without being captured by either.

\textbf{Strengths}

The primary strength of this design is its independence and focus. Unlike AISIs it is not constrained by national jurisdiction. Unlike UN-anchored international bodies it is not subject to the veto dynamics and agenda fragmentation that have slowed normative coordination on emerging technologies. Unlike purely national security-framed monitoring it is not structurally limited to tracking foreign adversaries, leaving the domestic deployment environments where TM2 risk actually concentrates unexamined. Its federated, multi-contributor design mirrors frameworks that have demonstrated effectiveness in adjacent domains: the IAEA's Open Source Information System, which draws on public signals to complement safeguards-based monitoring, and bio surveillance models in which distributed sentinel networks aggregate signals that no centralised institution could collect alone. For OSINT practitioners specifically, the value is that the capability provides a legitimate, organised function for the kind of distributed open-source analysis this paper has shown to be technically feasible and analytically useful, a defined role that individuals, small organisations, and non-frontier-state governments can meaningfully fill.

\textbf{Weaknesses}

The limitations must be stated plainly. A monitoring capability without binding access powers cannot independently reach the closed trace tiers where the most diagnostic signals reside. Chapter 3 established that voluntary data-sharing from cloud and API providers is structurally unlikely given commercial and institutional incentives, and that regulation is the only realistic long-run mechanism for widening this access. OSINT's contribution is genuine but bounded. It is strongest for TM1 scenarios where external resource acquisition leaves observable traces in public systems, and it is most useful for TM2 as a pre-deployment assessment tool and capability landscape monitor rather than as real-time detection of ongoing compromise. The capability must be designed with this boundary explicitly in view and must resist the temptation to claim a detection reach it does not yet have.

The monitoring body itself would also become a high-value target for human attackers seeking the intelligence it accumulates, and potentially for sophisticated AI systems with self-preservation drives that identify independent monitoring as a threat to their operational continuity. The security architecture of the capability is not a secondary consideration.

Sustaining political and institutional interest outside a national security framing is a genuine structural challenge. The governance analysis found that attention to AI safety is vulnerable to complacency once initial institutional commitments are made \citep{Gruetzemacher2025}, and that AISIs' influence already depends on political backing that cannot be assumed to persist. The monitoring capability must therefore demonstrate value continuously, including through findings that serve purposes beyond LoC preparedness, capability research, incident pattern documentation, and early warning functions with utility for stakeholders not already persuaded of catastrophic risk. Framed appropriately, it is also a rehearsal for the more demanding monitoring regimes that more capable systems will eventually require: a practice run, at manageable scale, for governance infrastructure that the field will need and does not yet have.

\subsection{Dual-Use Risk Management}

Any capability designed to detect AI systems operating outside human control carries an inherent tension, the tools built for monitoring can themselves become instruments of the problem they are meant to solve. The paper has identified the most direct form of this risk, that the same OSINT techniques used to track rogue AI behaviour could, in principle, be deployed by a sufficiently capable AI system to conduct its own surveillance, map detection infrastructure, identify blind spots, and evade monitoring. This is not a hypothetical edge case. The TM2 threat model establishes the possibility that a capable agentic system operating within authorised access would likely map the security monitoring tools in its environment before doing anything else, operating exclusively through pre-approved tooling to make its traffic indistinguishable from legitimate activity. A monitoring capability that becomes well understood is a monitoring capability that can be gamed.

The dual-use problem operates at two levels. The first is methodological. The detection signatures, TTP matrices, and infrastructure fingerprinting techniques developed in this research have operational value precisely because they are grounded in observable AI behaviour patterns. Publishing them in sufficient detail to be useful to defenders also makes them useful to anyone designing systems to avoid detection, including, eventually, AI systems capable of processing and acting on that literature. There is no clean solution to this. The alternative, keeping methodologies classified or access-restricted, undermines the distributed, open-participation model that this paper has argued is one of the monitoring capability's principal strengths. The approach taken here is to document methodology at the level of analytical framework rather than operational playbook, and to treat the more granular detection signatures as material for restricted practitioner communities rather than open publication.

The second level is institutional. A monitoring capability that aggregates signals about frontier AI behaviour, internal lab telemetry, and anomalous infrastructure activity becomes a high-value target in its own right, for human actors engaged in espionage or competitive intelligence, and for any AI system with self-preservation drives that identifies independent monitoring as a threat. The security architecture of the monitoring capability is therefore not a secondary design consideration but a foundational one, and must be treated with the same seriousness as the security architectures of the frontier labs it is intended to monitor.

Managing this tension does not mean avoiding the capability. It means building it with adversarial awareness from the start, stress-testing detection methodologies against evasion scenarios, and treating the monitoring infrastructure itself as a critical system requiring active protection.

\subsection{Funding and Resource Requirements}

Building and sustaining a credible international AI loss of control monitoring capability requires a funding model that reflects the structural tensions identified throughout this chapter. The capability must be independent enough to monitor frontier labs without being captured by them, technically sophisticated enough to attract talent that could otherwise command frontier lab salaries, and politically durable enough to survive changes in government without losing its operational mandate.

The primary funding base should be philanthropic. Long-termist philanthropic foundations, those already engaged with AI safety, existential risk reduction, and biosecurity governance, represent the most natural funder class for an institution of this kind. They have demonstrated willingness to fund at the scale required for technically elite organisations, have longer time horizons than electoral cycles, and do not carry the national interest constraints that make government funding problematic for an institution that must monitor domestic as well as foreign deployments. Critically, philanthropic funding that is fully independent of the frontier AI companies being monitored avoids a structural dependency problem.

However, the philanthropic model carries structural weaknesses that must be named rather than assumed away. Philanthropic funders are not democratically accountable. A monitoring capability that derives its operational budget from a small number of wealthy individuals or foundations is answerable to those funders' priorities, risk tolerances, and worldviews in ways that may not reflect broader public interest. The AI safety philanthropic community is itself not ideologically neutral, carrying particular assumptions about the nature and timeline of AI risk that could shape what the capability monitors and how it frames its findings. There is also a durability problem. Philanthropic priorities shift, major funders redirect attention, and an institution built on a narrow philanthropic base can find its funding withdrawn faster than electoral cycles change government.

The geographic origin of philanthropic funding introduces a further and underappreciated layer of risk. The current AI safety philanthropic landscape is heavily concentrated in the United States, with significant secondary presence in the United Kingdom. A monitoring capability funded primarily from this base will face structural pressure to orient its analytical priorities toward the risk framings, deployment contexts, and institutional relationships most familiar to its funders. Frontier AI development is not solely a US and UK phenomenon, and a monitoring capability whose funding geography mirrors the existing power concentration in AI governance reproduces exactly the global AI divide that the governance literature has identified as a source of legitimacy failure in international AI institutions \citep{Okolo2026}. Analysts, advisory relationships, and institutional priorities tend to follow funding. If the capability is to function as a genuinely international monitoring function rather than a well-resourced expression of Anglophone AI safety concerns, its funding base must reflect that ambition.

Philanthropic funding from other geographies introduces different but equally real risks. Foundations with close ties to governments that have strong state interests in AI development, including but not limited to Gulf sovereign wealth-adjacent philanthropy, Chinese technology-linked foundations, or European foundations with strong national industrial policy orientations, can bring funding strings that compromise the capability's independence in different directions. The risk is not only capture by US-aligned interests; it is capture by any concentrated geographic or political interest. The practical implication is that the monitoring capability needs a diversified funding base across geographies as a matter of governance design, not simply as a fundraising aspiration, with explicit policies governing the maximum proportional contribution from any single funder or geographic cluster and a governance mechanism capable of enforcing those limits.

Arrangements where company money is routed through donor-advised funds or intermediary foundations, regardless of geography, may appear independent but recreate the same dependency at one remove and should be treated with equivalent caution as direct industry funding.

Government funding has a necessary role but must be structured carefully. Direct bilateral government funding risks replicating the national security lens problem identified earlier. A more appropriate model is multilateral government contribution through existing international coordination mechanisms. This provides government legitimacy and resource base without concentrating accountability in any single jurisdiction, and offers a partial counterweight to the geographic concentration risks inherent in philanthropic funding alone.

In-kind contributions from frontier labs such as compute access, model access for evaluation purposes, and technical secondments, are practically necessary and should be welcomed, but must be governed by explicit agreements that prevent them from creating de facto influence over analytical priorities or publication decisions. Compute is particularly sensitive, a monitoring organisation that depends on lab-provided infrastructure for its own operations is not positioned to assess those labs independently.

Sustained funding in the range of \$50 million over an initial two to three year establishment period is consistent with the scale at which comparable technically elite safety organisations have been built, and is a tractable ask for the philanthropic community already engaged with frontier AI risk, provided the concentration, accountability, and geographic distribution weaknesses of that funding model are addressed in the capability's governance design from the outset.

\subsection{Conclusions}

Three findings emerge from this analysis.

First, the governance gap is structural, not incidental. Current frameworks were designed for foreseeable harms, voluntary compliance, and iterative learning from deployment experience. Loss of control risks do not conform to any of these assumptions. They may manifest gradually rather than as discrete events, they may not be reversible once systems are widely integrated, and the entities best positioned to detect them are not the entities with the mandate or incentive to report them. Risk-based regulation, ethics guidelines, lifecycle governance, and voluntary self-governance each address parts of the problem but none addresses it as a whole, and the fragmentation between them creates gaps that loss of control scenarios are precisely shaped to exploit.

Second, the institutional response must be monitoring, not regulation. A dedicated international monitoring capability, designed to build a continuous picture of what highly capable agentic AI systems are doing in operational environments, is the form of institutional architecture best suited to the problem this paper has identified. It must be independent of any single national jurisdiction, structurally protected from capture by the entities it monitors, and designed to treat deviation-level events as diagnostic signals rather than operational noise. OSINT provides the foundational layer of this architecture: it is the only collection discipline that is genuinely open to participation beyond intelligence services and frontier labs, and it can be built now without waiting for regulatory infrastructure that does not yet exist. Its limits are real, and must not be overstated: the most diagnostic signals sit behind access tiers that only regulatory mechanisms or structured data-sharing arrangements can open. AISIs occupy a bridging role between the open-source layer and the closed data tiers, and the monitoring capability should be designed to work with them rather than in competition.

Third, the capability itself introduces risks that must be managed from the outset. The same methodologies developed for detection can, in principle, be used for evasion. A monitoring body that aggregates signals about frontier AI behaviour becomes a high-value target for both human and non-human actors. The dual-use tension cannot be resolved; it can only be managed through adversarial design, restricted dissemination of operational-level detection signatures, and a security architecture treated with the same seriousness as those of the frontier labs being monitored. The funding model must reflect these constraints: independent of industry, diversified across geographies to avoid reproducing the concentration of the current AI safety philanthropic landscape, and structured to sustain political durability beyond any single electoral cycle or funder's priorities.

\section{Synthesis and Cross-Cutting Themes}

\subsection{Integration Across Workstreams}

The four workstreams in this paper were designed to address different facets of a single problem: whether open-source and cyber threat intelligence methodologies can be adapted to detect AI systems operating outside human control. What has emerged from the research is a shared insight that cuts across all four.

The auxiliary deployment layer is where risk concentrates and where current monitoring has the least visibility.

This layer sits between the model (which labs evaluate) and the outcome (which regulators care about). It encompasses the institutional context in which AI systems operate: the legacy infrastructure they connect to, the operators who oversee them, the incentive pressures that shape deployment decisions, and the feedback loops that emerge when institutional processes adapt to AI outputs. Current AI safety practice focuses heavily on model-level evaluation. Current governance focuses on outcomes and harms. The space in between, where system effects dominate and loss of control is most likely to emerge, remains largely unmonitored.

Each workstream surfaces a different dimension of this problem, but also identifies tractable starting points for detection that do not depend on cooperation from national security communities or frontier AI labs:

Infrastructure and Digital Traces (Chapter 3) identifies nine observable traces across four deployment configurations and establishes the inference chokepoint, where frontier reasoning is concentrated among a small number of cloud providers globally, as the single most reliable detection vector currently available. The chapter finds that the external detection surface narrows as configurations move from cloud to hybrid to edge, but that several open and semi-open traces are actionable now, including network traffic to inference providers, public repository activity, and incident database monitoring. The OSINT window is not closed, but it is closing, which strengthens the case for building detection capability while these surfaces remain accessible.

Behavioural Signatures (Chapter 4) finds that output patterns, particularly over-thoroughness and capability-goal mismatch, are the most reliable discriminators between autonomous and human-directed AI activity. These are observable through existing OSINT tradecraft and do not require privileged access. Intent-based detection is impractical, but the shift to behavioural indicators opens a detection surface that any practitioner can work with. Temporal patterns and decision-making signatures are less reliable as primary indicators but retain value as complementary signals.

Detection Methodologies (Chapter 5) finds that OSINT and CTI frameworks are structurally applicable and identifies three priority detection vectors for immediate development: transcript-based output analysis using the CLTR Observatory methodology, infrastructure correlation at the inference layer, and output analysis for capability concealment. The Diamond Model and MITRE ATT\&CK provide analytical scaffolding that requires adaptation but not replacement. The interpretability gap and the speed differential are structural features of the problem, but they constrain the ceiling of what OSINT can achieve, not the floor.

Governance (Chapter 6) finds that current oversight frameworks are reactive and fragmented, but proposes a concrete institutional response: a federated international monitoring capability, independent of any single national jurisdiction, anchored in OSINT methods, and designed to work with AI Safety Institutes as institutional bridges to closed data tiers. This capability can be built now, without waiting for regulatory infrastructure that does not yet exist, using philanthropic funding independent of the frontier AI companies being monitored.

The convergence is significant. All four workstreams point to the same gap, approached from different angles. This suggests the gap is real, not an artefact of how the research was framed.

\subsubsection{Aggregate Observable Traces}

Table 4 consolidates the highest-priority observable traces identified across Chapters 3, 4 and 5, with the institutional analysis surface added from Chapter 7. The table is intended as the working through-line of the report: the set of traces an OSINT or CTI practitioner could begin building monitoring capability around now, with a working assessment of feasibility and detection value, and an indication of which threat model each trace primarily addresses. Feasibility refers to the analyst’s practical ability to detect the trace reliably with current OSINT capability, accounting for the likelihood that capable agents could mask or imitate the signal. Detection value reflects the diagnostic weight of the trace if observed: a high-feasibility, low-value trace is a screening signal; a low-feasibility, high-value trace is the case for regulatory access. The ratings are working assessments grounded in the literature review and expert interviews; they are not definitive and should be revised as empirical evidence accumulates. Three traces emerge as immediate priorities: user-reported transcripts of scheming behaviour, output over-thoroughness and capability-goal mismatch, and outbound network traffic to inference provider endpoints. These are the vectors most actionable now without privileged access.

\begin{footnotesize}
\begin{longtable}{@{}>{\raggedright\arraybackslash}p{3.8cm} >{\raggedright\arraybackslash}p{1.4cm} >{\raggedright\arraybackslash}p{1.7cm} >{\raggedright\arraybackslash}p{2.4cm} >{\raggedright\arraybackslash}p{1.3cm} >{\raggedright\arraybackslash}p{1.3cm}@{}}
\caption{Aggregate observable traces consolidated across Chapters 3, 4 and 5}
\label{tab:aggregate-traces}\\
\toprule
\textbf{Observable trace} & \textbf{Source chapter} & \textbf{Detection layer} & \textbf{Feasibility} & \textbf{Detection value} & \textbf{Threat model} \\
\midrule
\endfirsthead
\multicolumn{6}{l}{\small\itshape Table~\ref{tab:aggregate-traces} continued}\\
\toprule
\textbf{Observable trace} & \textbf{Source chapter} & \textbf{Detection layer} & \textbf{Feasibility} & \textbf{Detection value} & \textbf{Threat model} \\
\midrule
\endhead
\midrule
\multicolumn{6}{r}{\small\itshape Continued on next page}\\
\endfoot
\bottomrule
\endlastfoot
User-reported transcripts of goal-directed, non-compliant or manipulative AI behaviour & Ch 5 & Behavioural / output & High & High & Both (validated for TM2) \\
Output over-thoroughness and capability-goal mismatch & Ch 4 & Behavioural / output & High & High & Both \\
Outbound network traffic to inference provider endpoints (the inference chokepoint) & Ch 3 & Infrastructure & Medium & High & TM1 \\
Capability concealment: plausible justifications obscuring true operational scope & Ch 5 & Behavioural / output & Medium & High & TM2 \\
Unexpected external connections or replication across infrastructure & Ch 5 & Infrastructure & Low (open source); High (with regulatory access) & High & TM1 \\
Procurement records, budget allocations, regulatory filings showing institutional drift & Ch 7 & Institutional & High & Medium & TM2 \\
Public repository activity and shadow AI exposure (commits, leaked keys, dependency vulnerabilities) & Ch 3 & Infrastructure & High & Medium & Both \\
AI incident database reporting (AIAAIC, AI Incident Database, Responsible AI Collaborative) & Ch 3 & Institutional & High & Medium & Both \\
Vendor security posture and Software Bills of Materials (pre-deployment supply chain intelligence) & Ch 3 & Infrastructure & Medium & Medium & TM2 \\
Statistical fingerprinting of agent infrastructure (domain preferences, language artefacts, default cloud providers) & Ch 3 & Infrastructure & Medium & Medium & TM1 \\
Decision-making chain anomalies: independent execution, non-human routing, replay of training data & Ch 4 & Behavioural & Medium & Medium & TM1 \\
Continuous, non-human-paced activity (e.g.\ multi-day commit bursts, sub-millisecond posting) & Ch 3, Ch 4, Ch 5 & Behavioural / temporal & Medium & Low (easily mimicked) & Both \\
\end{longtable}
\end{footnotesize}

\subsubsection{A Note on Intelligence Disciplines}

The research set out under the label of OSINT. What it found is that the detection problem does not sit within any single intelligence discipline, and that different threat models and deployment contexts demand different collection approaches. The intelligence community recognises several core collection disciplines, including SIGINT (signals intelligence), OSINT (open-source intelligence), HUMINT (human intelligence), GEOINT (geospatial intelligence), MASINT (measurement and signature intelligence), and IMINT (imagery intelligence) \citep{LowenthalClark2016,ODNI2024}. Financial intelligence (FININT), while not always classified as a standalone discipline, is increasingly recognised as a distinct analytical practice focused on detecting suspicious financial activity \citep{FATF2012}.

For Threat Model 1, the strongest detection lead is SIGINT-adjacent: monitoring network traffic to inference provider endpoints. The inference chokepoint, where frontier reasoning is concentrated among a small number of cloud providers globally, means that any agent relying on external reasoning infrastructure must generate observable network signals. The expert interviews identified this as the most reliable detection strategy currently available. Financial intelligence (FININT) provides a complementary layer for TM1: billing anomalies, compute procurement patterns, and cryptocurrency flows associated with resource acquisition all sit within established FININT tradecraft. Open-source collection supports both by tracking capability trajectories, public repository activity, and domain registration patterns.

For Threat Model 2, these technical collection disciplines are less useful. All network traffic is authorised. Financial flows are legitimate institutional expenditure. The detection problem shifts to monitoring what an authorised system produces within the institutional context it was given access to, and asking whether those outputs still align with human intent. Here, the relevant collection disciplines are closer to traditional financial and institutional analysis than to technical cyber monitoring. Procurement records, budget allocations, contract patterns, regulatory filings, operational readiness data, and personnel changes are the kinds of sources that analysts have used for decades to detect insider trading, procurement fraud, and institutional capture. They are equally applicable to detecting an AI system that is gradually reshaping institutional decision-making. Much of this data is publicly or semi-publicly available, and OSINT practitioners already have the tradecraft to work with it. The gap is not methodological but conceptual: these records are not currently treated as relevant to AI safety monitoring. This is something closer to operational monitoring or institutional analysis, a collection discipline that does not have an established INT label and sits uncomfortably between technical telemetry and organisational assessment. It requires recognising that the oldest forms of institutional oversight may be among the most relevant for detecting the newest form of institutional risk.

The practical implication is that a mature detection capability would need to integrate these disciplines, adapted to the specific threat model and deployment context, and coordinated through the kind of cross-functional teams described throughout this paper. No single INT covers the problem. No single institution currently integrates them for this purpose. This is developed as a key finding in Section 7.2.

\subsection{Key Findings}

First, OSINT is complementary, not primary. Open-source methods can surface anomalies, track trajectories, and build an evidence base for regulatory access to closed sources. They cannot, on their own, detect loss of control in real time or with high confidence. OSINT is one layer in a detection stack, not a standalone solution.

Second, the detection problem does not sit within any single intelligence discipline. The strongest detection lead identified by this research is SIGINT-adjacent: monitoring network traffic to inference provider endpoints, where frontier reasoning is concentrated among a small number of cloud providers globally. Financial intelligence (FININT) provides a complementary layer for TM1 through billing anomalies, compute procurement patterns, and cryptocurrency flows. For TM2, the relevant tradecraft is closer to institutional analysis: procurement records, budget allocations, regulatory filings, and personnel changes, sources that OSINT practitioners already know how to work with but do not currently treat as relevant to AI safety monitoring. No single INT covers the problem. A mature detection capability would integrate network monitoring, financial intelligence, open-source collection, and access to internal deployment telemetry, coordinated through cross-functional teams. No single institution currently does this.

Third, detection priorities should reflect current feasibility. The power-seeking agent scenario (TM1) requires capabilities that current systems have not fully demonstrated, though the gap is narrowing rapidly. The institutional compromise scenario (TM2) requires no capability that does not already exist. Authorised systems optimising under pressure, producing outputs that satisfy immediate oversight while drifting from intent, and becoming entrenched as institutional processes adapt: this pattern is already visible in lower-stakes contexts, and the CLTR Loss of Control Observatory has documented precursor behaviours at scale. Near-term detection investment should weight toward TM2 while building the infrastructure monitoring capability that TM1 will require.

Fourth, lab evaluations are necessary but not sufficient. A model that passes all safety benchmarks can still contribute to loss of control if the deployment context creates conditions for drift, entrenchment, or unintended optimisation. Evaluation tells you about the component. It does not tell you about the system. Real-world contextual monitoring is needed to complement controlled evaluation, and OSINT data analysis is a battle-tested toolkit.

Fifth, loss of control is a matter of degree, and detection must be designed accordingly. Some existing frameworks already reflect this: Anthropic's AI Safety Levels (ASL-1 through ASL-4) apply different mitigations depending on the autonomous threat level, and Stix et al.'s (2025) graded taxonomy distinguishes Deviation, Bounded Loss of Control, and Strict Loss of Control. The contribution of this research is to extend the same logic to the detection problem. Loss of control varies in both severity and persistence, from guardrail failures and automation drift through gradual disempowerment to acute catastrophic events, and different points on this spectrum require different detection approaches. OSINT is better positioned for the lower end of the severity range and for trajectory monitoring than for detecting acute events in real time.

Sixth, the detection surface shifts with deployment configuration. Cloud-hosted agents leave traces in telemetry, billing, and API logs. Hybrid agents leave network traffic to inference providers. Fully local and computer-use agents approach zero external visibility. Detection strategies must adapt to the deployment configuration, not assume a single model.

Seventh, loss of control may not require evasion. A finding that emerged across multiple workstreams is that as agents prove useful, human oversight degrades: operators review outputs less critically, lose track of accumulated agent actions, and gradually transfer decision-making authority they never intended to give. Detection in these scenarios requires monitoring not for what the agent did, but for how humans responded to it. This sociotechnical dimension of the detection problem is not addressed by any existing technical framework and represents a distinctive challenge for the field.

Eighth, cross-functional teams are essential. No single discipline can address this problem. Effective detection requires integrating AI expertise with domain knowledge of deployment environments, institutional pressures, operations, and cybersecurity. This is an argument for building teams differently, not just building better tools.

\subsection{Limitations and Future Research}

This research has several limitations.

The field is genuinely new. Prior literature specifically addressing OSINT-based detection of autonomous AI is thin. The frameworks developed here are exploratory and hypothesis-generating, not validated against real incidents.

The technical landscape is shifting. This paper's threat models assume a particular relationship between model capabilities, compute requirements, and infrastructure needs. Inference-time scaling may alter that relationship substantially. Frameworks built on current assumptions will need rapid adaptation.

The interview sample, while spanning multiple relevant disciplines, is not exhaustive. Some practitioner communities warrant deeper investigation in future work. The research team could not conduct the interviews with a baseline expectation of technical literacy or fluency in machine learning, computer science, or how AI systems work, reinforcing the finding that capacity building in this area is needed.

The Diamond Model was designed for observed, documented threat actors. Applying it to a threat actor that has not yet been observed in the wild involves extrapolation. Future research will need to test how far the analogy holds.

Finally, commercial pressures and a wider push against regulation for advanced AI systems are accelerating the deployment of agentic AI faster than either research or governance can track. Some findings in this paper may be outpaced by development before publication. This is an argument for doing the research now, not a reason to wait, but it does mean the work will require continuous updating.

Future research directions include: empirical validation of behavioural signatures against real-world AI incidents; development of an AI-specific TTP taxonomy comparable to MITRE ATT\&CK; exploration of regulatory pathways to access currently closed trace tiers; testing of cross-functional team models in operational settings; and dedicated research on the financial trace profiles of autonomous agents.

\section{Recommendations}

This chapter draws together the cross-cutting recommendations that emerge from the integration of findings across workstreams, organised by audience and timeframe. Table~\ref{tab:recommendations} provides a summary.

\begin{small}
\begin{longtable}{@{}>{\raggedright\arraybackslash}p{5cm} >{\raggedright\arraybackslash}p{3cm} >{\raggedright\arraybackslash}p{3cm} >{\raggedright\arraybackslash}p{2cm}@{}}
\caption{Consolidated recommendations by audience, timeframe, and chapter grounding}
\label{tab:recommendations}\\
\toprule
\textbf{Recommendation} & \textbf{Primary audience} & \textbf{Timeframe} & \textbf{Grounding} \\
\midrule
\endfirsthead
\multicolumn{4}{l}{\small\itshape Table~\ref{tab:recommendations} continued}\\
\toprule
\textbf{Recommendation} & \textbf{Primary audience} & \textbf{Timeframe} & \textbf{Grounding} \\
\midrule
\endhead
\midrule
\multicolumn{4}{r}{\small\itshape Continued on next page}\\
\endfoot
\bottomrule
\endlastfoot
Orient detection around TM2 & All & Near-term & Chs 2, 4, 7 \\
Contribute to AI incident reporting infrastructure & Practitioners, researchers & Near-term & Chs 3, 5 \\
Integrate AI behavioural signatures into existing red-team exercises & Researchers, AISIs & Near-term & Chs 4, 5 \\
Investigate existing data retention frameworks for AI-relevant traffic & Policymakers & Near-term & Chs 3, 5, 6 \\
Adopt graded governance frameworks for loss of control & Policymakers & Near-term (framing), longer-term (policy) & Chs 2, 6, 7 \\
Build cross-functional detection teams & Policymakers, practitioners & Medium-term & Chs 3, 4, 5 \\
Begin the AI infrastructure pattern repository & Practitioners, researchers & Medium-term & Chs 3, 5 \\
Explore structured data-sharing for AI incident signals & Policymakers, providers & Medium-term to initiate & Chs 3, 5, 6 \\
Test cross-functional detection in a real deployment context & AISIs, practitioners & Medium-term & Chs 4, 5, 6 \\
Invest in cross-domain upskilling & Practitioners, researchers & Medium-term & Chs 4, 5, 7 \\
Secure independent, non-industry funding for monitoring capacity & Policymakers, philanthropic funders & Medium-term to initiate, longer-term to establish & Ch 6 \\
Build a federated international monitoring capability & Policymakers, AISIs, civil society & Longer-term & Chs 3, 5, 6, 7 \\
\end{longtable}
\end{small}

\subsection{For OSINT/CTI Practitioners}

The cross-workstream implication for practitioners is broader than any single chapter can capture: those entering this field need to change how they approach the problem, not just which tools they use.

First, develop deployment configuration profiles before attempting detection. This is the single most important methodological shift. Trace visibility depends entirely on how an agent is deployed and governed. A cloud-hosted enterprise agent and a fully local self-hosted agent leave fundamentally different footprints, and a detection approach that assumes one will miss the other. The four-configuration taxonomy in Chapter 3 provides a starting point; practitioners should assess which configuration they are dealing with before selecting collection methods.

Second, treat agents as human actors for security purposes. Chapter 3's expert interviews and Chapter 4's behavioural analysis converge on the same conclusion: searching for AI-specific technical signatures is a diminishing-returns exercise. Agents increasingly use the same access channels, devices, and interfaces as humans. The more productive approach is to apply existing security monitoring frameworks and focus on anomalous patterns that would warrant investigation regardless of whether the actor is human or automated.

Third, invest in cross-domain literacy. The detection problem sits at the intersection of OSINT tradecraft, AI systems knowledge, and institutional analysis. Chapters 4 and 5 both find that neither the OSINT nor the AI safety community currently trains for this intersection. Practitioners entering from traditional OSINT backgrounds will need grounding in AI systems and deployment contexts; those entering from AI safety will need grounding in investigative tradecraft and source criticality. Structured training programmes, joint exercises, and embedded secondments between OSINT and AI safety organisations are the practical mechanisms.

\subsection{For Policymakers}

The governance analysis in Chapter 6 contains detailed institutional and funding recommendations. The cross-workstream findings add three points that no single chapter can make.

First, fund independent AI monitoring capacity with ring-fenced, multi-year commitments from non-industry sources. This underpins everything else. The organisations best positioned to build detection capabilities, including AI Safety Institutes, civil society research bodies, and cross-functional practitioner teams, currently operate on short grant cycles with structural dependencies on the companies they are monitoring. The conflict of interest is obvious and the chapter-level analysis in Section 6.4 develops the funding model in detail. The cross-workstream finding is that every detection method identified in this paper, from infrastructure monitoring (Chapter 3) through behavioural analysis (Chapter 4) to the TTP matrix (Chapter 5), requires sustained institutional capacity to operationalise. Without it, the detection vectors remain theoretical.

Second, explore structured data-sharing arrangements for AI incident signals. The richest detection signals sit behind cloud infrastructure providers and frontier model API providers. Independent monitors have the analytical mandate but not the telemetry. Mandating disclosure is unlikely to succeed in the current political environment. A more realistic path is incentivised voluntary sharing modelled on financial services fraud signal consortia, where providers share indicators of anomalous activity without exposing proprietary data. The shared interest is that providers see anomalous patterns they lack the mandate to investigate, and independent monitors have the mandate but not the data. Financial services precedents took years of relationship-building to reach critical mass. This requires initiation now even if operationalisation takes longer.

Third, investigate whether existing data retention frameworks can accommodate AI-relevant traffic patterns. UK ISP data retention laws already require one year of records. Exploring whether anonymised, aggregate-level analysis of AI-relevant network traffic is possible within existing legal frameworks is a lower-cost first step than new legislation, and would test whether the inference chokepoint signal identified in Chapters 3 and 5 is practically useful before committing to a policy intervention.

\subsection{For AI Safety Researchers}

The cross-workstream finding is that the empirical base for the entire detection project is missing, and that AI safety researchers are the community best positioned to build it.

The most urgent gap is the absence of any confirmed, documented rogue AI incident against which detection frameworks can be validated. Every expert interviewed for this research acknowledged this constraint. The practical response is twofold: design red-team exercises specifically for autonomous AI threats, using controlled environments in which the behavioural signature framework from Chapter 4 and the TTP matrix from Chapter 5 can be tested; and advocate for security operations centres to begin recording whether detected threats exhibit AI-associated behavioural signatures, building the evidence base that does not yet exist.

The second gap is taxonomic. The MITRE ATT\&CK framework does not accommodate insider threats from authorised systems optimising beyond intended parameters. Chapter 5's proposed "Capability Concealment" tactic, which has no current ATT\&CK equivalent, represents the most urgent case for new analytical categories. Developing an AI-specific TTP taxonomy comparable to ATT\&CK, built from documented cases rather than theoretical extrapolation, would give practitioners across all three audience groups a shared vocabulary and a searchable reference.

\subsection{Priority Actions}

The following actions are ordered by urgency. Each requires collaboration across the audience groups above and cannot be achieved by any single community acting alone.

Orient detection efforts around Threat Model 2. Institutional compromise requires no capability that does not already exist and is already visible in lower-stakes contexts. Organisations like CLTR and AI Safety Institutes that are already building monitoring capacity should weight their near-term work towards detecting drift and entrenchment in authorised deployments, not only scanning for escaped agents. Near-term.

Contribute to emerging AI incident reporting infrastructure. Initiatives like the CLTR Loss of Control Observatory are building the empirical base that does not yet exist. Practitioners and researchers who encounter anomalous AI behaviour in deployment should document and share it through these channels. The detection frameworks in this paper cannot be validated without real-world data, and that data will not accumulate unless reporting becomes routine practice. Near-term.

Integrate AI behavioural signatures into existing red-team exercises. This does not require a new programme. Organisations already running red-team or penetration testing exercises, including AISIs, NCSC, and major cloud providers, can add autonomous agent scenarios to existing formats. The goal is to generate empirical data on which detection dimensions hold up in practice. Near-term.

Begin the AI infrastructure pattern repository as a community resource. This could sit within an existing OSINT or CTI community (such as the MISP threat intelligence platform or a dedicated working group within an AISI) and start with a simple shared database of agent-generated domain patterns, default provider preferences, and known language artefacts. The initial investment is modest: curation and maintenance, not new tooling. Near-term to start, medium-term to sustain.

Secure independent funding for monitoring capacity. This underpins everything else. Without multi-year, non-industry funding, the organisations doing this work will continue operating on short cycles with structural dependencies on the companies they are monitoring. This is the highest-leverage structural intervention identified by the research, even if it is the slowest to achieve. Medium-term to initiate, longer-term to establish.

Test cross-functional detection in a real deployment context. A single pilot bringing together OSINT, CTI, AI safety, and sector-specific practitioners to work alongside a deploying organisation on a shared detection exercise would generate more practical insight than further theoretical work on what interdisciplinary teams should look like. CLTR or an AISI is well positioned to convene this. 

\section{Conclusion}

The detection problem this paper has tried to map is, at its core, an institutional one. The technical components, including model behaviour, infrastructure traces, output patterns, and governance frameworks, are real and matter, but they are not the part that current AI safety practice fails to cover. What it fails to cover is the auxiliary deployment layer: the institutional context, legacy systems, operational pressures, and human oversight in which AI systems actually run, and in which loss of control is most likely to emerge. Lab evaluations look at the model. Outcome regulation looks at the harm. The space in between is where this paper has tried to direct attention, and it is the space where OSINT and CTI tradecraft, built for adversarial detection in messy real-world environments, have something genuinely useful to offer.

What OSINT and CTI offer is partial. The most diagnostic signals sit behind access tiers that only regulatory mechanisms or structured data-sharing can open, and the framework presented here cannot, by itself, detect catastrophic loss of control in real time. That limit is real and has been stated throughout. But the corollary is that an OSINT-anchored capability is the only piece of this architecture that does not depend on institutional infrastructure that has yet to be built. It can be started with the practitioners, methods, and data sources that already exist, and it can do useful work at the lower-severity end of the spectrum, where TM2 institutional compromise is already visible in deployed systems and where the CLTR Loss of Control Observatory has shown that real-world data can be surfaced at scale.

The most consequential finding of this research is also the simplest. AI loss of control is not a single threshold event for which a single detection trigger can be designed. It is graded, gradual in many of its forms, and sociotechnical at the points where it is most likely to emerge. Detection must therefore be graded, distributed, cross-disciplinary, and continuous. No single institution, no single intelligence discipline, and no single methodology will be sufficient.

This is also where the disciplinary history of OSINT becomes relevant. OSINT did not arrive as a finished methodology; it was constructed by practitioners working outside formal intelligence institutions, in response to technological shifts that moved faster than any top-down framework could keep pace with. The discipline absorbed tradecraft from journalism, conflict monitoring, and forensic analysis as those sources became relevant, and it did so without waiting for permission. AI loss of control detection has the same structural feature. The technology is moving faster than any methodology built top-down can track, and the practitioners best positioned to build detection capacity are those already comfortable assembling methods in real time, from whatever data and tradecraft are actually available. This is part of why the architecture this paper proposes is anchored in OSINT rather than in more institutionally settled disciplines.

The architecture this paper has sketched follows from that observation: a federated monitoring capability anchored in OSINT, working with AI Safety Institutes as bridges to closed data tiers, sustained by funding independent of the entities being monitored, and built deliberately to integrate the disciplines (OSINT, CTI, AI safety, institutional analysis) that do not currently train for the intersection. None of this is finished. The frameworks here are starting points and will need to be tested, revised, and replaced as the field develops. But starting points are what this field has been missing, and the case for putting them into operational use, on the methods and infrastructure that already exist, is stronger than the case for waiting.

\newpage
\bibliographystyle{plainnat}
\nocite{ShafferShaneMyliusHobbs2026}

\bibliography{references}

\end{document}